\documentclass[twocolumn]{aastex631}
\usepackage[utf8]{inputenc}
\usepackage{subfigure}
\usepackage{enumerate}
\usepackage{float}

\begin{document}


\title{Evidence of Ultra-faint Radio Frequency Interference in Deep 21~cm Epoch of Reionization Power Spectra with the Murchison Widefield Array}

\author{Michael J. Wilensky}
\affiliation{Jodrell Bank Centre for Astrophysics, University of Manchester, Manchester M13 9PL, UK}
\affiliation{Astronomy Unit, Queen Mary University of London, Mile End Road, London, E1 4NS, United Kingdom}
\affiliation{Department of Physics, University of Washington, Seattle, WA 98195, USA}
\correspondingauthor{michael.wilensky@manchester.ac.uk}

\author{Miguel F. Morales}
\affiliation{Department of Physics, University of Washington, Seattle, WA 98195, USA}
\affiliation{Dark Universe Science Center, University of Washington, Seattle, 98195, USA}

\author{Bryna J. Hazelton}
\affiliation{Department of Physics, University of Washington, Seattle, WA 98195, USA}
\affiliation{eScience Institute, University of Washington, Seattle, WA 98195, USA}

\author{Pyxie L. Star}
\affiliation{Department of Physics, University of Washington, Seattle, WA 98195, USA}

\author{Nichole Barry}
\affiliation{International Centre for Radio Astronomy Research, Curtin University, Perth, WA 6845, Australia}
\affiliation{ARC Centre of Excellence for All Sky Astrophysics in 3 Dimensions (ASTRO 3D)}

\author{Ruby Byrne}
\affiliation{Astronomy Department, California Institute of Technology, 1200 E California Blvd, Pasadena, CA, 91125, USA}
\affiliation{Department of Physics, University of Washington, Seattle, WA 98195, USA}

\author{C. H. Jordan}
\affiliation{International Centre for Radio Astronomy Research, Curtin University, Perth, WA 6845, Australia}
\affiliation{ARC Centre of Excellence for All Sky Astrophysics in 3 Dimensions (ASTRO 3D)}

\author{Daniel C. Jacobs}
\affiliation{Arizona State University, School of Earth and Space Exploration, Tempe, AZ, 85008, USA}

\author{Jonathan C. Pober}
\affiliation{Department of Physics, Brown University, Providence, RI 02912, USA}

\author{C. M. Trott}
\affiliation{International Centre for Radio Astronomy Research, Curtin University, Perth, WA 6845, Australia}
\affiliation{ARC Centre of Excellence for All Sky Astrophysics in 3 Dimensions (ASTRO 3D)}

\begin{abstract}
    We present deep upper limits from the 2014 Murchison Widefield Array (MWA) Phase I observing season, with a particular emphasis on identifying the spectral fingerprints of extremely faint radio frequency interference (RFI) contamination in the 21~cm power spectra (PS). After meticulous RFI excision involving a combination of the \textsc{SSINS} RFI flagger and a series of PS-based jackknife tests, our lowest upper limit on the Epoch of Reionization (EoR) 21~cm PS signal is $\Delta^2 \leq 1.61\cdot10^4 \text{ mK}^2$ at $k=0.258\text{ h Mpc}^{-1}$ at a redshift  of 6.5 using 14.7 hours of data. By leveraging our understanding of how even fainter RFI is likely to contaminate the EoR PS, we are able to identify ultra-faint RFI signals in the cylindrical PS. Surprisingly this signature is most obvious in PS formed with less than an hour of data, but is potentially subdominant to other systematics in multiple-hour integrations. Since the total RFI budget in a PS detection is quite strict, this nontrivial integration behavior suggests a need to more realistically model coherently integrated ultra-faint RFI in PS measurements so that its potential contribution to a future detection can be diagnosed.
\end{abstract}

\section{Introduction}
\label{sec:intro}

The Epoch of Reionization (EoR) is one of few periods in the Universe's history that is still largely unconstrained. Since the neutral Hydrogen 21-cm line can directly track the remaining neutral Hydrogen at a given redshift, high redshift intensity maps of 21-cm radiation will allow us to directly map the reionization process. For reviews, see \citet{Furl2006}, \cite{Morales2010}, and \citet{Pritchard2012}. A first step towards understanding cosmic reionization is to measure the power spectrum of redshifted 21-cm brightness temperature fluctuations, which is a particularly natural task for radio interferometers. Examples of interferometers that have produced upper limits on the reionization power spectrum signal are the Giant Metrewave Radio Telescope (GMRT; \citet{Paciga2013}), the LOw Frequency ARray (LOFAR; \citet{vanHaarlem2013}), the Murchison Widefield Array Phase I (MWA Phase I; \citet{Tingay2013}) and Phase II (MWA Phase II; \citet{Wayth2018}), the Precision Array for Probing the Epoch of Reionization (PAPER; \citet{Parsons2010}), and the Hydrogen Epoch of Reionization Array (HERA; \citet{DeBoer2017}). Lessons learned from analysis of early results have been specifically incorporated into newer generations of EoR-focused telescopes, such as the Hydrogen Epoch of Reionization Array, and the MWA Phase II. As the knowledge of 21-cm power spectrum analysis has grown, upper limits at various redshifts have gradually improved \citep{Beardsley2016, Patil2017, Barry2019b, Li2019, Mertens2020, Trott2020, Rahimi2021, HERA2022, HERA2022b}. 

Theoretical models of the reionization signal suggest that it is 4-5 orders of magnitude fainter than typical astrophysical radio sources. Due to the extreme dynamic range requirements of these experiments, a myriad of data analysis challenges exist. Most prominently, power spectrum analyses require exquisite understanding of the spectral variations in the data in order to make a high quality measurement. Such variations have numerous origins: incomplete or inaccurate calibration models \citep{Barry2016, Patil2016, Ewall-Wice2017, Byrne2019, Dillon2020, Byrne2021}, intrinsic instrument bandpass response \citep{Ewall-Wice2016b, Trott2016, Barry2019a, Li2019, Fagnoni2021}, internal cable reflections \citep{Ewall-Wice2016a, Kern2019, Kern2020a}, and digital nonlinearities \citep{Benkevitch2016}. Additionally, failing to model refraction of incoming radiation through the ionosphere can significantly disrupt measurements relying solely on direction-independent calibration \citep{Morales2009, Jordan2017, Trott2018}. These effects can be mitigated using direction-dependent calibration techniques \citep{Vedantham2016, Hurley-Walker2018}, however such techniques can sometimes induce signal loss \citep{Sardarabadi2019, Mevius2022}. 

Even if other instrumental effects can be handled, an interferometer is naturally chromatic because interferometric baselines measure different angular wave modes at different frequencies. This gives rise to a well-studied wedge-shaped contamination in cylindrical power spectra known as the foreground wedge \citep{Datta2010, Trott2012, Morales2012, Vedantham2012, Hazelton2013}. Generally the spectral resolution of typical reionization-focused instruments allows for sampling of larger line-of-sight wave modes compared to perpendicular wave modes, where sampling of the latter is determined by baseline lengths and orientations. Consequently, there is only a narrow region in this space that is both outside the wedge and accessible by typical baselines that is free of foreground contamination. However, systematic effects may still contaminate it. This area is known as the EoR window \citep{Liu2014a, Liu2014b}.

An exhaustive review of systematic effects is outside the scope of this work. We refer the reader to the comprehensive review of data analysis in 21-cm cosmology presented in \citet{Liu2020}. The purpose of this work is to present an upper limit of the cosmological 21-cm power spectrum signal derived from the second observing season of the MWA Phase I. Of primary importance in the analysis leading to this limit was the nature of a particular systematic effect ubiquitous to all radio experiments called radio frequency interference (RFI), considered here as any non-astrophysical radio signal that is observed by the telescope. Typical RFI sources in EoR and cosmic dawn experiments are anthropogenic, such as frequency modulated (FM) radio broadcasts, digital television broadcasts (DTV; \citet{Wilensky2019}), and ORBCOMM satellite transmissions \citep{Line2018}. In concept, RFI poses a unique threat to EoR measurements in that most sources demonstrate sharp spectral variation over the band used for an EoR power spectrum measurement. \citet{Wilensky2020} found that the contamination in the power spectrum produced by as little as 1 mJy apparent RFI flux can be comparable to the EoR signal on some wave modes. Bright RFI can be easily identified and removed in the raw visibilities. This process is called ``flagging." A profusion of flagging algorithms exist in the radio astronomy literature. Two that have been applied to MWA phase I data are \textsc{AOFlagger} \citep{Offringa2015} and \textsc{SSINS} \citep{Wilensky2019}. In \citet{Barry2019b}, it was found that removing observations identified by \textsc{SSINS} to contain RFI that went undetected by previous excision methods made a measurable difference in the power spectrum upper limit, particularly for the most highly affected observations. 

In our analysis, we find it valuable to distinguish between different types of RFI based on the level of sensitivity required to statistically identify it. We separate it into four categories:

\begin{enumerate}[I]
    \item \textbf{Visibility level}: RFI that can be detected in the visibilities of a single baseline, where most 21-cm EoR analyses do their flagging.
    \item \textbf{Full array}: RFI that requires the sensitivity of the full array (combined coherently or incoherently). More analyses are moving towards flagging at this level, e.g. \citet{Barry2019b, Li2019, Rahimi2021}, \citet{HERA2022, HERA2022b}, and this one. The \textsc{SSINS} algorithm, used in the aforementioned MWA limits and this work, is designed to find full array RFI.\footnote{In previous literature where \textsc{SSINS} has been used, its class of RFI excision capability has been described as ``ultra-faint," whereas here we use the term ultra-faint to refer to RFI fainter than what \textsc{SSINS} detects.}
    \item \textbf{Ultra-faint}: RFI that requires coherent averaging of some significant subset of data in consideration for an upper limit, but significantly less than the entire volume of data. It may also require the full array sensitivity in order to detect it. In this work we observe this in the Fourier domain, i.e. in power spectra, when coherently combining approximately 2-5\% of the full analysis data set.
    \item \textbf{Unexcisable}: RFI that is only detectable when coherently combining an extremely large subset of the data in a full-array manner, such that exclusion of the affected data would significantly alter the sensitivity of the analysis (e.g. half the analysis volume). We propose this type of RFI as a distinct possibility that has not been ruled out, but do not claim to have necessarily observed it in this analysis.
\end{enumerate}

\begin{table*}[]
    \centering
    \begin{tabular}{c|c}
        Selection State & Amount of Data Remaining (hours) \\ \hline
        Initial pointing selection & 98.56 \\ \hline 
        Ionospheric Cut & 39.98 \\ \hline 
        Full Array RFI Cut (Absolved Observation cut) & 18.39  \\
        \hline 
        Final RFI Cut (Wall of Shame cut) & 14.68 \\ \hline
    \end{tabular}
    \caption{Amount of data remaining \textit{after} each selection listed in the left-hand column.}
    \label{tab:cut_levels}
\end{table*}

In this work we show that RFI yet fainter than that detected by \textsc{SSINS} (ultra-faint RFI) exists within MWA observations, and its effect on the power spectrum measurement is observable when less than an hour of data is coherently integrated. We accomplish this by splitting the data into subsets that are either highly likely or highly unlikely to contain this ultra-faint RFI, and comparing their cylindrical power spectra. We confirm that the distinct signature is indeed associated with the RFI by constructing power spectra with and without RFI flags applied, and noting an enhancement of the signature when RFI flags are not applied. We also note, however, that this signature is not necessarily observed in our deepest integrations, suggesting that RFI may have favorable integration properties for EoR measurements. This leaves the status of RFI in EoR science unclear, since it may yet appear at a level fainter than our current integrations, but brighter than the cosmological 21-cm signal. 

This paper is laid out as follows. In \S\ref{sec:daq} we describe the instrument and data acquisition, pre-processing, calibration, and power spectrum estimation pipeline. From the flagged and subsequently calibrated data, we derive statistics about the RFI content and conduct a jackknife test to determine the effect of ultra-faint RFI on power spectrum measurements, presented in \S\S\ref{sec:RFI}-\ref{sec:subint}. In \S\ref{sec:limit}, we show our final power spectrum upper limit, and in \S\ref{sec:conc} we present our conclusions.

\section{Instrument and Observation Description}
\label{sec:daq}

Here we briefly describe the properties of the MWA that are relevant to our analysis and provide some details about the choice of observational parameters. For a more thorough description of the MWA Phase I, see \citet{Tingay2013}. We also describe some preliminary data cuts based on metrics that have been used similarly in previous MWA upper limits.

\subsection{Instrument and Observations}

To construct this upper limit, we used data from the second season of the MWA Phase I. This is a radio array located in the radio-quiet Murchison Radio Observatory in Western Australia. The Phase I configuration correlates 128 pseudorandomly located receiving elements to form interferometric visibilities. Each receiving element of the MWA is a square tile of 16 crossed-dipole antennas that have been beamformed to produce a single voltage signal. Using analog delays between individual dipoles within a tile, we can coarsely point the tile to different locations on the sky. 

We select MWA observations of the ``EoR0" field, which is centered on a right ascension of 0 hours and a declination of $-27$ degrees. Throughout each observing night, the array is repointed roughly every 30 minutes in order to track this field. This field has been studied in previous MWA EoR limits from the first observing season \citep{Beardsley2016, Barry2019b}, as well as in other seasons, array configurations, and observing bands \citep{Li2019, Trott2020}. We focus on five pointings centered on zenith as the field transits from East to West. This is based on analysis in \citet{Beardsley2016}, where it was found that the galaxy's presence in the array sidelobes produces extreme contamination in the EoR window. This analysis was reconfirmed with a similar quality metric using the Phase II configuration in \citet{Li2019}.  

The voltage stream of each tile is channelized by a two-stage polyphase filter bank (PFB; \citet{Prabu2015, McSweeney2020}). This divides the 30.72 MHz observing band into 24 coarse channels of width 1.28 MHz, each of which is divided into 32 fine channels of width 40 kHz. The observing band for this analysis extends from 167.1-197.8 MHz. Each observation is 112 seconds in length, with an integration cadence of 2 seconds. 

In total we analyze 3168 observations from the second season of the MWA Phase I, which amounts to 98.56 hours of data. However, the vast majority of these are removed from the final limit by a strict ionospheric cut and subsequent RFI-based cuts. We describe these cuts in significantly more detail in the following sections and summarize the results in Table \ref{tab:cut_levels}. In brief, about 1/3 of the observations have less-than-excellent ionospheric quality, another 1/4 could not have its ionospheric quality collected (usually a sign of poor observational quality), and the rest were cut strictly based on inferred RFI content. Among observations whose ionospheric quality could be gathered, we do not find any obvious correlation between total RFI occupancy and the magnitude of the ionospheric metric. We do observe a slight excess of observations with high RFI occupancy among observations whose ionospheric metric could not be gathered, however this only accounts for a tiny minority of such observations. It could be that excess RFI is causing the failure in the metric gathering in some instances (e.g. one night of extremely bad RFI exists within this subset), however there exist numerous examples within the data for high-occupancy observations with collectable ionospheric metrics, so this hypothetical relationship is tenuously supported at best. 

Previous limits using a highly similar analysis pipeline on the MWA Phase I analyzed the first season of data \citep{Beardsley2016, Barry2019b}. We expected the second season of data may be better for power spectrum measurements due to the the correction of the ``digital gain jump" from the first season as well as reduced digital nonlinearities. Both of these effects arise in the receiver chain. The digital gain jump is caused by an interaction between non-linearities in the digital signal chain (arising from quantization) and the choice of digital multipliers (called “digital gains") used to flatten the bandpass between the receivers and the correlator \citep{Prabu2015}. The digital gains are removed (by dividing out by their value) after correlation, but in some cases the digital non-linearities of the signal chain make it impossible to perfectly remove the digital gains. In the first season of MWA data, the digital gains that were chosen resulted in a single large jump at 188.16 MHz. This large jump interacted strongly with the digital non-linearities and as a result there was a residual discontinuity (called the “digital gain jump”) in the data even after the digital gains correction. Data above the digital gain jump were excluded in \citet{Barry2019b}. For the second season, the digital gains were chosen to have much smaller changes across the band, resulting in significant improvements in the bandpass smoothness after the digital gains correction. In addition we correct for the remaining digital non-linearities using a Van Vleck correction \citep{Benkevitch2016}, explained more in \S\ref{sec:enhance}. See \citet{Barry2018} for a detailed description and analysis of these effects. Finally, we will likely benefit in future analyses by combining multiple seasons of data, and therefore stand to gain from analyzing multiple seasons with a similar pipeline.


\begin{figure*}
    \centering
    \includegraphics[width=\linewidth, page=2]{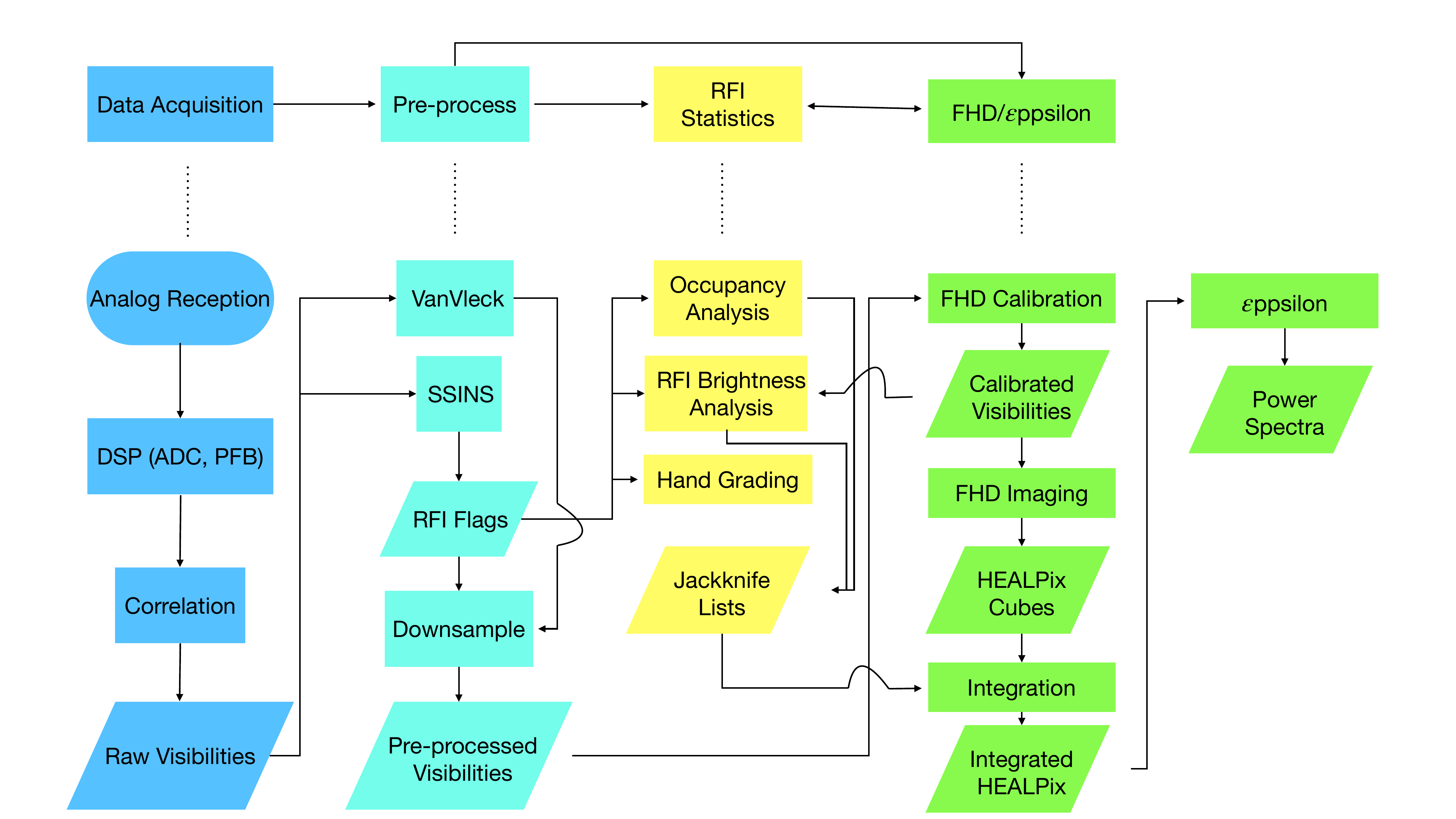}
    \caption{Flowchart of most salient steps in the analysis of this data set. The top row of nodes shows the major analysis steps. Analysis stages are color coded as follows; Blue: Data Acquisition, Cyan: Pre-processing, Yellow: RFI Analysis, Green: Calibration, Imaging, and Power Spectrum Estimation. The lower flowchart shows each major analysis step in more detail and how the outputs of each step feed into subsequent steps.}
    \label{fig:flow}
\end{figure*}

\subsection{Analysis Overview}

As with all 21-cm power spectrum estimation pipelines, ours involves several analysis steps. A flowchart describing the process from data acquisition to power spectrum estimation is shown in Figure \ref{fig:flow}. Before calibration, imaging, coherent averaging, and power spectrum estimation, the raw visibilities must be pre-processed (cyan nodes in Figure \ref{fig:flow}). Compared to previous MWA limits that use the same calibration, imaging, and power spectrum software as this analysis \citep{Beardsley2016, Barry2019b, Li2019}, we use an enhanced pre-processing pipeline. We describe specific enhancements to the pipeline in the next subsection. 

Pre-processing involves several corrections to the raw visibilities. The phases of the visibilities are adjusted according to the differences in cable length between the tiles and receivers, and also to specify a phase center on the sky. Since the alteration to the bandpass resulting from the PFB is theoretically known, this shape and the digital gains for each coarse channel are divided out. Due to quantization nonlinearities, this does not fully correct the bandpass, and residual frequency structure is left over. We address nonlinearities due to quantization error during pre-processing using a Van Vleck correction \citep{Benkevitch2016}, which in general corrects for quantization artefacts by solving an integral equation that relates the analytic correlation coefficient to the output of a digital correlator. This is a new technique in our pipeline, and we describe it in more detail in \S\ref{sec:enhance}. We also flag RFI using Sky Subtracted Incoherent Noise Spectra\footnote{\url{https://github.com/mwilensky768/SSINS}} (\textsc{SSINS}; \citet{Wilensky2019}), as well as 80 kHz around each coarse channel edge. Finally, we downsample to 80 kHz frequency resolution to save disk space and processing time.

The pre-processed visibilities are then calibrated, gridded, and imaged (at each frequency) in HEALPix format \citep{Gorski2005} using Fast Holographic Deconvolution\footnote{\url{https://github.com/EoRImaging/FHD}} (\textsc{FHD}; \citet{Sullivan2012, Barry2019a}). The calibration strategy is almost identical to that used for \citet{Barry2019b}. However, we use the autocorrelations to establish a bandpass shape in the manner of \citet{Li2019}, rather than how they were used in \citet{Barry2019b}. Cable reflection systematics are handled during calibration by fitting the amplitude, phase, and delay in a hyperresolved Fourier basis \citep{Barry2019a}. We use a modified gridding kernel, as described in \citet{Barry2019a, Barry2019b}. We also grid model visibilities generated from the \textsc{GLEAM} catalog \citep{Hurley-Walker2017}, allowing us to form ``dirty" (just the data), model, and residual (data minus model) HEALPix cubes, which are downsampled to 160 kHz frequency resolution. We can also propagate weights and variance information about the measurement by gridding as if each visibility is equal to 1 using the beam and the square of the beam, respectively. HEALPix cubes from multiple observations can then be coherently averaged before passage to Error Propagated Power Spectrum with InterLeaved Observed Noise\footnote{\url{https://github.com/EoRImaging/eppsilon}} ($\varepsilon$\textsc{ppsilon}; \citet{Barry2019a}), where power spectrum estimates are formed. We form separate cubes for the even and odd time integrations within each observation, so that $\varepsilon$\textsc{ppsilon} can form a power spectrum estimate without a thermal noise bias. Important features of $\varepsilon$\textsc{ppsilon} include the use of a generalized Lomb-Scargle periodogram as well as multiple forms of noise metrics, described in \citet{Barry2019a}.

The in-situ simulation capabilities of \textsc{FHD} allow for robust tests of signal loss. By simulating visibilities for a fiducial EoR signal and estimating a power spectrum from them under various conditions, \citet{Barry2019b} largely verifies that there is no appreciable signal loss in the power spectrum estimation pipeline used in this work. More specifically, in the absence of direction-independent calibration errors (note there is no direction-dependent calibration in the pipeline, an effect known to cause signal loss in some instances), we are able to recover an input EoR signal in an \textit{in situ} simulation i.e. residuals from gridding and other effects are beneath the expected signal. This and our end-to-end error propagation allows us to confidently place upper limits on the EoR power spectrum signal in our measured power spectra.

\begin{figure*}
    \centering
    \includegraphics[width=\linewidth]{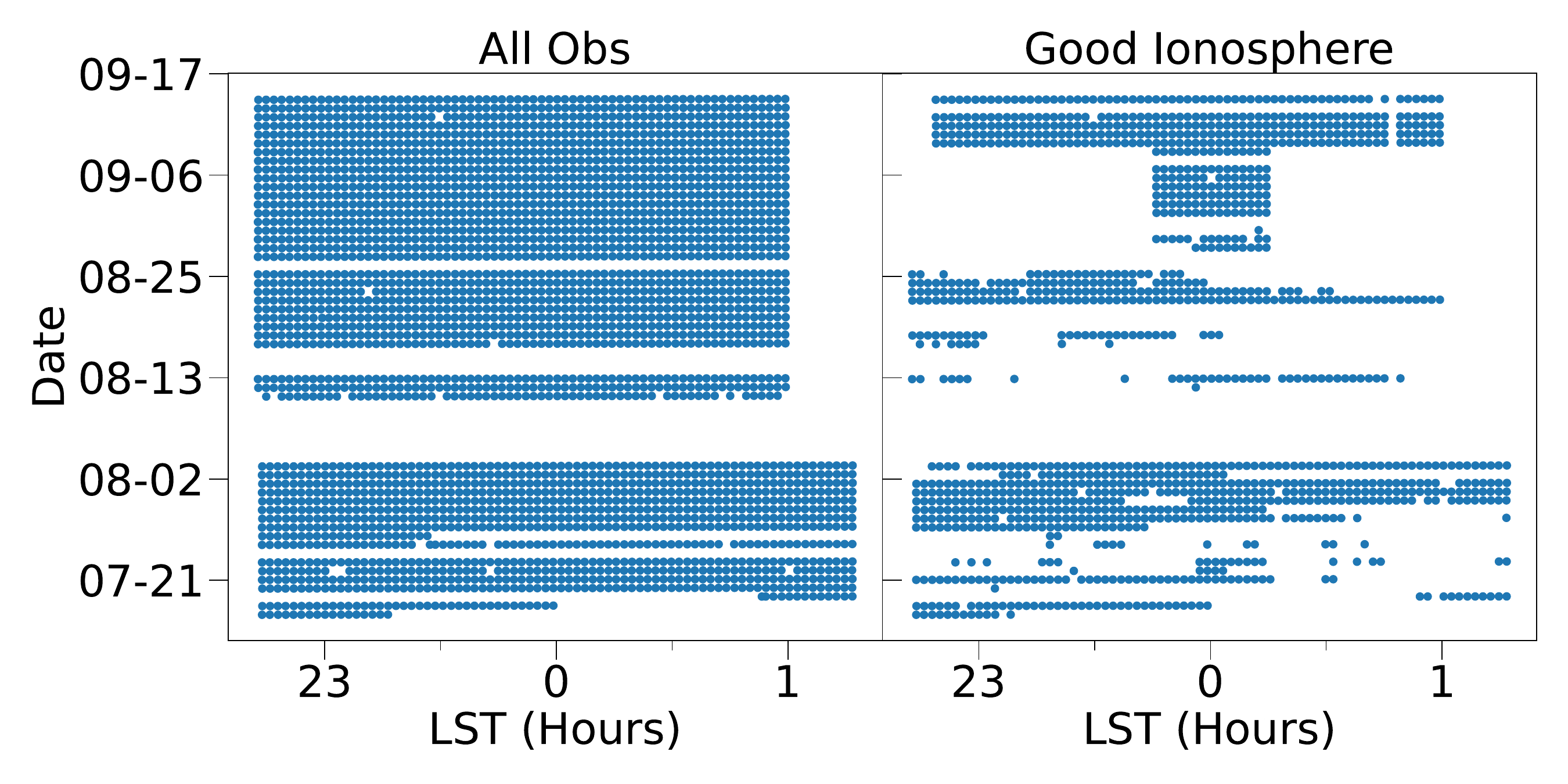}
    \caption{Scatter plot of observations included in the analysis in only the central five pointings (left), and after an ionospheric quality cut (right). Vertical axis shows the date of the observations (blue dots), and horizontal axis is the Local Sidereal Time.}
    \label{fig:iono_scat}
\end{figure*}

We also selected data based on ionospheric quality using a metric described in \citet{Jordan2017}. This metric is based on a combination of the median ionospheric offset as well as a PCA-based measure of offset isotropy to determine ionospheric quality. Lower scores indicate less active ionospheres. We perform a relatively harsh cut, removing all observations from the analysis with an ionospheric quality assurance metric of 5 or greater, cf. \citet{Trott2020}, which amounted to 33.26 hours. We also cut any observation for which the ionospheric metric could not be gathered, of which there were 25.32 hours. This usually results from a failure of the observation to calibrate in the MWA Real-Time System (RTS; \citet{Mitchell2008}), which is a requirement for assessing the ionosphere. In some instances, this can indicate poor observational quality. Since we wanted to isolate the effect of RFI on power spectrum measurements, we resolved to remove such observations outright, thus avoiding a potential source of confusion. Combined, these cuts remove 58.58 of the 98.56 hours of data from the original selection that was based solely on field, observing band, and pointing. We show the result of this cut in Figure \ref{fig:iono_scat}. 

\subsection{Pipeline Enhancements}
\label{sec:enhance}

The majority of the pipeline enhancements in this analysis are related to RFI flagging and analysis of RFI statistics to inform data cuts. We briefly describe our flagging enhancements and changes here, and devote \S\ref{sec:RFI} to describing the statistical analysis. We also describe our digital nonlinearity correction and calibration enhancements here.

In \citet{Barry2019b}, a major improvement to the limit came from excluding observations that contained DTV RFI. These observations were identified by \textsc{SSINS}, which is designed to identify full-array RFI (\S\ref{sec:intro}). \textsc{SSINS} is described in full detail in \citet{Wilensky2019}, but we also briefly describe it here. As implemented in this work, it begins by time differencing the uncalibrated visibilities of a single 2-minute observation at the 2-second integration cadence, which subtracts out all slowly varying components such as the cosmological and astrophysical signals. It then incoherently averages these visibility differences (averages their amplitudes, discarding their phases) over all the baselines to produce a single dynamic spectrum. In the absence of RFI, this remaining spectrum should consist of Gaussian noise. Assuming weak stationarity (unchanging mean and covariance function), the mean of this Gaussian at each frequency is determined by a time-average of each channel. Due to the properties of the thermal noise in the visibility, the standard deviation of the noise at this stage is proportional to its mean, so we can obtain a $z$-score for each sample in this way. We then use an iterative match filter to search for pre-determined shapes within the spectrum that belong to known sources of interference. This can be thought of in a Bayesian setting as generating flags using the maximum a posteriori decision rule amongst a small dictionary of rectangular signals (top-hat power spectral density) of a given strength. The resulting flags in this spectrum are propagated across all baselines and in time according to the visibilities that formed the corresponding differences.

While \textsc{SSINS} is capable of producing flags at the time and frequency resolution of the visibilities, the actual RFI flags from \textsc{SSINS} were not applied in the \citet{Barry2019b} limit.  Rather, observations with known RFI contaminants were excluded from the analysis altogether, including data within those observations that were not classified as contaminated by \textsc{SSINS}. In this analysis, we apply the \textsc{SSINS} flags to the data, extending them in frequency so as to avoid excess EoR window power generated by chromatic flags \citep{Offringa2019a, Ewall-Wice2021, Wilensky2022}. Actually applying the \textsc{SSINS} flags could allow for a measurement with reduced thermal uncertainty compared to \citet{Barry2019b} since it may allow for more data to be used. However, in \S\ref{sec:subint} we analyze power spectra with these flags applied at small integration depths and resolve to fully remove any observations identified as containing RFI in developing the final limit, as was done in \citet{Barry2019b}. 


In this limit, we do not apply \textsc{AOFlagger} flags, unlike every other MWA EoR limit so far. Upon examining the outputs of each flagging pipeline, we found a modest number of RFI events that were caught by \textsc{AOFlagger} and not by the pre-extended \textsc{SSINS} flags. After frequency extension, many of these events were coincidentally excised. Additionally, some samples that were flagged by \textsc{AOFlagger} and not by \textsc{SSINS} may be false positives that result from \textsc{AOFlagger}'s morphological detection algorithm, which can overextend flags for broad features \citep{Offringa2012a}. The \textsc{AOFlagger} flags have a small random false-positive rate \citep[percent-level, depending on the night;][]{Offringa2015} that appears uniformly distributed throughout the times and frequencies of the data. If we extend these flags in frequency to avoid excess power in the window, we incur massive data loss due to the false positives. In principle some of these data could be recovered by implementing some of the strategies in \citet{Offringa2019a}, but we did not do this. Since we cannot generically extend the \textsc{AOFlagger} flags in frequency without causing massive data loss, we opt to entirely forego them at the possible expense of a few extra false negatives in the entire season.

We apply a Van Vleck correction for the nonlinear distortion present in the raw visibilities during pre-processing.\footnote{For implementation details, see \path{007_Van_Vleck_A.pdf} and \path{007_Van_Vleck_B.pdf} in \url{https://github.com/EoRImaging/Memos}} The MWA digital signal pipeline involves several stages of quantization. The Van Vleck correction employed in this work corrects the nonlinearity associated with the final round of quantization. Corrections for other quantization stages are forthcoming. The actual correction involves solving an integral equation that relates the quantized and analog visibilities. This equation has no closed-form inverse for the MWA bit depth. In order to generate fast solutions, the equation is approximately inverted using Chebyshev polynomials. The Van Vleck correction reduces calibration errors due to spectral structure from the digital nonlinearities that cannot be handled by a calibration that assumes a linear gain.

Most of calibration proceeds identically to \citet{Barry2019b}. To speed calibration convergence  by roughly a factor of 2 relative to what was used in \citet{Barry2019b}, we employ a Kalman filter during calibration. Within iterations of the calibration solver, a new guess is determined based on a weighted combination of the previous guess and the current one. The filter is a Bayesian optimization technique where a prediction about the likely location of the solution guides the weighting of the two guesses so as to speed up calibration (as opposed to e.g. a strictly gradient-based approach). The 2013 MWA data had a digital gain jump approximately 2/3 the way through the EoR highband, which required calibration to operate separately on the sections of the band to either side of this discontinuity. The data we selected in 2014 did not require separating the band into two pieces for calibration since the digital gain jump has been corrected in data acquisition. For estimation of the bandpass, we use the autocorrelations in the style of \citet{Li2019}, described thoroughly in \S5 of that reference. This method avoids pitfalls associated with using a particular reference antenna, which may cause idiosyncrasies of that antenna (e.g. imperfect cable reflection fitting) to be imprinted on the other tiles.

\section{Analysis of RFI Statistics}
\label{sec:RFI}

We perform an occupancy analysis based on the \textsc{SSINS} flags. We find several interesting features about the time dependence of the RFI occupancy. We also use these flags in order to study the brightness statistics of RFI after calibration. We then use these time and brightness properties in order to conduct jackknife tests with cylindrical power spectra. We find that there is a definite signature of ultra-faint RFI in the cylindrical power spectrum that is most visible in sub-hour length integrations. The effect on the power spectrum is the main topic of \S\ref{sec:subint}, however we comment briefly here on possible reasons for an enhanced RFI signature in the power spectrum at certain integration depths.

Our sub-hour integrations tend to consist of observations from the same night or relatively few nights, and oftentimes observations that are contiguous with one another. If these contiguous observations are flagged due to a common RFI reflector, then it is possible that the RFI signal might integrate coherently for those observations. For example, if an airplane moving due N-S is reflecting DTV interference, then the E-W baselines in particular will observe no fringing of this source and thus average coherently. In longer integrations, we are likely to combine more RFI sources since we include more nights of data. If different RFI sources tend to average incoherently, then deep integrations may dilute the RFI signal and therefore produce smaller power spectrum contamination \citep{Wilensky2020}. Our observations therefore seem to support the idea that independent RFI sources tend to average incoherently. This is our leading hypothesis as to why we seem to see more contamination in sub-hour integrations compared to longer ones.

\subsection{Summary of Flagging Settings}

The match-filter in \textsc{SSINS} can be programmed to search for occupants that span arbitrary contiguous frequency ranges in the observing band. We generally reserve this feature for physically or empirically motivated frequency ranges, since searching every possible frequency range creates massive computational overhead. The most common occupants for this observing band are digital television and whole-band ``streaks" of uncertain origin \citep{Wilensky2019}.\footnote{The uncertainty stems from the fact that the contaminants are not band-limited, and can therefore come from a number of hypothetical sources. Some occurrences have been clearly associated with extremely bright ORBCOMM \citep{Wilensky2021}, but this only accounts for a small fraction of instances.} Hereafter, we will refer to these whole-band streaks as ``broadband" RFI. Based on the Western Australian digital television allocations, we search for four 7-MHz-wide DTV signals, all adjacent, starting at 174 MHz; these are designated as channels 6-9. Channel 9 begins at 195 MHz, so we only observe 2.8 MHz of its allocated bandwidth. We also search for narrowband occupants and broadband RFI. 

Recall from \S\ref{sec:enhance} that the \textsc{SSINS} $z$-scores are calculated from the 2-minute time-average of incoherently baseline-averaged visibility differences, and reflect a probability under a null hypothesis for pure thermal noise to achieve a deviation from the mean at least as large as the data's. Under that null hypothesis, we expect less than 1 datum with $z$-score greater than about 4.2 in a single observation's SSINS. For all shapes other than broadband RFI, we use a significance threshold of 5, i.e. a (potentially frequency-averaged) sample whose $z$-score is $5$ or greater is considered as contaminated.  For broadband RFI, we use a significance threshold of 10. We use a greater significance threshold for broadband RFI since the data exhibit lightly nonstationary thermal noise as a result of the receiver refrigeration cycle, which results in a significant number of false broadband detections at $z$-scores less than $10$ \citep{Wilensky2021}. 

The \textsc{SSINS} match filter proceeds iteratively in that when it detects an RFI event, it excises that data and recalculates the $z$-scores of the remaining data before searching for the next strongest RFI event. Before recalculating the $z$-scores, we flag a frequency channel for the entire observation if its occupancy is greater than 60\% and extend flags in frequency. This combination means that if any RFI occupant is found for 60\% of an observation, the entire observation is flagged. We summarize these settings in Table \ref{tab:flag_settings_table}.

\begin{table}[]
    \centering
    \begin{tabular}{c|c|c}
        Shape Label & Frequencies (MHz) & Significance Threshold  \\ \hline
        TV6 & 174-181 & 5 \\ \hline 
        TV7 & 181-188 & 5 \\ \hline 
        TV8 & 188-195 & 5 \\
        \hline 
        TV9 & 195-202 (197.8) & 5 \\ \hline
        Narrowband & Various & 5 \\ \hline
        Broadband & 167.1-197.8 & 10 \\ 
    \end{tabular}
    \caption{Association of shapes in the match filter, their frequencies, and the significance thresholds used in the filter. The TV9 shape is allocated all the way to 202 MHz, but our observing band cuts off at 197.8 MHz. Narrowband shapes are 1 fine channel wide and allowed anywhere in the band.}
    \label{tab:flag_settings_table}
\end{table}

\subsection{RFI Occupancy Analysis}
\label{sec:occ_analysis}

Any time data is excised by the \textsc{SSINS} match filter, the time, frequencies, and shape label of the most recently identified RFI event are recorded. We analyze this flagging metadata to understand the RFI content of the data set. First we discuss a few important caveats about this flagging metadata related to misclassification, which are explored in substantially more detail in \citet{Wilensky2021}. Then we present the occupancy results.

\subsubsection{Flagging Nuances}
\label{sec:nuance}

Below we list a few nuances about the \textsc{SSINS} flags that are important to understand when interpreting trends within the flagging results.

\begin{enumerate}
    \item \textbf{Prioritizing excision over classification}: While \textsc{SSINS} does associate each RFI event with a label, there can be misclassifications that produce identical flags, and we do not formally check for this in every possible instance. For example, 3-4 DTV events occurring simultaneously may be classified as a broadband event since combinations of DTV events are not checked in the match filter. This will produce the same flags after frequency extension. We judge the rate of this misclassification to be low based on hand-grading the SSINS flags for each observation (prior to frequency extension).

    \item \textbf{Assumption of stationary noise}: The $z$-scores in \textsc{SSINS} are calculated assuming weak stationarity (specifically constant mean and variance in this case). When bright sources enter or exit sensitive parts of the beam, the variance of the thermal noise can change significantly on timescales associated with the beam crossing time, which is insignificant for the 2 minute observations in this work. In the MWA Phase I, there are 16 receivers, each of which are responsible for digitizing and channelizing the voltage signals of 8 tiles. A temperature-dependent gain in the amplifiers \citep{Barry2018} causes smooth temporal variations in the noise variance on time scales associated with the receiver refrigeration cycle \citep[a few minutes;][] {Wilensky2021}. Since \textsc{SSINS} flags are generated on uncalibrated data, this cycle tends to impart a gradient in the $z$-scores that is minuscule but coherent in the SSINS over the entire observing band. The coherency leads to an overreporting of broadband events at a given significance. We counter this by increasing the significance threshold for broadband events. We expect this produces an average false positive rate of less than 2\%.

    \item \textbf{Insufficient time samples}: In heavily contaminated observations, having fewer time integrations with which to estimate the mean per frequency and polarization prevents the $z$-scores from being approximately Gaussian. What we generally find is that the $z$-scores calculated by \textsc{SSINS} for simulated stationary Gaussian noise are more concentrated than a standard normal random variable, which can lead to false negatives. For this reason, we completely flag all observations where greater than 60\% of integrations have been flagged. 
\end{enumerate}

\subsubsection{Occupancy Results}

In this section we discuss the RFI occupancy of the data set as reported by \textsc{SSINS}. We find that the overall distribution of RFI occupancy is relatively unaffected by the ionospheric cut, and therefore only show distributions before the ionospheric cut so that we have a larger sample size. 

\begin{figure*}
    \includegraphics[width=\linewidth]{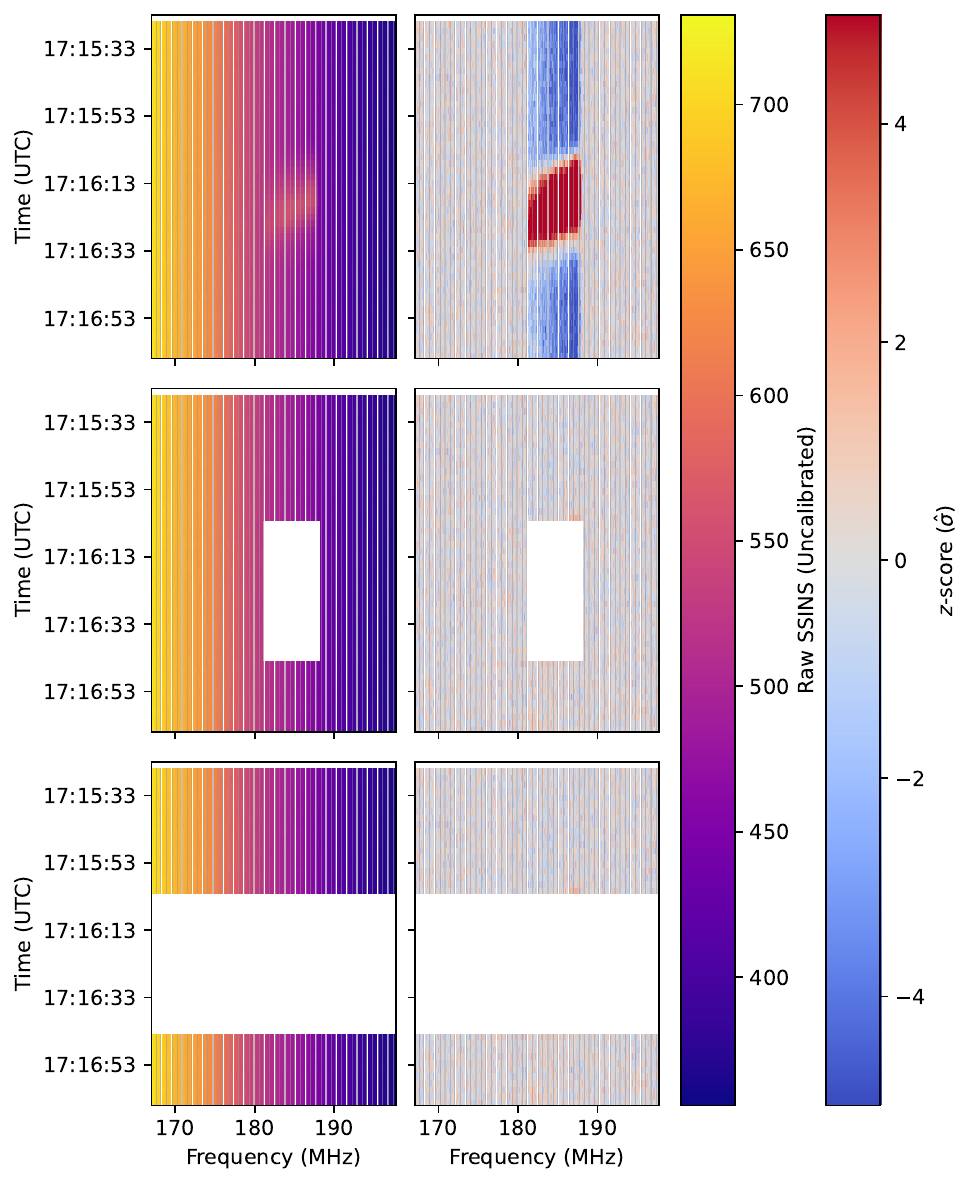}
    \caption{Left: Raw SSINS of a 112-second observation at various stages of flagging. Right: Mean-subtracted SSINS. Top: Data with initial flagging mask. Middle: Additional flagging mask after match filtering. Bottom: Flags after frequency extension.}
    \label{fig:SSINS_triple}
\end{figure*}

As an example of fairly typical flagging results for  moderately bright DTV contamination, we show Figure \ref{fig:SSINS_triple}. This shows the raw SSINS and its $z$-scores for a 112-second observation in the data set. There is an initial flagging mask that flags the edges of the coarse channels as well as the first integration. This latter flagging set is a holdover from previous analyses, where the correlator would occasionally produce oddities in the data at the very beginning of the observation. We did not check whether this behavior was still present, and flagged the beginning of each observation just in case. The DTV contamination is most obvious in the mean-subtracted version, where it is clear that it lasts at least 20 seconds. The $z$-scores shown are based on a time-average of the unflagged data, which biases the $z$-scores low for the rest of the integrations in the channels occupied by the DTV. Since the flagger recalculates $z$-scores after each event is identified \citep{Wilensky2019}, this initial bias does not lead to extra false detections. The middle row shows the results of the \textsc{SSINS} match filter, indicating that the event lasted roughly 40 seconds. In imaging experiments for DTV events that appear in SSINS similarly to these, we find that they are usually some sort of aircraft reflection often appearing in the second southern sidelobe of the beam. The bottom row shows flags after frequency extension. We use these extended flags to develop the flagging time series used in the statistical analysis in this section.

\begin{figure}
    \centering
        \includegraphics[width=\linewidth]{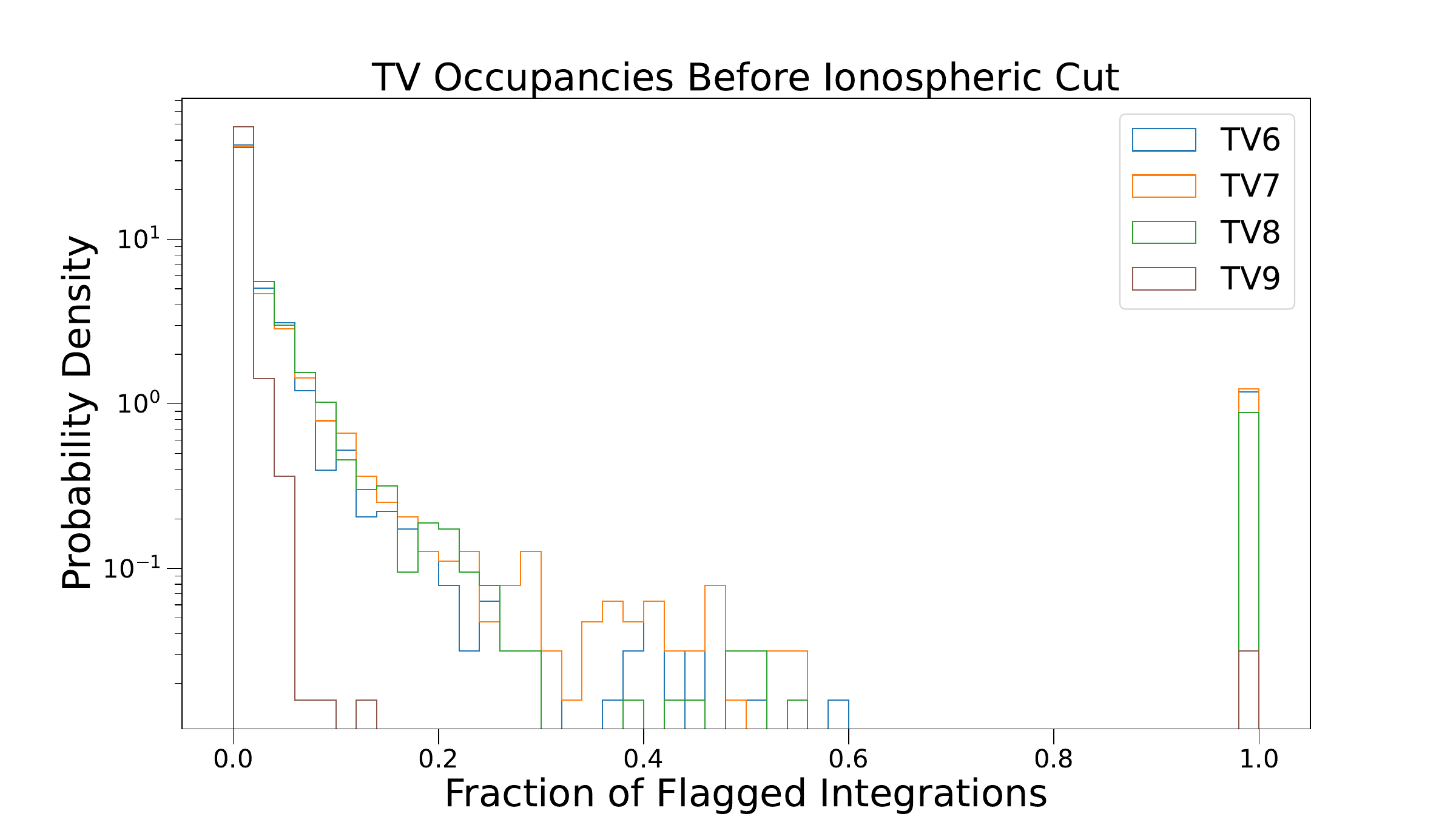}
    \caption{Histogram of DTV occupancies per observation, before the ionospheric cut. Since observations with more than 60\% occupancy are entirely flagged, there is a cutoff in the histogram at 0.6 and a buildup at 1. Only a small fraction of the allocation for TV9 is in the observing band, so the \textsc{SSINS} match filter is less sensitive to it. Data about TV broadcasting stations also suggests this allocation may be used less often.}
    \label{fig:TV_occ_hist}
\end{figure}

In Figure \ref{fig:TV_occ_hist}, we show a histogram of the TV occupancies for all of the 112-second observations before the ionospheric cut. We observe that channels 6, 7, and 8 have very similar occupancy distributions, while channel 9 is much rarer to observe, appearing in only 3.5\% of observations. Examining broadcasting stations in Western Australia, we find that channels 6, 7, and 8 are used relatively equally, while channel 9 is used less often. Since we only observe 2.8 MHz of channel 9's allocation, \textsc{SSINS} is also less sensitive to its presence. 

Since the majority of the observing band is allocated for DTV, we expect very few narrowband events, which is consistent with what we observe. Curiously, when we do observe narrowband events that are clearly true positives, they are often within the DTV allocation. Some of these events are extremely bright, and may be a result of out-of-band RFI clipping the ADC or causing intermodulation products. Alternatively, there could be locally generated RFI.

\begin{figure}[b!]
    \includegraphics[width=1\linewidth]{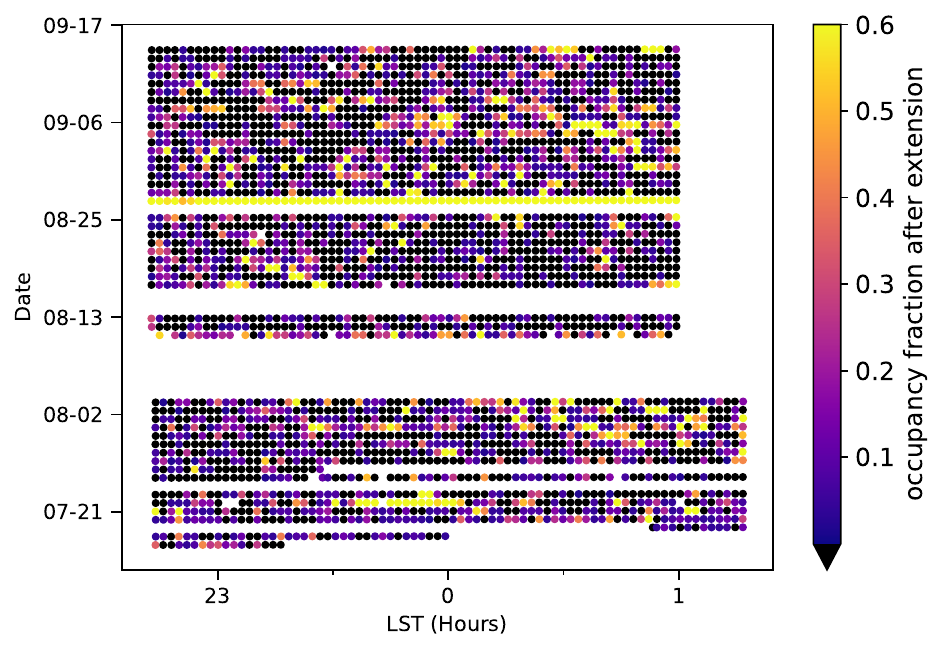}
    \caption{Total Occupancy scatter plot after extending flags across frequency, which roughly doubles the total occupancy in the data set. Each marker represents a 2-minute observation, where its horizontal position is its starting sidereal time and its vertical position is its beginning civil time. Clusters of flags are apparent.}
    \label{fig:total_occ_broadcast}
\end{figure}

Since we suspect a significant number of the RFI events recorded by \textsc{SSINS} are due to reflections from aircraft  \citep{Wilensky2019}, we seek to understand the timelike properties of the RFI occupancy. First, we notice that the different TV channels are sometimes flagged simultaneously, and flags tend to appear in clusters within a night. By constructing the correlation function of the time series of flags belonging to different channels, we sometimes observe that flags for different TV channels are strongly correlated within a night, however not all nights have such correlations. Not all broadcasting stations in Western Australia transmit at the same frequencies, and there is probably a range of aircraft trajectories that intercept these various stations. Furthermore, direct reception of the signals is possible. For instance, there is a night of times with extremely high TV occupancy on August 27, clearly visible in the occupancy scatter in Figure \ref{fig:total_occ_broadcast}. The SSINS of this night show extremely strong, persistent DTV interference. Since the RTS could not calibrate these observations, no ionospheric metric could be gathered to assess the possibility of long-range ionospheric-based reception such as ``sporadic-E" propagation. This could be a tropospheric ducting event allowing for long range direct reception of the RFI \citep{Sokolowski2016}, however we have not cross-referenced any sort of weather database or performed any other type of analysis that allows us to understand the exact origin of this extreme event. Combining all of the aforementioned properties, we expect that we sometimes observe different DTV signals simultaneously, but not always. The flags reflect this in their two-point statistics and this expectation is corroborated by manual inspection of the SSINS.

\begin{figure}
    \centering
    \includegraphics[width=\linewidth]{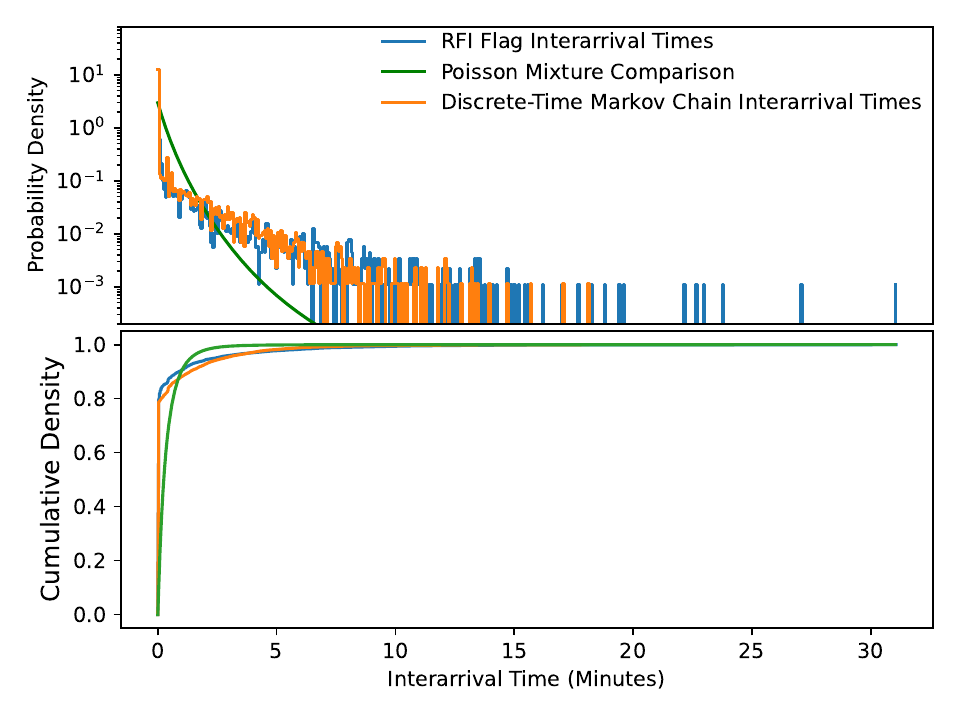}
    \caption{Top: Histogram of flag interarrival times over all nights for the broadcast \textsc{SSINS} flags. The ``Poisson Mixture Comparison" (green) is a hyperexponential distribution where each component has consistent mean interarrival time as each night of data (and is weighted according to the amount of data in the corresponding night). The Markov chain histogram is built from Markov chain realizations of each night with consistent transition probabilities as the night of RFI flags they are built from. Interarrival times from the Markov chain realizations are roughly consistent with those from the RFI flag time series, whereas the Poisson mixture model tends to underpredict short and long interarrival times, thereby failing to predict clumps of flags. Bottom: Cumulative density for what is shown in the top plot. The data and Markov chain cumulative densities are inconsistent with a Poisson model. }
    \label{fig:ia_times}
\end{figure}

In Figure \ref{fig:total_occ_broadcast}, we show the total occupancy of each observation with the flags extended. Here, the clustering behavior of the flags is more apparent. To also illustrate this clustering, we show a histogram of interarrival times between \textsc{SSINS} flags (time elapsed between flagged integrations) in Figure \ref{fig:ia_times}. We do not include the night with strong, persistent interference. We also show two models with parameters estimated from the data: a Poisson process based on the mean flagging rate and a Markov model based on the observed transition rates between flagged and unflagged data. 
Over 80\% of the mass of this histogram is contained at the sample spacing. If we model each night as its own Poisson process (i.e. each night has a different false positive rate, and the flags are dominated by false positives), Poisson mixture model fails to predict the abundance of mass at the sample spacing and also decays much faster than the data, which means it underpredicts long gaps between flagged data. This suggests clustered flagging with a range of gap sizes rather than a simple mixture of Poisson processes. If we perform a Kolomogorov-Smirnov test, we obtain a test statistic value of about 0.7, which corresponds to the discrepancy at short interrarival times, highlighting the Poisson model's failure to predict clumps. This corresponds to a $p$-value that is basically 0 in floating point precision, i.e. there is almost no probability that the empirical CDF of a collection of random samples from the Poisson mixture model would appear at least as discrepant as the data. 

To probe the clumping properties of the flags, we examine the transition probabilities in the time series. What we find is that unflagged data is almost always followed by more unflagged data at a rate of 98\%, while flagged data is followed by more flagged data at a rate of 80\%. Since the flag occupancy is relatively low ($\sim 10\%$), this manifests as long runs of unflagged data with small-medium sized clumps of flagged data. We find that such clumps can be modeled using a discrete time Markov chain with transition rates matched to the data. Each night we generate a time series using that night's transition rates, and then we histogram the resulting interarrival times. Specifically, we begin with an initial seed state, $x_0$, which takes the value 1 or 0 (1 indicates that it represents a flagged datum, while 0 indicates otherwise). We then sample from a Bernoulli distribution with probability of success equal to $P(X_1=x_0 | X_0=x_0)$. For example, if $x_0=1$ and the Bernoulli sample is a success, then $x_1=1$ as well, otherwise $x_1=0$. We then repeat this process to generate the chain for each night. While the realization of the season of interarrival times is noisy, it appears to emulate the interarrival time distribution. This quantitatively supports that the apparent clumping of the flags is indeed real.

\begin{figure}
    \centering
    \includegraphics[width=\linewidth]{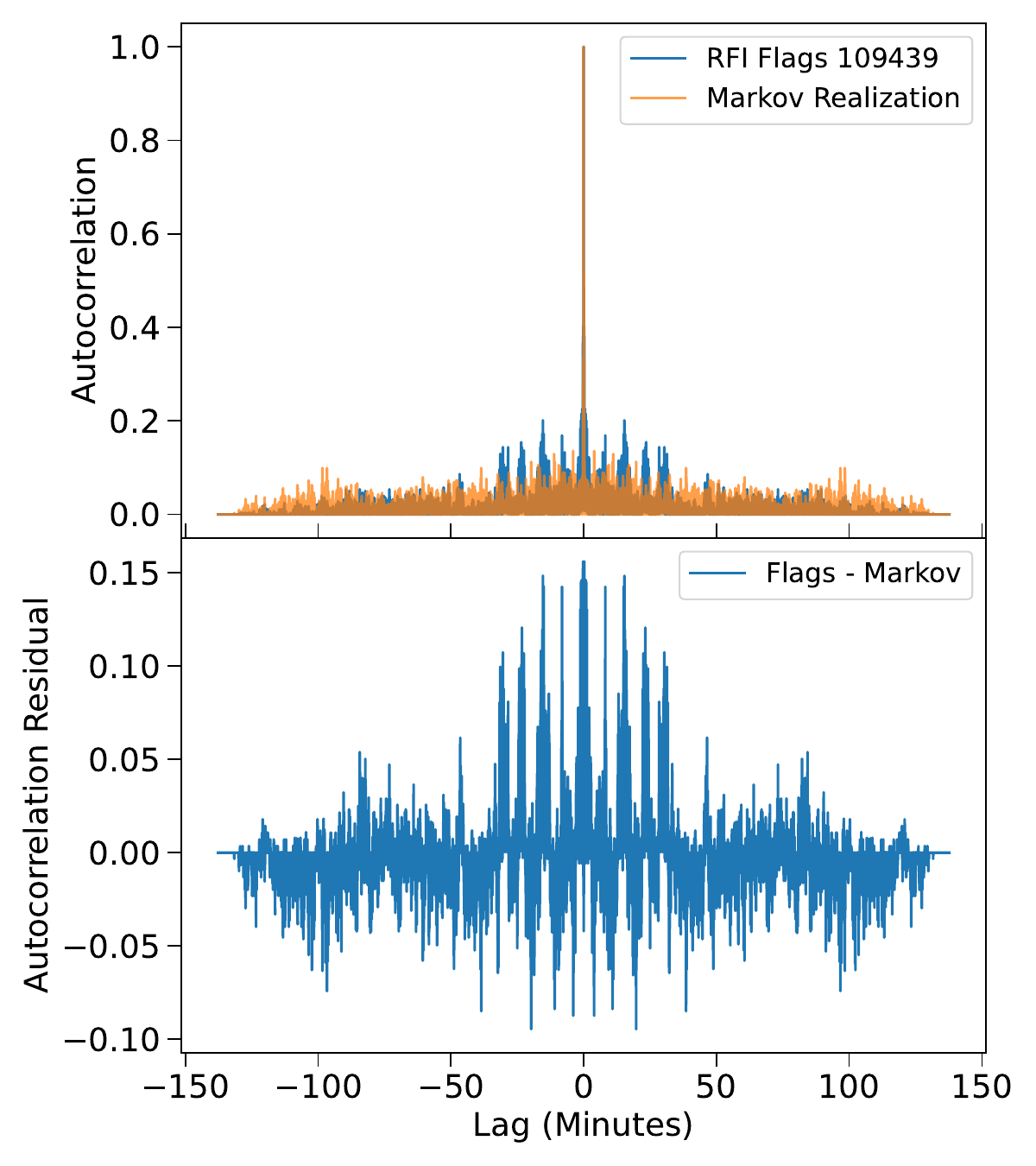}
    \caption{Autocorrelation of RFI flag time series for a particular night (blue) along with the autocorrelation function for a Markov chain realization (transparent orange). Bottom Panel: Residual of the top panel. The Markov chain realization is clearly missing the peaks at lags between 0 and 50 minutes, meaning that the Markov model does not fully capture the timelike properties of the RFI events, and so a more detailed model is required in future efforts.}
    \label{fig:autocorr_markov}
\end{figure}

The RFI flag time series for each night is not entirely consistent with the Markov chain; they differ in their two point statistics. The Markov chain is memoryless, and so any given realization generally does not produce an autocorrelation function with strong peaks away from zero lag (difference in time between samples). On the other hand, many of the RFI flags exhibit discernible peaks at a range of lags, though some nights with comparatively few flags appear somewhat consistent with the Markov realization. We show an example in Figure \ref{fig:autocorr_markov}. This suggests that flags from \textsc{SSINS} are not entirely random point processes with only short-range correlations. These additional peaks in the autocorrelation function may occur when reflectors transit different sidelobes of the beam, or may suggest multiple reflectors with some lag between them such as multiple aircraft on the same flight path. The additional peaks also appear to have a characteristic width from night-to-night, possibly indicating a typical clump size related to the speed of reflector transit through the beam.

It is clear from the transition probabilities, interarrival times, and two-point statistics of the RFI flag time series that RFI events tend to temporally cluster. This motivates a particular style of jackknife test, where observations with no RFI flags are compared to observations that contained some RFI flags. The logic is that an observation with RFI flags is more likely to contain RFI just beneath the sensitivity of \textsc{SSINS} than observations with no flags at all. In the next section, we develop a series of jackknife tests based on the brightness of the flagged emitters with this physically motivated assumption in mind.

\subsection{RFI Brightness Analysis}
\label{sec:brightness_analysis}

The excess contamination in the power spectrum from RFI responds quadratically to its flux. Since \textsc{SSINS} uses incoherently averaged time-differenced visibilities, it is not immediately apparent how to exactly relate the brightness in the SSINS to the brightness in the visibilities. However, the amplitude of a visibility difference can be no more than the sum of the brightnesses of the two samples that were differenced. In more detail, if the two samples are out of phase by exactly $\pi$, this upper bound is achieved. However, if the difference in phase and amplitude is small, then the amplitude of the difference will also be small. This relationship may not be constant over the set of baselines, however the samples identified by \textsc{SSINS} will fall more often in the former case than the latter with the exception of extraordinarily bright RFI (i.e. \textsc{SSINS} will select for large phase or amplitude differences).  Therefore, we can use the SSINS amplitudes as a proxy for the brightness in the visibilities. 

\begin{figure*}
    \centering
    \includegraphics[width=\linewidth]{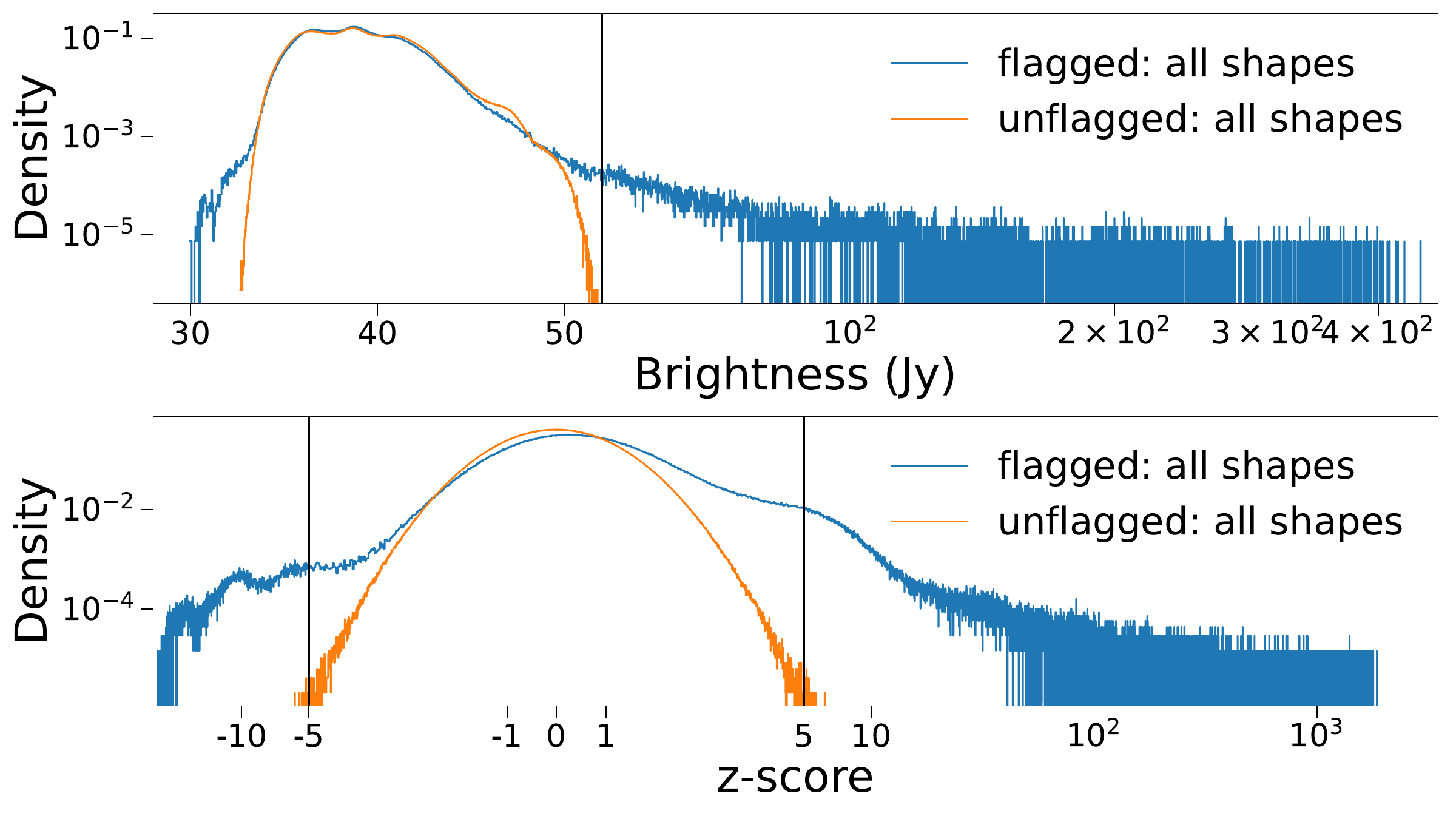}
    \caption{Histograms of SSINS amplitudes and z-scores for the entire season, separated by flagged and unflagged data. These $z$-scores are calculated per fine 40 kHz channel using the time-average of 112s of data as a reference. The horizontal axis is a hybrid linear-logarithmic scale, where the boundary between scales is demarcated by the black vertical lines. The z-scores of the unflagged samples (bottom orange) are highly Gaussian, as expected from the assumptions about the thermal noise. The flagged samples have a highly non-Gaussian z-score distribution. Their brightness distribution has significant overlap with the unflagged brightness distribution, and cannot be separated solely by drawing an amplitude cut.}
    \label{fig:cal_SSINS_all_shapes}
\end{figure*}

\begin{figure*}
    \centering
    \includegraphics[width=\linewidth]{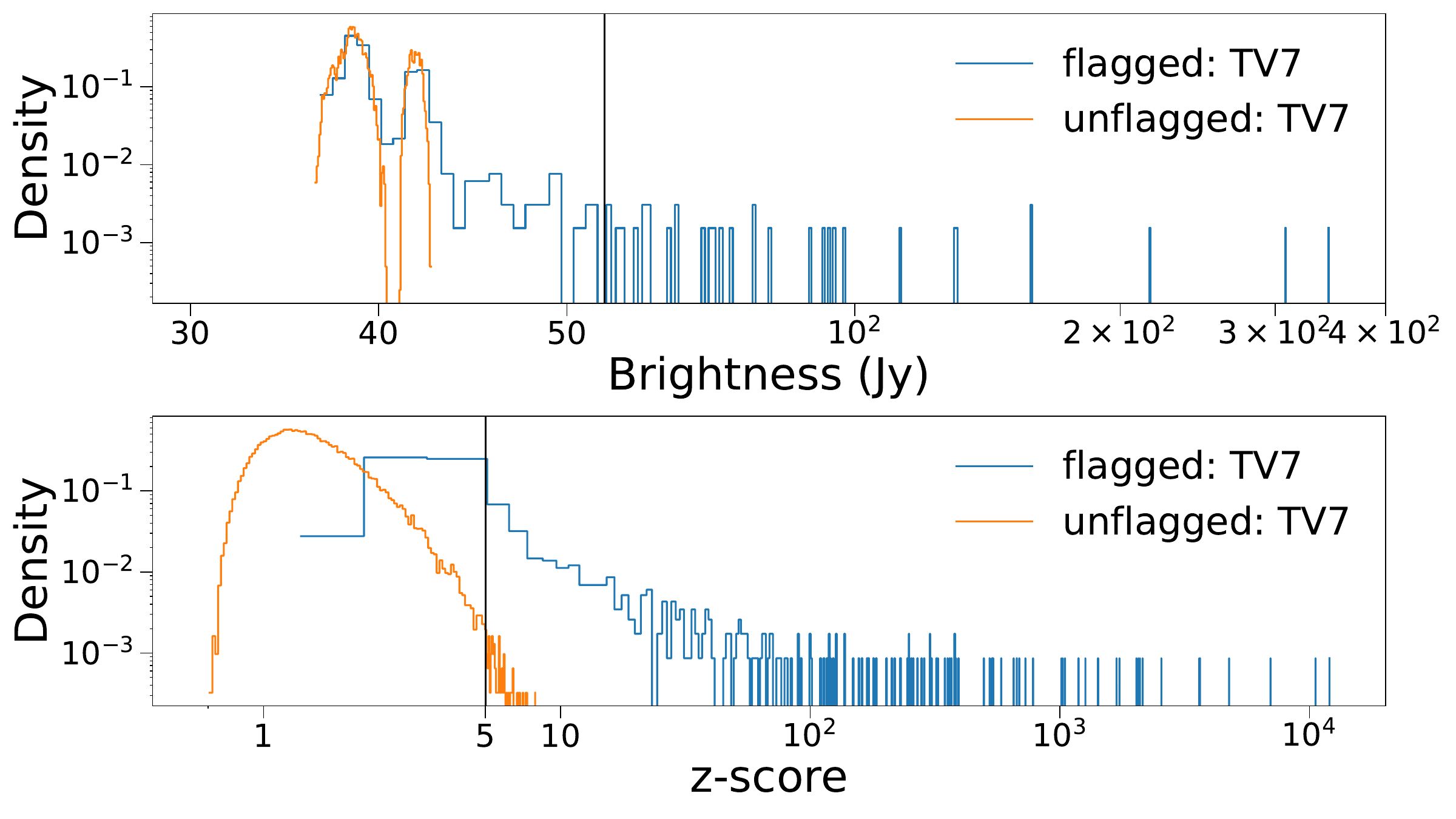}
    \caption{Same as Figure \ref{fig:cal_SSINS_all_shapes}, but with brightnesses and z-scores calculated by doing the \textsc{SSINS} sub-band sum that is employed in the match filter over the TV7 frequencies. Interestingly, the unflagged and flagged brightness samples appear bimodal. Examination of this plot on a per-pointing basis shows that the mode above 40 Jy belongs to the eastern-most pointing in the data, while the other mode is a combination of all other pointings.}
    \label{fig:cal_SSINS_TV7}
\end{figure*}

In Figure \ref{fig:cal_SSINS_all_shapes}, we show histograms of SSINS amplitudes as well as z-scores for the entire season, separated by those samples identified as contaminated (flagged) and those that were not (unflagged). In the top panel we show SSINS amplitudes. Due to the central limit theorem, we expect that the per-frequency distribution of the purely thermal samples in the SSINS follow a Gaussian distribution. However, when all frequencies and many observations are combined in mixture, we observe a highly non-Gaussian distribution. In the bottom panel, the unflagged z-scores, which are calculated per frequency and per observation, are highly Gaussian, reflecting the statistical assumptions underlying the \textsc{SSINS} pipeline. The unflagged samples have brightness between 33 and 52 Jy. On the other hand, the flagged samples have amplitudes that range from 30 to 450 Jy. The z-scores of the flagged samples exhibit a highly non-Gaussian distribution. 

Many of the flagged z-scores are far below the significance threshold. This occurs when the sample was flagged as a part of a shape in the match filter. These shapes are determined from a combination of official allocations from the Australian Communications and Media Authority\footnote{\url{https://www.acma.gov.au/publications/2004-04/guide/technical-planning-parameters-and-methods-terrestrial-broadcasting-tpps}} and empirical verification of the presence of such RFI shapes in our data \citep{Offringa2015, Barry2018, Wilensky2019}. Some shapes are significantly broader than others. Broader RFI can be found at a lower average flux density compared to narrower shapes due to the fact that broad clusters of positive (or negative) outliers would exist only with vanishing probability in a purely thermal SSINS. Thus, the $z$-scores at the single channel level can be very low when their corresponding measurements are identified as part of a broad shape.

These data are perhaps better visualized in 
Figure \ref{fig:cal_SSINS_TV7}, where we show the same quantities as in Figure \ref{fig:cal_SSINS_all_shapes}, but emulate the match filter process by summing over frequency and histogramming the maximum absolute deviation across the polarization axis of the averaged scores. The top panel now appears obviously bimodal for both the flagged and unflagged data. Remaking this plot per pointing shows that the mode above 40 Jy comes from the eastern-most pointing we selected (designated -2 in this work), while the other mode is a combination of the other four pointings. The thermal background in the bottom panel now takes the form of an extreme value distribution for a Gaussian of sample length equal to 4 (number of polarizations), and the recalculated flagged z-scores are offset from zero. Many still exist below the significance threshold. This happens because we flag uncalibrated data, but these data have been calibrated before being histogrammed. This can slightly change the $z$-score calculation since calibration corrects relative amplitude variations between different tiles, ultimately affecting the time dependence of the SSINS and therefore the $z$-scores \citep{Wilensky2021}. 

We argue that these are still likely to be true positives. The frequency-summing step in the \textsc{SSINS} match filter reduces the sample size in consideration by the number of frequency channels in the observation. This means that substantially fewer extreme values are expected in sub-band sums, and so a significance threshold of 5, which was set based on the sample size before summing, gives a wider margin of outliers compared to what is expected from the thermal background. More importantly, we find that observations with only these low significance values are still correlated with excess EoR window power compared to their clean counterparts. We show these in the next section by constructing jackknife tests designed to investigate the possibility of residual RFI in observations flagged by \textsc{SSINS}. 

\section{Power Spectrum Subintegrations: Evidence of Ultra-faint RFI}
\label{sec:subint}

The level to which residual RFI contaminates a power spectrum is determined by its apparent brightness and frequency structure. The apparent brightness is determined by the transmitter brightness, the circumstances of propagation, and the relative position of the telescope primary beam. The larger population centers of Geraldton and Perth are located to the Southwest and South of the array, respectively. Either of these locations may take in flight from countries Northwest of Australia, which would put reflectors in a more sensitive part of a Westerly pointed beam. Smaller cities and towns are also located due West, such as Carnarvon and Denham. Since transmitters tend to be located close to population centers, we expect that RFI contamination should generally increase when the telescope is pointed more to the West. 

To probe these different variables, we establish a series of jackknife tests where we group observations according to different RFI-related variables and examine their integrated power spectra. Some observations in the data set have no flags reported by \textsc{SSINS}, while some are reported to contain RFI. In some of our tests, we purposefully disable flags before calculating power spectra. It is standard to refer to data that has been identified as suspicious as ``flagged" and data that has not been flagged as ``unflagged." The latter term has a particularly confusing linguistic structure that seems to imply ``undoing flags that were already present" but more often refers to data that was found to be without a need for flagging i.e. deemed non-suspicious after a rigorous inspection (and indeed this is how we meant it in \S\ref{sec:RFI}). Since we need to make use of both senses of the word ``unflagged," we instead use the following nomenclature in this section when categorizing observations based on their SSINS:
\begin{itemize}
    \item \textbf{Pure}: Observations for which \textsc{SSINS} found no RFI.
    \item \textbf{Absolved}: Observations for which \textsc{SSINS} found RFI, and we have applied the flags so that the contaminated data has been excised.
    \item \textbf{Repentant}: Observations for which \textsc{SSINS} found RFI, but for investigative purposes we have elected not to use the calculated flags so that the contamination is still present. 
\end{itemize}

Because the EoR power spectrum is sensitive to very faint RFI contamination we want to test whether the data immediately adjacent in time to detected RFI is also contaminated but at a brightness below \textsc{SSINS}' sensitivity. To make this comparison, we compare the power spectra of absolved observations (RFI detected and excised to the best of our ability) to those of LST-matched pure observations with no detected RFI. At this stage, we only consider observations that pass the ionospheric cut. We jackknife by dividing the absolved observations into integration subsets separated by RFI shape reported by the match filter (narrowband, broadband, or any TV channel), telescope pointing (integers from -2 to +2), and reported z-scores of the RFI events (bins with edges 3, 5, 10, 100,  1000, 10000). For each absolved observation in a subset, we find a matching pure observation at a similar sidereal time. We then integrate these subsets, form power spectra, and examine them side by side.

\begin{figure*}[t]
    \centering
    \includegraphics[width=\linewidth]{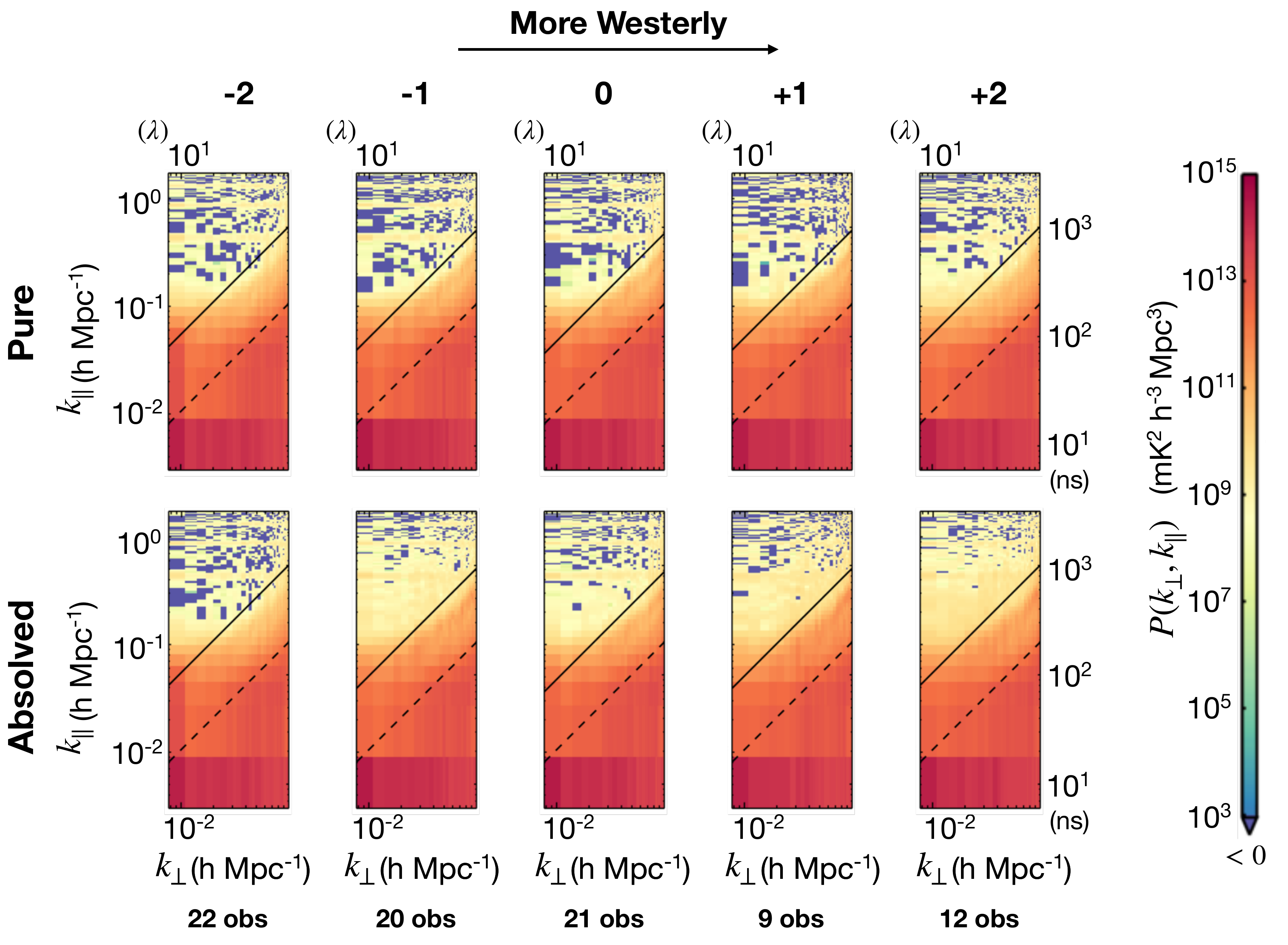}
    \caption{Top: East-West polarized power spectra for pure observations over the five pointings present in the data. The array points more Westerly towards the right of the figure. The number of observations in each integration is annotated at the bottom of the figure. The depth of integration ranges by about a factor of 2.5. All of these power spectra appear noise-dominated in the EoR window (region above the solid black line) in that there is a roughly even speckling of positive and negative data. Bottom: Corresponding absolved power spectra, where any RFI type is included so long as the observation has RFI z-scores that lie between 10 and 100. We see a signal-like component in the EoR window. Rather than an even speckling of positive and negative data, we see consistent, smoothly varying positive power up to and occasionally past the first coarse band harmonic. This suggests residual RFI that was uncaught by \textsc{SSINS}. This contamination is generally worse in the later pointings that point more West. There is no clear correlation between integration depth and contamination levels.}
    \label{fig:pointing_sequence}
\end{figure*}

This produced 160 power spectra per polarization (80 pure/absolved pairs) over the 1285 observations that passed the ionospheric cut. Before examining our jackknife axes, we examined each power spectrum individually, paying special attention to the EoR window where we expect RFI contamination to be most obvious if at all present. Of these 160 observation sets, we identified 46 observation sets, 32 absolved and 14 pure, that demonstrated clear excess power in the lower left corner of the EoR window. The excess power is ``signal-like" in the sense that it appears to vary smoothly as a function of $k_\perp$ and $k_\parallel$ until it seemingly becomes noise dominated at very high $k_\parallel$. This feature can vary in power anywhere from $10^9$ to $10^{11}$ $($ mK$^2$ $h^{-3}$ Mpc$^3$ $)$, with stronger features generally appearing in shallower integrations. Since most of these integration subsets are relatively short (approximately 1 hour or less), a signal-like component in the window is surprising and indicative of systematic contamination. 

\begin{figure*}
    \centering
    \includegraphics{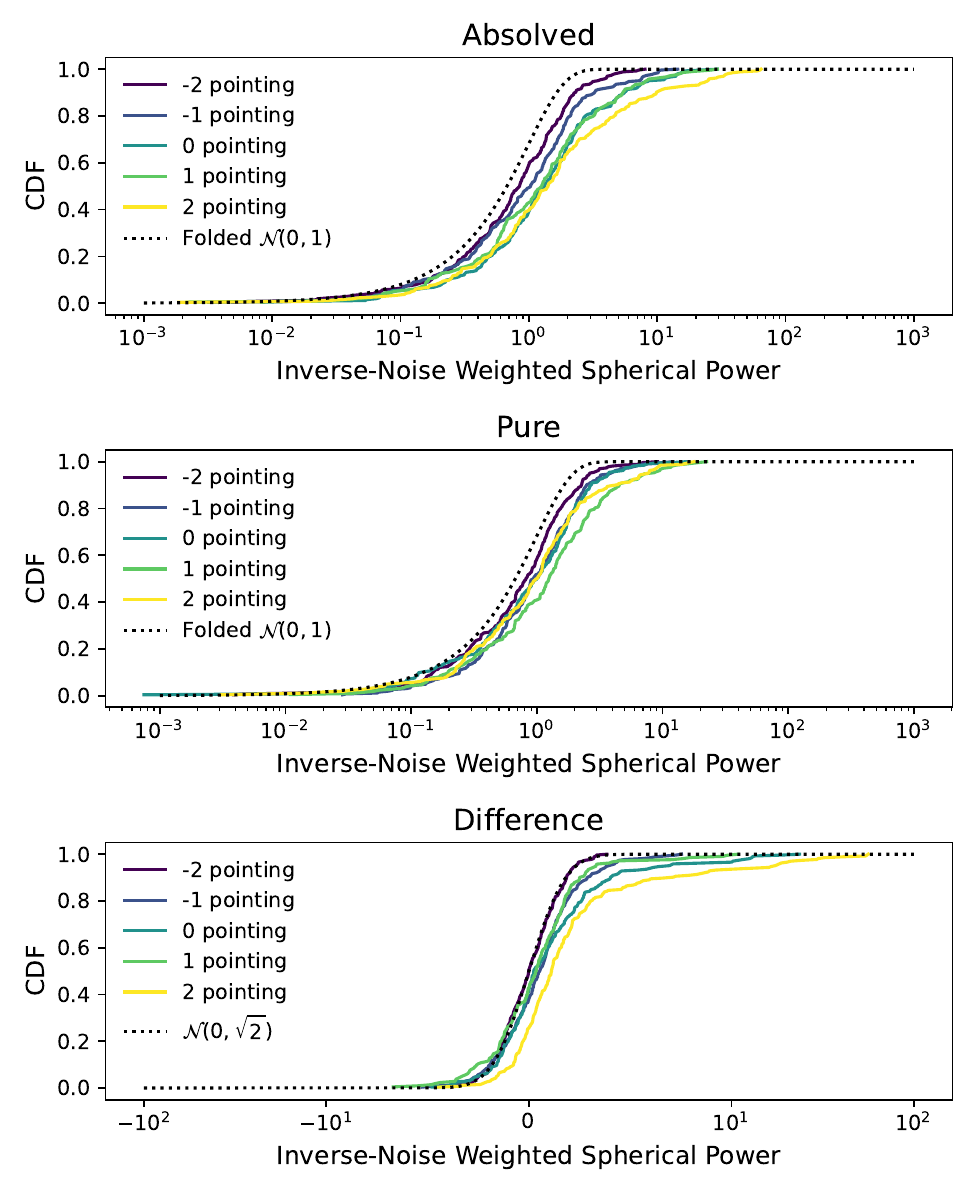}
    \caption{Top, middle: Cumulative density functions for the absolute value of the inverse-noise weighted spherical power neglecting the foreground wedge and coarse band harmonics. The top row, which shows measurements from absolved observations, appears ordered by pointing from East to West for most of the domain. The pure observations are not as clearly ordered or distinct from one another. Neither data set looks consistent with pure noise, however. Bottom: Cumulative density function for the difference between corresponding inverse-noise weighted spherical power of the top two rows (absolved minus pure). The preference of the distributions to lie towards the right of the dotted curve, which reflects a null hypothesis, shows that absolved observations generally produce power spectra that are more positively outlying than their pure counterparts.}
    \label{fig:sphere_by_pointing}
\end{figure*}

This feature almost exclusively appeared in power spectra made from only E-W polarized dipoles and has similar morphology across all identified sets. For the 14 pure observation sets we identified, 12 of their absolved counterparts were also identified in the 46 identified sets (that is, 24 of the 46 sets were LST-matched pure/absolved pairs). The power is usually substantially worse in the absolved sets. The significantly larger severity and prevalence of this feature in the absolved sets suggests that this signature is likely to be residual RFI unidentified by \textsc{SSINS}.\footnote{Recall from \S\ref{sec:occ_analysis} that we flag all frequencies whenever we identify RFI at any frequency, so this cannot be excess power from spectral variations introduced by flagging.} 

\begin{figure*}[t!]
    \centering
    \includegraphics[width=1\linewidth]{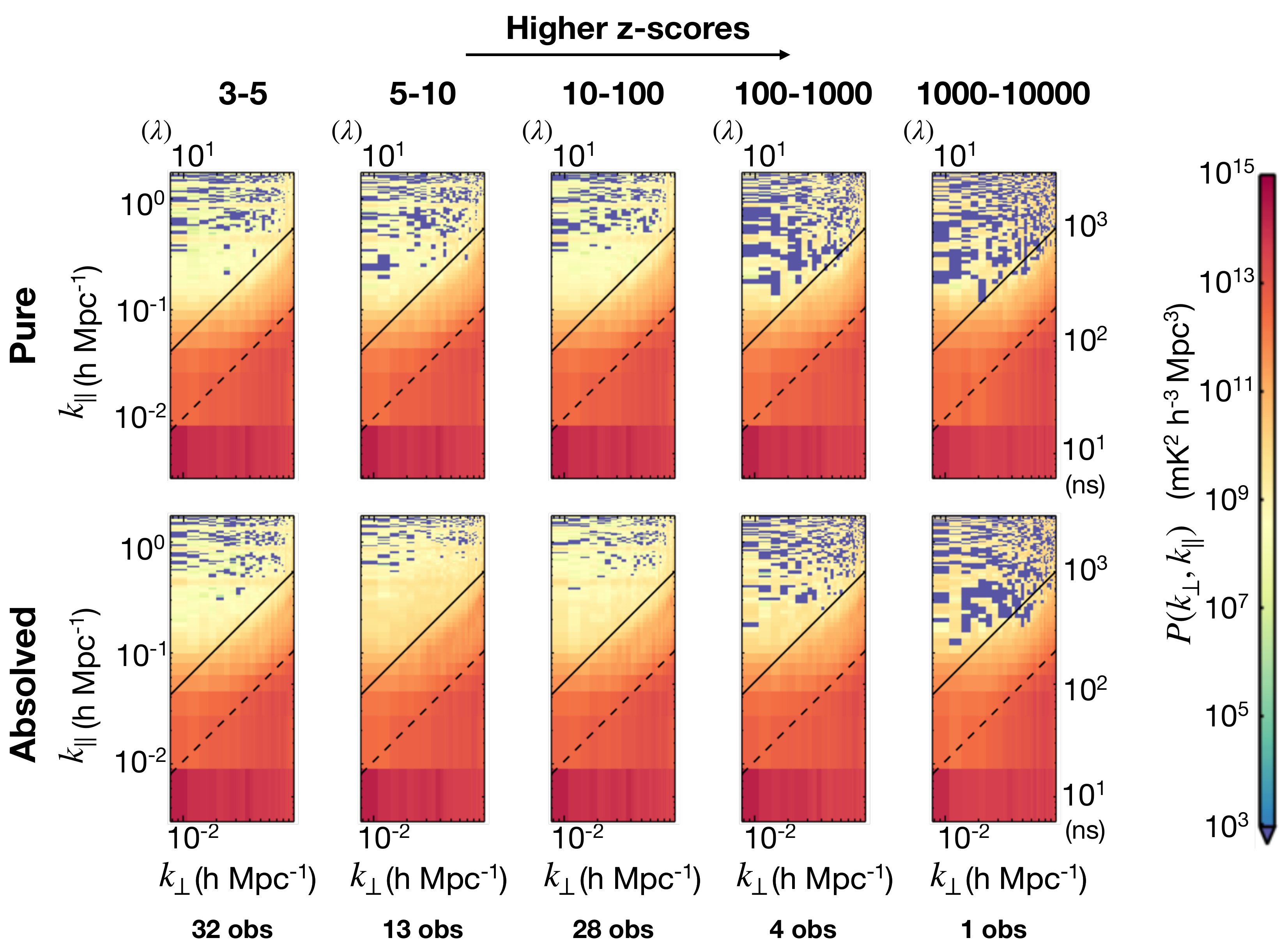}
    \caption{Similar power spectrum jackknife test as Figure \ref{fig:pointing_sequence}, except this time comparing only the +2 (most Westerly) pointing for RFI of any type in the annotated z-score bins at the top of the figure. This comparison demonstrates two features that were a common among our 2d power spectrum jackknife tests. First, excess power is more noticeable for observations in which the identified contaminants had relatively low SSINS z-scores. Since the number of contributing observations ranges from 1 to 32, there is significant variation in the noise levels of each integration, which makes the significance of excess power from residual RFI harder to discern from this Figure alone. Figure \ref{fig:sphere_SSINSz} makes this claim more evident. Second, the +2 pointing appears to be the most obviously contaminated pointing, and contamination can sometimes even be seen in the pure observations. For example, the corresponding pure integration for the power spectra whose SSINS $z$-scores fall between 10 and 100 as well as the one for $z$-scores between 3 and 5 appear to have moderate excess window contamination, though less than their absolved counterparts.}
    \label{fig:brightness_sequence}
\end{figure*}

To further assess the probability of this hypothesis, we co-examined pure/absolved pairs across axes of our predetermined subsets. If the severity or prevalence of the effect is enhanced for choices of RFI parameters that should enhance them, then the evidence for the RFI hypothesis is increased.

\subsection{Power Spectrum Jackknife Test Analysis}

\begin{figure*}
    \centering
    \includegraphics{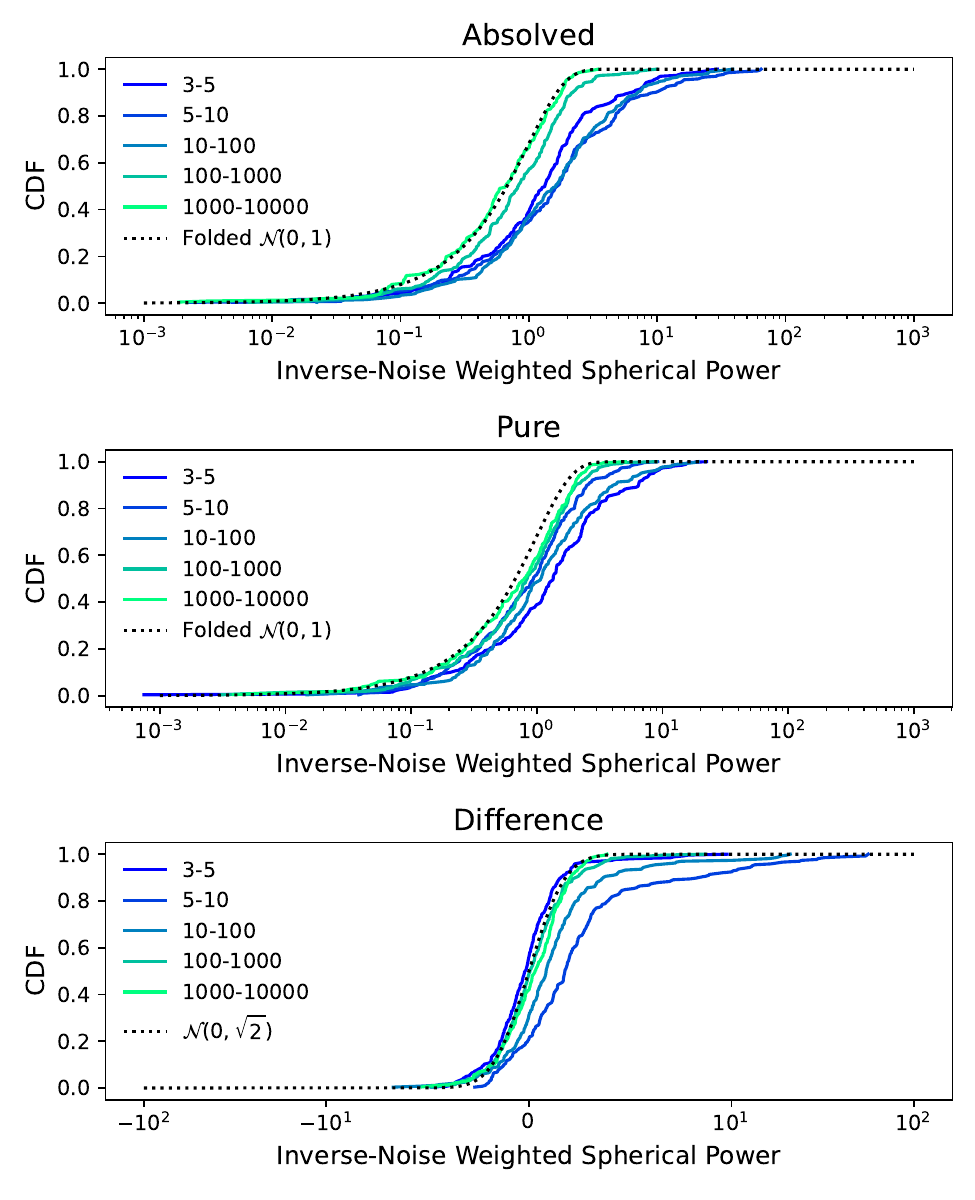}
    \caption{Cumulative distribution functions in the same style as Figure \ref{fig:sphere_by_pointing}, but this time collated by the $z$-score of the events found by \textsc{SSINS}. In this case, the absolved observations appear to fall into two classes. The observations with extremely bright events do not appear to produce significantly outlying power spectra, whereas observations with only faint or moderately bright RFI appear to produce more strongly outlying power spectra. The pure power spectra also seem divided in this way, although the distinction is less obvious. Since observations with lower SSINS $z$-scores are more plentiful, this trend that is present in both the top and middle panels may indicate a systematic that integrates somewhat coherently over the time scales used in these jackknife tests. We again see that absolved power spectra are more often positively outlying than pure ones.}
    \label{fig:sphere_SSINSz}
\end{figure*}

In Figures \ref{fig:pointing_sequence}-\ref{fig:narrow_obs_ps}, we show several representative examples of such comparisons across each of our jackknife axes. In each comparison, we hold two of the three jackknife axes fixed and compare across the third (e.g. Figure \ref{fig:pointing_sequence} holds $z$-score and RFI shape fixed and lets pointing vary). While some comparisons clearly support our premise, others are more anomalous though not inexplicable. We show one more test in Figure \ref{fig:no_flag_flag_diff} in which we compare the absolved and repentant forms of a single DTV-affected observation set (i.e. we make power spectra with and without flags for the same observations), and note an obvious enhancement of the contamination in the repentant power spectrum. The enhancement has the same morphology as the feature we identify in our other jackknife tests, even those with more dubious results. Overall we find the results of these tests to be highly compelling and supportive of the idea that this power spectrum feature is the signature of ultra-faint RFI. In what follows, we walk through the specific comparisons associated with each figure in more detail.

\subsubsection{Pointing Test}

In general, we found that this excess power becomes more noticeable as the array points more Westerly. As an example, we show a sequence of pointings for integrations over observations with any RFI type in them whose z-scores were all between 10 and 100 in Figure \ref{fig:pointing_sequence}. The excess window power suggests that absolved observations possess residual RFI uncaught by \textsc{SSINS}, i.e. ultra-faint RFI, that makes a noticeable effect on the power spectrum measurement. In the full set of integrated power spectra, we see the pointing trend occur regardless of RFI type, and in most z-score bands. We also note that 13 of the 14 pure subsets we initially identified as containing the excess window contamination all came from the 2 most Western pointings. Supposing this is caused by ultra-faint RFI, we expect more prevalence for Western pointings physically since there are more transmitter sites and population centers towards the West of the array than towards the East, and so pointing the telescope beam more West will put associated RFI events in a more sensitive location in the beam.

Since showing every integration comparison available to us in the manner of Figure \ref{fig:pointing_sequence} would be impractical, we show a summary statistic in Figure \ref{fig:sphere_by_pointing} to demonstrate our claimed trends. For each integration subset, we take the spherically averaged power spectra, excluding contributions from the foreground wedge and coarse band harmonics, and divide the power at each wave mode by its noise level as propagated by $\varepsilon$\textsc{ppsilon}. This choice of weighting allows us to more easily compare integrations with varying amounts of contributing observations. 

For the top two panels, we calculate the empirical cumulative distribution function for the absolute value of this quantity, collated by pointing and whether or not the observation set is pure or absolved. In other words, for each distribution function, the pointing and pure/absolved status is fixed, and the brightness within the SSINS and RFI type are unrestricted. We expect that the noise is, at least approximately, zero-mean and Gaussian distributed \citep{Wilensky2023}. Therefore if the data were strongly noise-dominated, we would expect this distribution to resemble a folded Standard Normal distribution (i.e. a normal distribution with mean zero and standard deviation 1, folded over the vertical axis since we are looking at the absolute value). What we observe is that neither the pure nor absolved observation sets appear noise-dominated; deviations from zero are greater than expected under that hypothesis. Importantly, the absolved data (top row) are clearly ordered by pointing from East to West for weighted power values greater than about 1 (though zenith and +1 track each other almost perfectly). The pure data, on the other hand, are not clearly ordered by pointing until the very highest power values. In summary, absolved power spectra from more Westerly pointed data appear consistently more outlying in the EoR window than spectra from more Easterly pointed data, and power spectra from pure data do not show this as obviously or for as great a range of powers.

In the bottom panel, we compute the (signed) difference between the inverse-noise-weighted spherical power between corresponding absolved and pure power spectra and form the cumulative distribution functions collated in the same way as the top two panels, but without taking the absolute value. In this case, we would expect that if the pure and absolved sets were biased by the same amount relative to the noise (potentially indicating similar residual systematics, though this comparison is complicated), then the difference would be noise-like and appear normally distributed with standard deviation equal to $\sqrt{2}$. We find in this panel that the mass of the data is almost exclusively to the right of this null hypothesis, except in the -2 pointing, indicating that absolved power spectra are generally more positively outlying than pure power spectra. Thus, there is a clear statistical difference between the pure and absolved data, and we suggest that this difference is caused by residual RFI. 

We generally find that the excess power is far more noticeable in the East-West polarization than the North-South. We know from inspection of the SSINS that RFI is often seen as unpolarized, i.e. roughly equal strength in all four instrumental polarizations. In some specific instances it can appear polarized. Usually if it appears polarized, it appears stronger in the East-West dipoles. Occasionally it will appear stronger in the North-South dipoles. The apparent polarization of the RFI source is a combination of its broadcast polarization, some propagation effects, and perhaps most importantly, its geometric location relative to the array. 

Since a dipole has no sensitivity along its axis, RFI from the Southern or Northern horizon will appear to the array as if it is East-West polarized, and vice-versa for the Western and Eastern horizons. There are a number of digital television transmitter sites in Western Australia, and therefore a range of possible propagation directions for DTV RFI. The strongest transmitters are in Perth, which is to the South and slightly West. However, direct reception is extraordinarily unlikely due to the extreme remoteness of the MWA. It is more likely that reflections off of aircraft or other transient phenomena make up the bulk of DTV receptions. For example, the largest population center near the MWA is Perth, and most flights that could reflect DTV to the MWA are probably headed towards there (or perhaps towards Geraldton, which is slightly Southwest of the array). Some of these reflections occur when the aircraft is near its destination, and so the RFI source will appear in the Southern sidelobes, e.g. \citet{Wilensky2019}. Since this is also where the strongest transmitters are, we expect more noticeable power spectrum contamination in the East-West polarization.

\subsubsection{SSINS $z$-score Test}

In Figure \ref{fig:brightness_sequence}, we show power spectra in the +2 pointing, which is the most commonly contaminated pointing in our jackknife tests, separated by the SSINS z-scores of the identified events without discriminating between RFI types. Interestingly, we see that integrations of observations with high SSINS z-scores do not display the characteristic excess power in the window. However, this may be due to the fact that such high z-scores are relatively rare, and that there is significant variation in the noise-levels between different members of the jackknife test. It may be that the noise must be integrated down before an obvious RFI signal appears in the window. We also remark that all integrations with RFI z-scores below 100 have a corresponding power spectrum made from pure observations that demonstrates some excess power in the window, though not as severely as the power spectrum for absolved observations. This suggests the presence of false negatives in the \textsc{SSINS} pipeline that can have a significant effect on the power spectrum measurement.

\begin{figure*}[ht!]
    \centering
    \includegraphics[width=\linewidth]{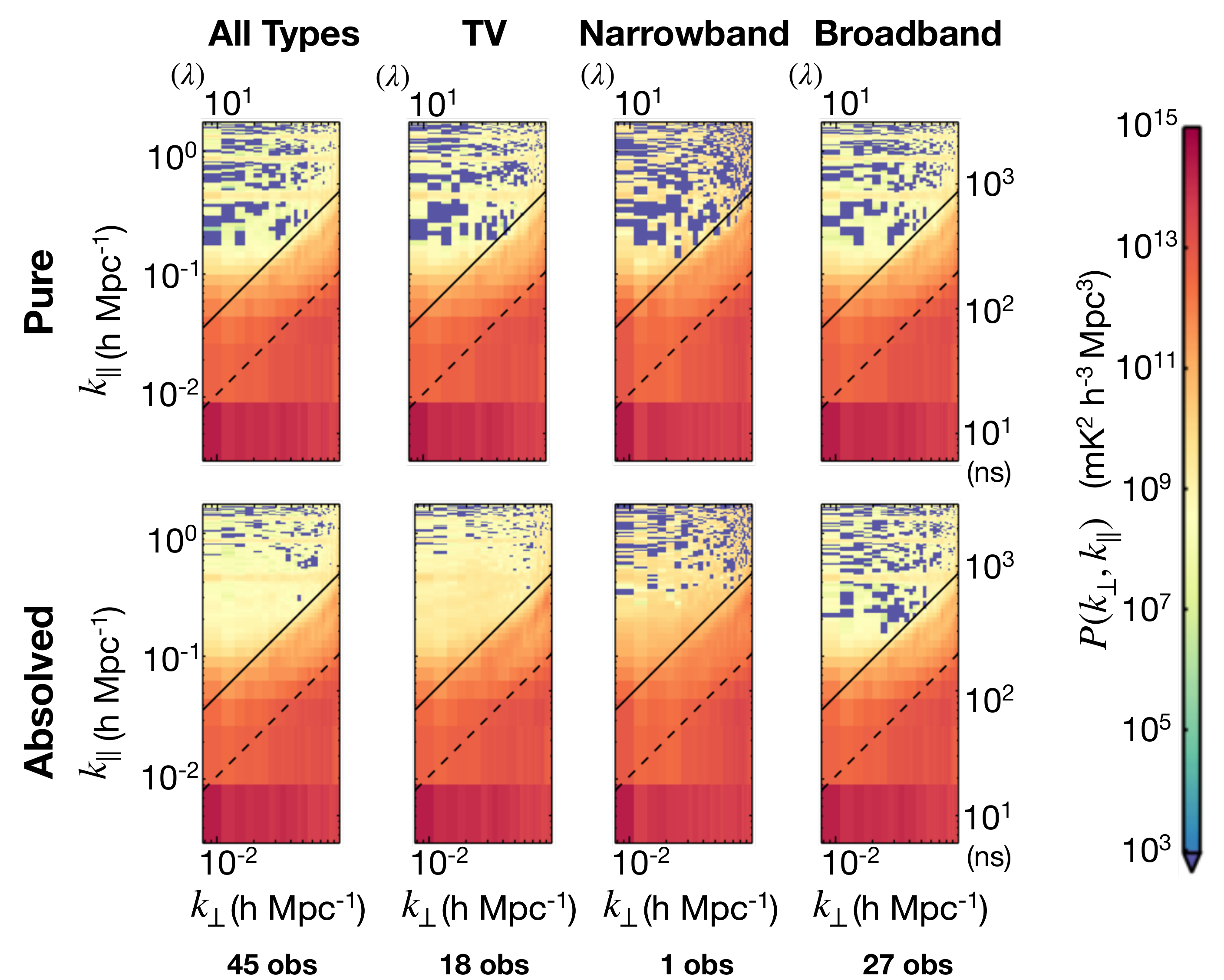}
    \caption{A power spectrum jackknife test for zenith-pointed observations, where spectra are separated by RFI type. These all have RFI events with z-scores between 10 and 100. We find that the region of contamination seems unrelated to the type of RFI. This may be due to multiple RFI types cohabiting these observations.}
    \label{fig:shape_sequence}
\end{figure*}

To summarize the jackknife tests across the SSINS $z$-score axis, we show another collection of cumulative distribution functions in Figure \ref{fig:sphere_SSINSz}. This is the same data as in Figure \ref{fig:sphere_by_pointing}, but it has now been collated according to the $z$-score of the events that were found by \textsc{SSINS}. We find that brighter events in the SSINS do not correlate with stronger outliers in the EoR window. It seems that, within the absolved sets, observations with faint to moderately bright events in the SSINS tend to produce the most strongly outlying power spectra. Interestingly, the LST-matched pure observations also seem to at least weakly display this trend. Since there are generally more observations with fainter events, the power spectra corresponding to the fainter $z$-score bins usually have more integration time. One might expect this behavior in the presence of a systematic that averages coherently at moderate integration depths. While the bottom panel of Figure \ref{fig:sphere_SSINSz} shows that window power is generally more outlying in the absolved power spectra, we speculate all of the observations are affected by a residual systematic that averages coherently in sub-hour length integrations. In a test discussed later in this section, we find that the window contamination displayed in the 2d power spectra so far is enhanced when RFI flags are turned off. Given the similarity of contamination between the two rows of Figure \ref{fig:brightness_sequence}, we suspect that this coherently integrating systematic is indeed residual RFI that is generally worse in the absolved observations but still present in the pure observations.

\begin{figure}
    \centering
    \includegraphics[width=\linewidth]{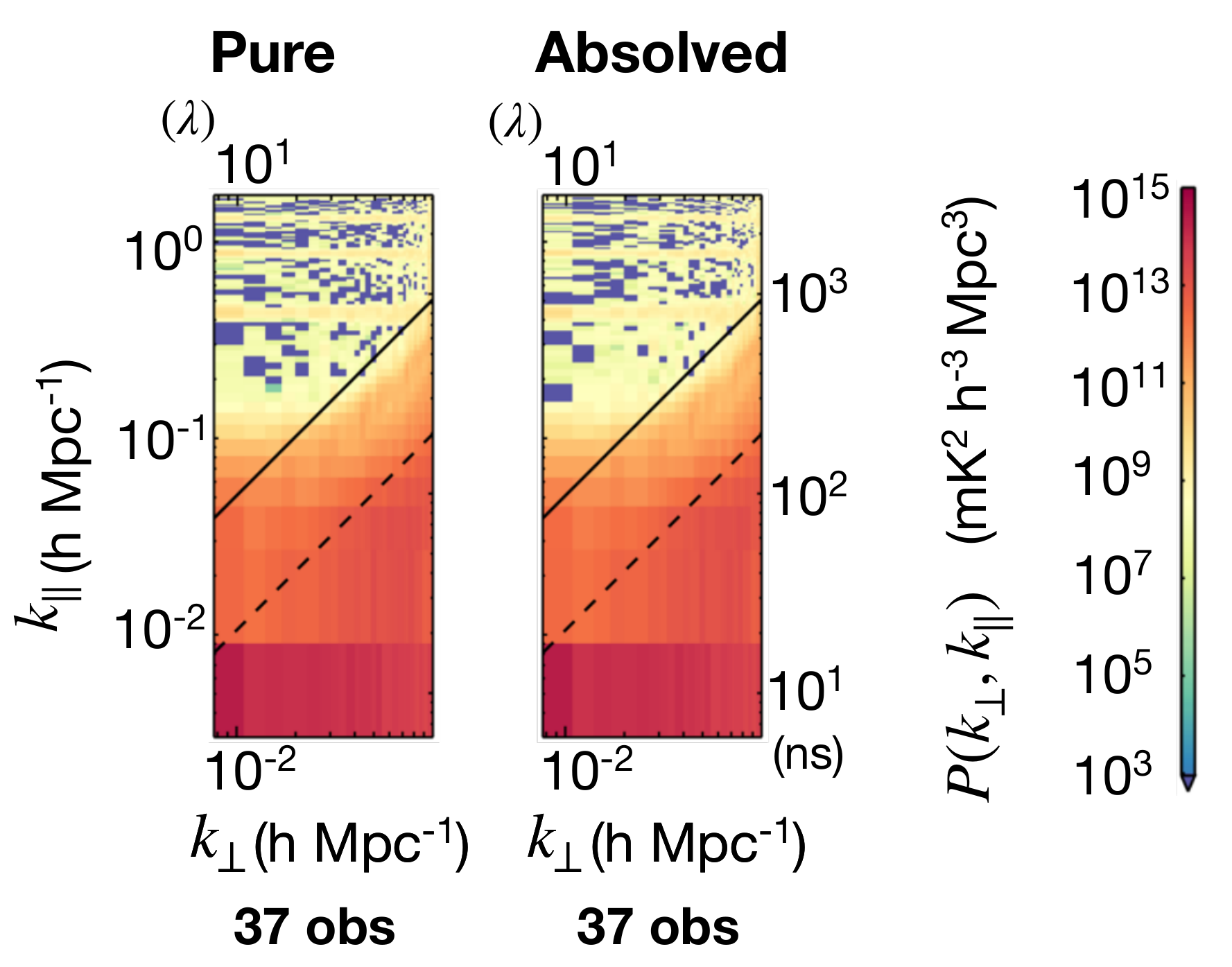}
    \caption{Power spectrum jackknife test for observations classified as containing broadband interference. Excess window power takes similar shape as RFI events in Figure \ref{fig:shape_sequence}.}
    \label{fig:narrow_obs_ps}
\end{figure}

\subsubsection{RFI Shape Test}

From the theoretical discussion in \citet{Wilensky2020}, we expect that different RFI types might have different power spectrum contamination shapes. In Figure \ref{fig:shape_sequence}, we show a jackknife test over RFI shape as determined by \textsc{SSINS}. In general, we find no obvious power spectrum difference among the shapes, contrary to expectation. For instance, we expect narrowband contamination to be approximately constant as a function of $k_\parallel$. Even in more dramatic cases with substantially more noticeable residual interference than is shown in Figure \ref{fig:shape_sequence}, the narrowband RFI does not produce a noticeable constant pattern. If we compare to \citet{Wilensky2020}, we see that narrowband RFI power spectra at a given flux density is generally lower compared to their DTV counterparts, but also that it is constant in $k_\parallel$. Since the noise levels are approximately constant in $k_\parallel$, we expect that if this were only due to narrowband contamination, then the entire EoR window would show clear contamination. 

\begin{figure*}
    \centering
    \includegraphics[width=\linewidth]{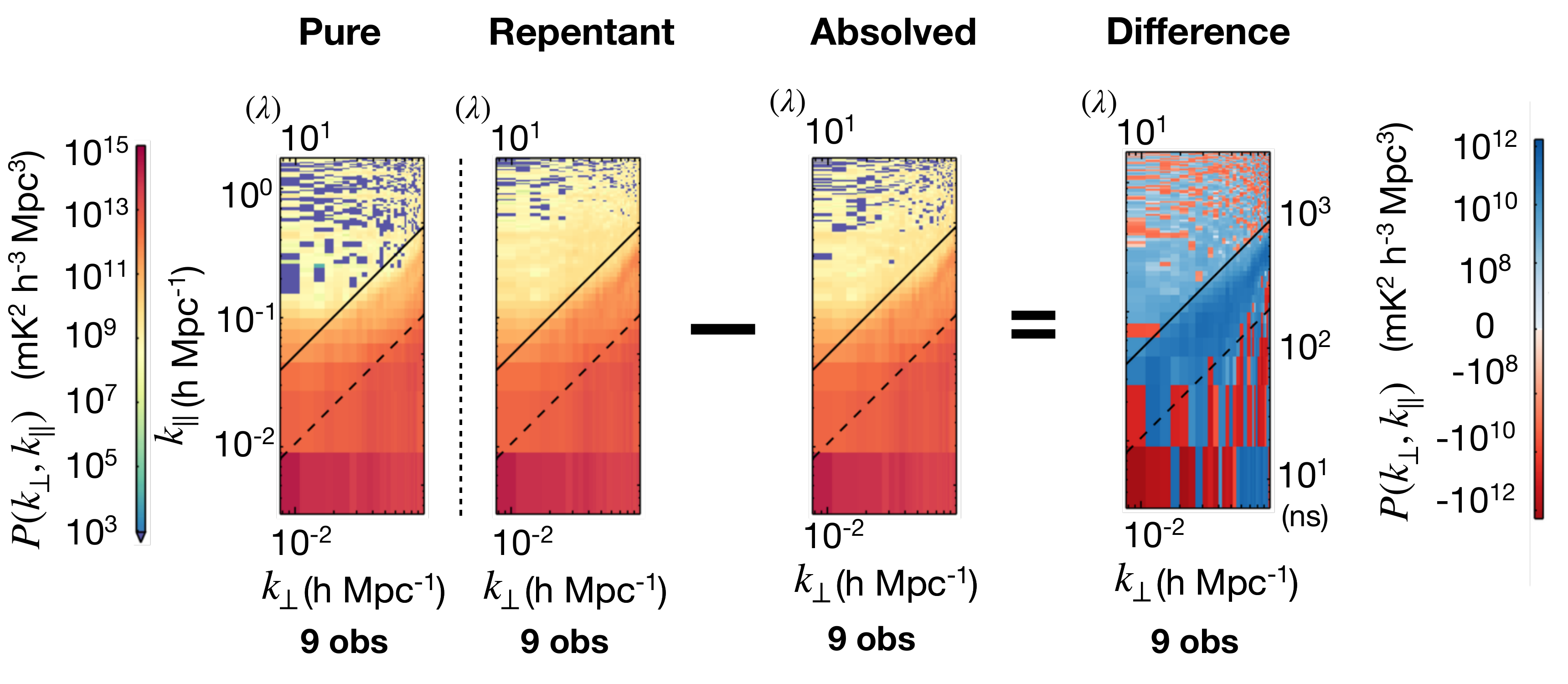}
    \caption{A power spectrum test in which two separate power spectra were made for the same observations, once with flags applied and once without flags applied. In the difference plot, blue bins indicate power is higher without flags. We find that applying the \textsc{SSINS} flags removes power from the window in the exact shape as seen in other Absolved power spectra. Additionally, power is removed in the region of the wedge corresponding to the sidelobes, which is where we observe DTV sources in our images. The clean LST-matched observations are shown on the far left as a reference.}
    \label{fig:no_flag_flag_diff}
\end{figure*}

If there is residual RFI in this power spectrum, it is not narrow in frequency. For this to be true, it must be something broader that \textsc{SSINS} pathologically misses altogether. It could be a relatively stationary source of DTV interference, such as a distant reflector that is coincidentally moving towards the array or nearly so. This reflector may be simultaneously reflecting multiple RFI signals, some of which are narrow in frequency and identified by \textsc{SSINS}. Alternatively, while the RFI appears narrow in frequency in the SSINS, it is possible that the signal has structure outside of the frequencies where most of its power is concentrated. This would lead to non-constant power spectrum contamination, but would require frequency sidebands bright enough to produce such structure, yet faint enough that \textsc{SSINS} cannot flag it as a broad shape. All of our narrowband subintegrations contained less than 16 minutes of data, and all subintegrations that showed window contamination appeared similarly to the narrowband panel of Figure \ref{fig:shape_sequence}, but had just 2-4 minutes of contributing data. This was the only type of RFI that showed any obvious smooth window contamination at such a shallow integration depth. Since so little narrowband interference is observed, it is hard to speculate confidently about its effect on the power spectrum measurements. However, given that this particular shape does seem associated with absolved observations that have narrowband RFI, we are inclined to think this is some sort of RFI effect. 

Since the broadband events in the jackknife test shown in \ref{fig:shape_sequence} showed no obvious excess power, we show a power spectrum for a group of absolved observations in the -1 pointing classified as containing broadband interference flagged by \textsc{SSINS}, along with the LST-matched observations in Figure \ref{fig:narrow_obs_ps}. These z-scores were between 10 and 100, indicating that the interference was moderately bright in the SSINS compared to the thermal background. The excess window power is similar to that for other shapes shown in Figure \ref{fig:shape_sequence}, but much fainter. We show this example to point out that we do see some excess power associated with broadband events. However due to the ambiguous nature of broadband events, as well as the distinct possibility that other RFI may be present in addition to the broadband events caught by \textsc{SSINS}, it is difficult to know if broadband interference in general poses as significant a problem as other types of RFI. We discuss these two points in more detail in what follows.

It is nontrivial to theorize about the expected contamination of broadband interference due to the fact that it could come from several different types of sources. For instance, we have evidence that it could be the sidebands of bright ORBCOMM interference, 40-70 MHz away from the central frequency. The ORBCOMM signal would need to be understood in extreme detail since these sidebands are so far from the allocation. If it were approximately flat-spectrum over the observing band, then it would give no excess power in the window, similar to the foregrounds. Therefore, supposing these signals are indeed responsible for the excess power, there must be some nontrivial structure in the sidebands of these signals. A related possibility is that the sheer brightness of an RFI source has caused clipping or other nonlinearities in the receiver chain. Another broadband emitter is lightning, which produces emissions over a range of radiofrequency scales \citep{Vine1987}. Some theoretical studies suggest that certain types of lightning events can produce relatively flat emissions over the bands considered in this work \citep{Luque2017, Shi2019}, although \citet{Luque2017} shows some nontrivial structure in this range. Observations of lightning emission at VHF radio frequencies are numerous \citep{Hare2018, Hare2020, Pu2021, Sterpka2021, Scholten2022}.  In summary, since the particular signal structure behind the broadband events in the MWA SSINS is as of yet undetermined, we find it difficult to hypothesize about the expected shape of power spectrum contamination.

The similarity in contamination between the different shapes could also be due to other factors. First, as noted in \S\ref{sec:occ_analysis}, \textsc{SSINS} is more optimized for flagging than classification, and the occasional misclassification does occur. Second, it is possible that a given reflector is reflecting multiple different transmissions simultaneously or with a short gap between them. This means that any given observation classified as containing some type of interference does not exclude it from containing another type of interference. In other words, there is some amount of overlap between the subsets shown in Figure \ref{fig:shape_sequence}, and this may be responsible for the similarity in contamination shape.  

\subsubsection{Flags On/Off Test}

To more directly investigate the effects of applying the \textsc{SSINS} flags, we examine an integration in which residual RFI was present in absolved observations by forming the power spectrum using the observations in their repentant form and then subtracting the absolved power spectrum. Note that we are using the same observations in each case, and only choosing whether or not to apply the flags. We show the result in Figure \ref{fig:no_flag_flag_diff}. While the change in the power spectrum is slight, when we plot the difference we see a clear signature in the window matching the shape of excess power that we have seen so far. The flags produce a 35\% difference in data volume, meaning that the noise levels are appreciably different. However, the power difference varies smoothly and is positive for a significant number of adjacent wave modes, indicating the effect is not due strictly to a difference in noise levels. We also see that the flags removed power in the region of the wedge between the dashed and solid lines. The corresponding LST-match of this observation set is shown on the far left of the figure to allow for a comparison in the style of the previous jackknife tests.

The dashed and solid lines in the wedge represent the extent in $k_\parallel$ to which sources at the edge of the primary beam and at the horizon throw power due to the chromatic point-spread-function. The region between these two lines corresponds to the sidelobes. In this example, we chose a handful of observations from the +1 pointing that were identified as containing DTV interference. We expect that a DTV source has a smooth component that appears in the power spectrum wedge, and a spectrally sharp component that throws power into the higher $k_\parallel$ modes. This power spectrum suggests that the RFI sources are consistently located in our sidelobes, corroborating the imaging experiments in \citet{Wilensky2019}. 

\subsubsection{Discussion}

These jackknife tests show that there is very likely a class of ultra-faint RFI uncaught by our current excision methods, and that this RFI will bias power spectrum measurements for a wide swath of spherical modes. The severity of this bias in a deep integration compared to its appearance in these jackknife tests depends on the integration properties of the RFI. We expect RFI will not average coherently over many hours of data, but we cannot reliably speculate about the strength of this bias since we do not have a model of unexcisable RFI. However, since we have identified subsets of the data that contain ultra-faint RFI beyond what is identified during pre-processing, we can at least compute the cleanest power spectrum upper limit possible with this data set and examine whether our efforts are likely to have made a difference. 

\section{Final Power Spectrum and Upper Limits}
\label{sec:limit}

Before extracting an upper limit on the 21-cm EoR power spectrum signal, we perform a final reduction based on the results of the jackknife tests. These final eliminations are based on a qualitative assessment of the EoR window in the power spectrum of various integrated subsets. Since our aim is to make a power spectrum measurement, we perform these cuts conservatively so as to avoid selection bias as much as possible.

\subsection{Final Cuts}

First, we remove all absolved observations. This is motivated by the fact that a substantial number of power spectra involving only absolved observations expressed excess window power compared to their pure, LST-matched counterparts. Furthermore, the shape of this excess power matches the shape of the power spectrum difference in the jackknife test shown in Figure \ref{fig:no_flag_flag_diff}. We term this shape as ``the RFI footprint." Finally, the idea that residual faint RFI might temporally neighbour \textsc{SSINS}-identified samples is motivated both by physical considerations and the occupancy study in \ref{sec:occ_analysis}. We did not remove any pure observations, even if they were immediately adjacent to absolved ones. This leaves 591 observations, which is about 18.4 hours of data. 

\begin{figure*}
    \centering
    \includegraphics[width=\linewidth]{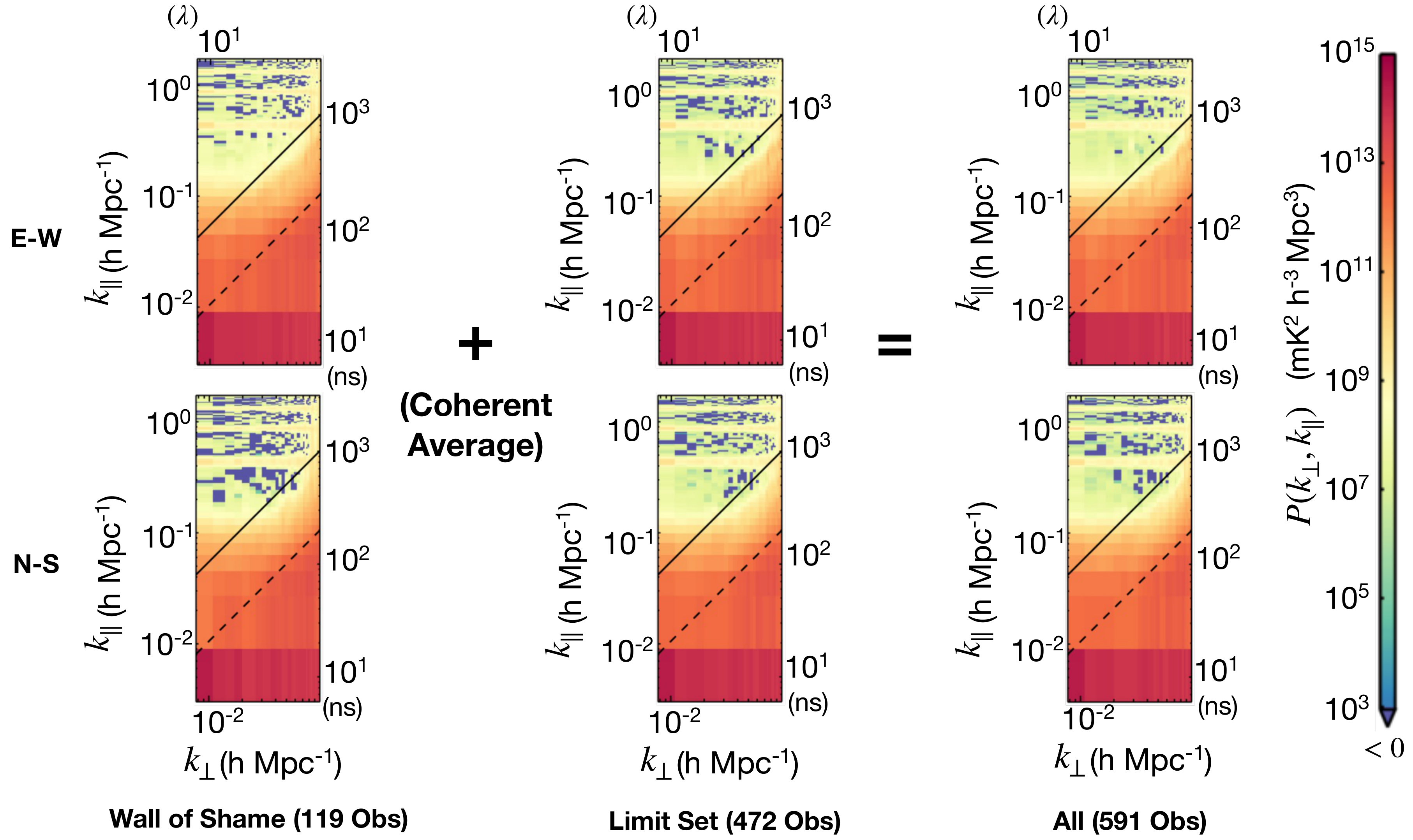}
    \caption{Deep 2d power spectra from three different sets of data, all of which were deemed uncontaminated by \textsc{SSINS}. As with all other 2d power spectra shown in this work, we show the ``dirty" power spectrum i.e. without model subtraction. The Wall of Shame set shows a clear RFI footprint in the spectrum made from the E-W dipoles, but appears noise-limited in the one from the N-S dipoles. On the other hand, the Limit Set power spectra appear similarly regardless of which polarization is used. Specifically both the E-W and N-S power spectra appear systematically dominated below the first coarse band harmonic in the limit set, in contrast to the Wall of Shame set and the majority of shorter power spectrum integrations we inspected. A key difference between this contamination and the RFI footprint is that it does not extend to large $k_\perp$. The coherent average of the two performs to expectation in that adding the Wall of Shame set does not strongly affect the appearance of the power spectrum from the N-S data, and imprints the characteristic RFI footprint in the spectrum made from the E-W data.}
    \label{fig:deep_2d_ps}
\end{figure*}

Next, we note that some integrations involving only pure observations also display this excess window power, although this is substantially less common than in their absolved counterparts. Hoping to remove as much RFI as possible before forming an upper limit, we perform cuts on these observations based on manual inspection of window power. From the integration subsets that were already constructed for the previous stages of the jackknife test, we remove all integrations with fewer than 20 observations that have obvious excess window power. We then reintegrate the remaining observations, separating them roughly by pointing and day, such that there are between 12 and 20 observations (24-40 minutes) per integration. This integration depth is based on balancing the interference-to-noise ratio in the EoR window with the desire to be as fine-grained as possible in removing observations with this metric. In other words, we want to use the smallest integration subset possible that allows us to identify ultra-faint RFI. From these new integration subsets, we again remove any whose resulting power spectra show obvious excess window contamination. Since we are wary of selection bias, we attempt to be as conservative as possible during this step. We show the power spectra of all integrations that are removed during this stage in Appendix \ref{app:wall_of_shame}. This removes a further 119 observations, which we designate as the ``Wall of Shame" set, and leaves 472 observations (14.7 hours), which we call the ``Limit Set." We also note that no observations from the +2 (most Westerly) pointing survive this test. 

\subsection{Deep Cylindrica Power Spectra}

We show deep 2d power spectra from each of the three sets (wall of shame, limit set, and coherent average of the two) calculated over the entire observing band in Figure \ref{fig:deep_2d_ps}. The power spectrum estimated from the East-West dipole data for the Wall of Shame set displays the characteristic RFI footprint. On the other hand, the power spectrum made using only the North-South dipole data appears noise-limited in the window. Thus, we see that even when we deeply integrate observations known to have RFI contamination, the footprint appears polarized. This can be understood by considering that the brightest DTV transmitters in Western Australia are nearly due South of the array. Since the North-South dipoles are not sensitive to the Southern horizon, we generally expect the RFI flux to be stronger in the East-West dipoles than the North-South ones. 

In contrast, the Limit Set displays power spectra that look very similar between the different dipoles. This is remarkable in the first instance since previous power spectrum upper limits with the MWA are preferentially deeper from the North-South dipoles compared to the East-West ones. It is possible that this long-standing preference is related to the difference in RFI content between dipoles. We could investigate this claim more rigorously by extending these RFI mitigation techniques to previously analyzed data that express the preference. Both polarizations appear to have a systematic source of contamination in the lower-left corner of the EoR window. What distinguishes this systematic from the characteristic RFI footprint is that it does not extend to higher $k_\perp$ modes. Furthermore, it is relatively equal in power across the polarizations, which would be peculiar for RFI given the Wall of Shame power spectra. We suggest that this is possibly a different systematic effect that is unaccounted for in this analysis. Note that when we coherently average the Limit Set and Wall of Shame, we observe the contamination extend to higher $k_\perp$ in the power spectrum from E-W data, but not in the one from N-S data. 

As a counterargument, we remark that there are many other DTV transmission sites in Western Australia than the aforementioned ones South of the array, distributed in varying directions. Most of these transmission sites are significantly less powerful than the one in Perth, some by multiple orders of magnitude. Since this is a complex scattering problem, it is difficult to deduce exactly how bright each transmitter should appear. However, some basic scaling arguments suggest that the RFI flux should still be dominated by the Southern transmitters. Therefore, it is possible that this could be extremely faint RFI from the dimmer transmitters, while the flux from the brighter transmitters was removed by our repeated selections. However, for this to be true, there would need to be an explanation for why the longer baselines (higher $k_\perp$ modes) display less power from these hypothetical RFI events. None of the transmitters are close enough for regular direct reception, and therefore the nature of the scatter would have to be different such that the long baselines do not respond as strongly to the RFI. This is physically unusual, since the scatterers are likely to be the same class of object regardless of where the transmitter is. One possibility is that the closer transmitters tend to scatter closer to the array. If the scatterer is sufficiently close, it might appear less point-like and thus the longer baselines might not respond as strongly to the interference. However, this would be an unusual coincidence and is therefore less likely to explain a longitudinal trend in a large data set.

Yet another possibility is locally generated RFI within the array e.g. from electronics. This type of RFI, depending on its source, might be seen longitudinally in a large data set and may appear roughly equally in the different dipole arms depending on its location. Furthermore, if it is close enough, we might expect it to preferentially affect shorter baselines compared to longer ones, and we might also expect it to preferentially affect some pointings compared to others. However, locally generated RFI is often bright and therefore eliminated early in the data analysis. This would have to be a particularly pernicious form of ultra-faint locally generated RFI. We cannot rule out this hypothesis from this analysis alone. Examining the near-field RFI environment is an important topic of ongoing investigation within the MWA collaboration, however we consider it outside the scope of this work.

\subsection{Spherical Power Spectrum Upper Limits}

\begin{figure*}
    \centering
    \includegraphics[scale=0.4]{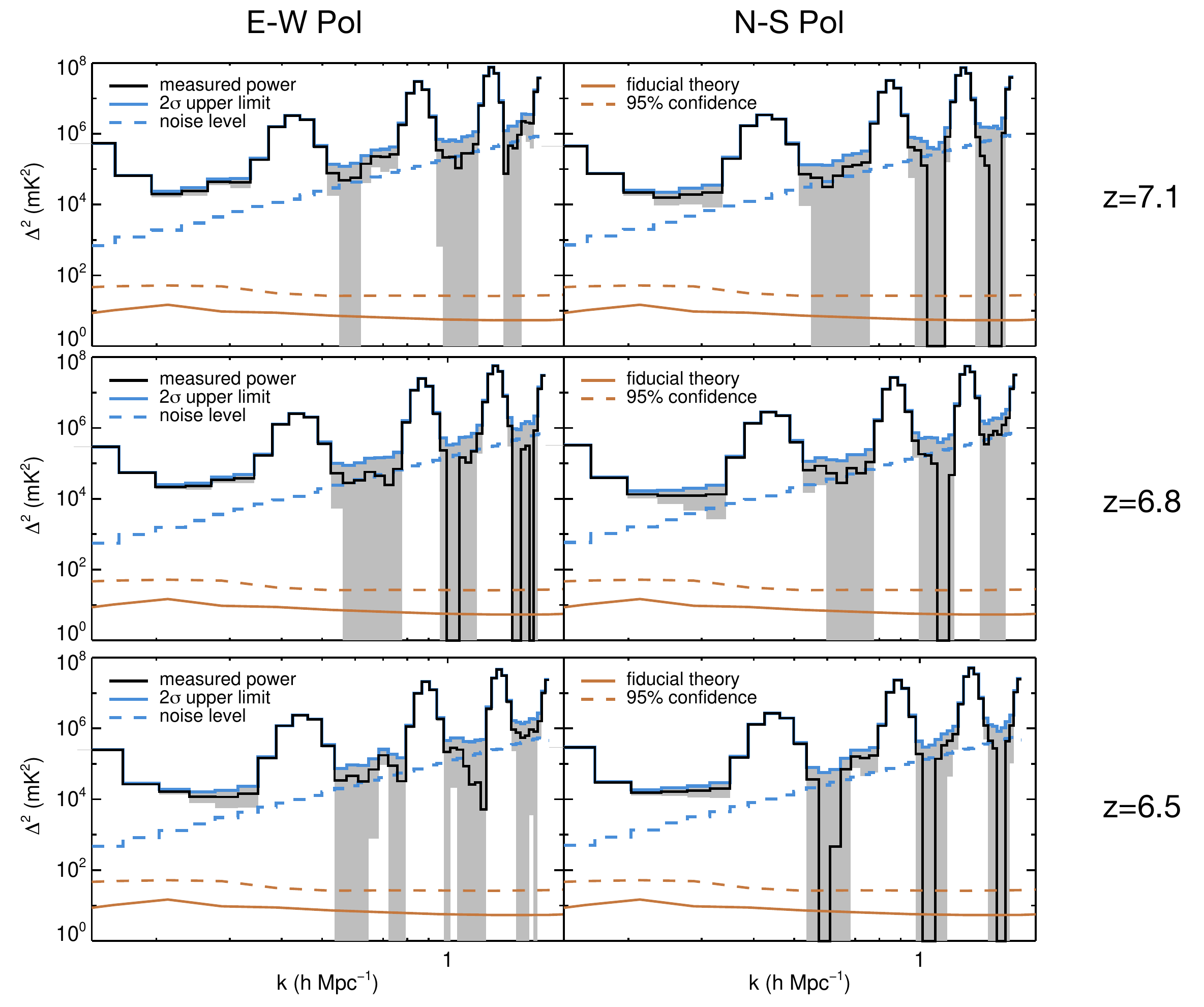}
    \caption{Spherical power spectra for the Limit Set, along with $2\sigma$ upper limits, a fiducial theory model for the EoR signal whose implementation is described in \citet{Barry2019b}. In short, the astrophysical constraints from \citet{Park2019} are used in combination with \textsc{21cmmc} \citep{Greig2015} to generate samples from a probability distribution of 21-cm power spectra. The fiducial model and theoretical confidence interval are then calculated from these samples (brown solid and dashed lines, respectively). We notice that the range of wave modes between the coarse band harmonics are almost all noise-dominated. This could be a direct consequence of the rigorous RFI cuts we have employed. Compare to limits at the same redshifts in \citet{Li2019}.}
    \label{fig:final_limits}
\end{figure*}

We report the final limits using the 472-observation set. We also integrate the 119 observation set separately to examine how the residual RFI integrates in the power spectrum. Finally, we make an integration with the entire set of 591 observations to see how this final cut affects the power spectrum. We draw limits for three different redshifts: 6.5, 6.8, and 7.1. These correspond to three different bands, each of which is exactly half the 30.72 MHz bandwidth of the MWA EoR highband. They are centered 1/4, 1/2, and 3/4 the way through the band. Since we use a Blackman-Harris taper, the effective bandwidth of each measurement is 7.68 MHz i.e. 1/4 of the band \citep{Harris1978}. This makes the measurements roughly independent.\footnote{More specifically, these frequency bands have 50\% overlap, and the minimum 4-term Blackman-Harris window function has a 50\% overlap correlation of 3.8\% \citep{Harris1978}.} We display the spherical power spectra along with the 2-$\sigma$ upper limits in Figure \ref{fig:final_limits}. Our lowest upper limit that we report is $\Delta^2 \leq 1.61\cdot10^4 \text{ mK}^2$ at $k=0.258\text{ h Mpc}^{-1}$ using the East-West polarized dipoles at a redshift  of 6.5. 

\subsection{Discussion}

These limits are calculated assuming a positive-defnite signal-like component and thermal noise, as in Appendix A of \citet{Li2019}. No distinction is made between systematic contributions to the signal-like term and the cosmological signal of interest. With information about the strength of various contaminating effects, one could produce an upper limit that takes these into account. For example, if we estimated an RFI excision algorithm to be effective down to some nominal flux level and had a statistical model of the remaining RFI sources e.g. \citet{Offringa2013b}, we could incorporate our uncertainty about this systematic into the upper limit calculation. In general, this is frustrated by the fact that we must extrapolate about RFI we did not detect based on RFI we detected. This is necessarily dependent on the flagging strategy. For \textsc{SSINS} this may be particularly complicated, since flags are generated on time-differenced data and therefore undetected RFI may belong to an entirely different class not necessarily distinguished by total brightness. One way to build up this model is to complement an existing RFI excision strategy with the following imaging strategy.

While some RFI might be stationary within an observation, most RFI will not be stable in celestial coordinates from observation to observation, particularly if those observations are significantly lagged with respect to civil time. Therefore, one could difference images of appropriate observations and look for strongly outlying pixels. This could be done at varying integration depths for relatively cheap computational cost within the current pipeline. In other words, this could be used to find full-array RFI within an observation, and then again at a deeper integration depth to find ultra-faint RFI. This would simultaneously allow for RFI identification and localization of the RFI sources, which will be important for developing models of the RFI statistics of the observatory. These models can ultimately be used for determining the depth to which RFI excision needs to be performed. In fact, a similar strategy for RFI identification was implemented in \citet{Prabu2020} and further tested with the Engineering Development Array in \citet{Tingay2020}, though only at a few seconds of integration depth rather than combining multiple observations. Thus, there is some promise that such a method could work, although clearly some experimentation would be required to determine exactly how to implement it in deeper integrations and between different observations. 

In addition to time differencing strategies, we can imagine incorporating significantly more specific prior information regarding the RFI signals than is currently used by common flaggers. The details about reflector trajectories, details about the signal properties that are known a priori by the transmitting parties, and an atmospheric model could be combined to produce an accurate physical model of a received RFI signal. For example, \cite{Prabu2022} demonstrates that RFI detection can be greatly enhanced by two different methods that incorporate more specific prior information about the RFI. The first method, called ``shift-stacking," coherently averages in image space over a region expected to be occupied by a known reflector. The second method implements a post-correlation refocusing of the MWA to a distance at which we expect a reflector. In brief, at these distances, there is still non-negligible curvature in the wave fronts of the reflected signal, and the visibility phase can be adjusted to account for this. One could imagine checking for catalogs of reflectors that ought to be present in the data using methods like these, or conversely, using methods like these to help build such catalogs.

A clear challenge with a physical modeling approach is the construction of a complete model, particularly when different propagation phenomena may produce spectral distortions in the RFI signal. In general, programming a flagger to identify a particular spectral distortion can implicitly make it less sensitive to other types of spectral distortion unless each class of distortion is included in the model.\footnote{For example, the match filter for a decreasing spectral tilt will be less sensitive to distortion that produces an increasing spectral tilt.} However, assuming such a model can be made, this will not only enhance flagging performance but also allow us to more accurately predict statistical properties about RFI that was left unflagged. This ultimately will allow for more rigorous uncertainty estimates on power spectrum measurements since there will be an explicit model for unmitigated systematics.

We observe that the range of wave modes between the first and second coarse band harmonics is noise-limited in most of the presented spherical power spectra. This suggests that adding more high-quality data would probably improve the limit. Interestingly, if we evaluate the upper limit using the coherent average of the Wall of Shame and Limit Set, we see that it degrades the limit systematically between the first and second coarse band harmonics (which was not the region of the EoR window used for the final cut) for redshifts 6.5 and 6.8 in the East-West polarization. The effect is slight, but significantly consistent across the stated bins. The lowest limit of any spherical mode that we observed, which we do not report, occurs in this coherent average below the first coarse band harmonic. We do not report that limit since the coherent average was known to be RFI-contaminated before we constructed it. 

\section{Conclusion}
\label{sec:conc}

In this work, we analyzed a season of data from the MWA phase I with a particular focus on the presence and potential effect of RFI in 21-cm power spectrum measurements. The goal was to examine to what extent ultra-faint RFI exists within deep power spectrum integrations despite the deployment of proven mitigation methods, as well as to see whether this ultra-faint RFI contamination was likely to deteriorate constraints on the 21-cm cosmic reionization signal. 

This RFI analysis, in combination with other quality metrics, ultimately resulted in the decision to cut about 85\% of the data from this season to produce the best upper limit. Even in the absence of an ionospheric quality cut, we still would have cut at least 54\% of the data based on RFI content alone. This level of data loss is catastrophic from a scientific standpoint. Starting from just 2.5 hours per night that we select due to contamination from the galaxy at other sidereal times, it would take dozens of seasons of data of the same quality, instrumental sensitivity, and cut decisions to reach a confident detection. While a more sensitive instrument such as the SKA might need fewer high-quality observing hours to reach the same detection confidence, we expect that the global RFI environment is worsening with the addition of extremely large satellite constellations in low Earth orbit, and so data cuts of this style may be even more severe for future experiments. 

From our RFI analysis, we identified a host of RFI that evaded post-correlation detection by observing a residual RFI footprint in measured cylindrical power spectra that is enhanced when RFI flags are turned off. Residual RFI within observations already known to contain RFI is motivated by the physical properties of RFI reception, which we know to be largely due to transient scatterers. We found the timelike properties of the RFI flags to be consistent with this physical picture. We also found that the RFI footprint was most prevalent the further West the array was pointed. This was true to such an extent that after all data cuts based on independent pre-power-spectrum metrics, all remaining subintegrations in the most Westerly pointed data showed the RFI footprint. The footprint also appeared preferentially in power spectra made from only the East-West aligned dipoles as compared to those made from the North-South aligned dipoles. These two observations are consistent with the fact that the brightest DTV transmitters are nearly due South of the array, where the North-South dipoles are less sensitive, and offset slightly West. 

We performed a deep integration, cutting all observations identified as contaminated by any amount as well as any subintegrations that clearly presented the RFI footprint. The resulting power spectra appeared consistent between the polarizations, and did not have an obvious RFI footprint, although some systematic contamination is visible in the lower-left corner of the EoR window. This similarity between the polarizations is a new feature of this limit compared to previous ones made with the MWA, and may be a result of the enhanced RFI mitigation. When we only use the subintegrations known to possess an RFI footprint, we obtain power spectra that are systematically limited in the East-West polarized data, but not in the North-South polarized data, and the presence of a footprint is more obvious. Finally, when we coherently average all the data together, we see systematic dominance in both polarizations, but the systematic dominance only extends to larger perpendicular wave modes in the East-West polarized data. This suggests that ultra-faint RFI is unlikely to disappear through dilution in a coherent average, and by corollary, that the systematic domination in the EoR window of the limit set may in fact be something other than RFI. However, since (a) the behavior of RFI in coherent averaging schemes is poorly understood, (b) the RFI budget for EoR detection is strict, and (c) we can willfully construct relatively deep integrations with noticeable RFI footprints, we ultimately suggest solving the problem of modeling ultra-faint RFI in deep power spectrum integrations. 

We use the subset of data least likely to be contaminated in order to set upper limits on the cosmological 21-cm signal. In total, about 85\% of the initial data selection was not used for the upper limit. Our deepest upper limit was $\Delta^2 \leq 1.61\cdot10^4 \text{ mK}^2$ at $k=0.258\text{ h Mpc}^{-1}$ and $z=6.5$.  While this is not the lowest upper limit set with MWA Phase I, we remark that these limits are noise dominated in the majority of modes higher than the first coarse band harmonic, not including the harmonics themselves. Since RFI produces significant power to extremely high line-of-sight wave modes, we suggest that this prominence of noise-dominated bins at high $k$ (in excess of previous limits at similar depths) may be a consequence of our extremely thorough RFI cuts. We note that this required cutting about 2/3 of the data spared by the ionospheric cut. Since the RFI environment is unlikely to improve in future seasons, obtaining enough high-quality, RFI-free data for a sensitive EoR measurement appears a daunting task. We therefore highlight not only a need to improve RFI mitigation, but to do so in a data-sparing way.

\section{Acknowledgements}

We acknowledge Jacob Burba, Hugh Garsden, Steven Tingay, and Zheng Zhang for helpful comments. We thank the development teams of \textsc{pyuvdata}, \textsc{FHD},
and $\varepsilon$\textsc{ppsilon} which enabled this work. This work was
directly supported by NSF grants AST-1643011, AST-
1613855, OAC-1835421, AST-1506024, AST-1907777, AST-2106510, AST-2107538, AST-2205471, and AST-2228990. This project
has received funding from the European Research Council
(ERC) under the European Union’s Horizon 2020 research
and innovation programme (grant agreement No. 948764). We acknowledge the University of Washington eScience Institute and UW Research Computing Azure Cloud Credits program for providing computational resources.

This scientific work makes use of the Murchison Radio-
astronomy Observatory, operated by CSIRO. This scientific work uses data obtained from Inyarrimanha Ilgari Bundara / the Murchison Radio-astronomy Observatory. We acknowledge the Wajarri Yamaji People as the Traditional Owners and native title holders of the Observatory site. Establishment of CSIRO's Murchison Radio-astronomy Observatory is an initiative of the Australian Government, with support from the Government of Western Australia and the Science and Industry Endowment Fund. Support for the operation of the MWA is provided by the Australian Government (NCRIS), under a contract to Curtin University administered by Astronomy Australia Limited. This work was supported by resources provided by the Pawsey Supercomputing Research Centre with funding from the Australian Government and the Government of Western Australia.

\bibliography{main}

\begin{thebibliography}{}
\expandafter\ifx\csname natexlab\endcsname\relax\def\natexlab#1{#1}\fi
\providecommand{\url}[1]{\href{#1}{#1}}
\providecommand{\dodoi}[1]{doi:~\href{http://doi.org/#1}{\nolinkurl{#1}}}
\providecommand{\doeprint}[1]{\href{http://ascl.net/#1}{\nolinkurl{http://ascl.net/#1}}}
\providecommand{\doarXiv}[1]{\href{https://arxiv.org/abs/#1}{\nolinkurl{https://arxiv.org/abs/#1}}}

\bibitem[{{Abdurashidova} {et~al.}(2022){Abdurashidova}, {Aguirre},
  {Alexander}, {Ali}, {Balfour}, {Beardsley}, {Bernardi}, {Billings}, {Bowman},
  {Bradley}, {Bull}, {Burba}, {Carey}, {Carilli}, {Cheng}, {DeBoer}, {Dexter},
  {de Lera Acedo}, {Dibblee-Barkman}, {Dillon}, {Ely}, {Ewall-Wice}, {Fagnoni},
  {Fritz}, {Furlanetto}, {Gale-Sides}, {Glendenning}, {Gorthi}, {Greig},
  {Grobbelaar}, {Halday}, {Hazelton}, {Hewitt}, {Hickish}, {Jacobs}, {Julius},
  {Kern}, {Kerrigan}, {Kittiwisit}, {Kohn}, {Kolopanis}, {Lanman}, {La Plante},
  {Lekalake}, {Lewis}, {Liu}, {MacMahon}, {Malan}, {Malgas}, {Maree},
  {Martinot}, {Matsetela}, {Mesinger}, {Molewa}, {Morales}, {Mosiane},
  {Murray}, {Neben}, {Nikolic}, {Nunhokee}, {Parsons}, {Patra}, {Pascua},
  {Pieterse}, {Pober}, {Razavi-Ghods}, {Ringuette}, {Robnett}, {Rosie}, {Sims},
  {Singh}, {Smith}, {Syce}, {Thyagarajan}, {Williams}, {Zheng}, \& {HERA
  Collaboration}}]{HERA2022}
{Abdurashidova}, Z., {Aguirre}, J.~E., {Alexander}, P., {et~al.} 2022, \apj,
  925, 221, \dodoi{10.3847/1538-4357/ac1c78}

\bibitem[{{Barry}(2018)}]{Barry2018}
{Barry}, N. 2018, PhD thesis, University of Washington

\bibitem[{{Barry} {et~al.}(2019{\natexlab{a}}){Barry}, {Beardsley}, {Byrne},
  {Hazelton}, {Morales}, {Pober}, \& {Sullivan}}]{Barry2019a}
{Barry}, N., {Beardsley}, A.~P., {Byrne}, R., {et~al.} 2019{\natexlab{a}},
  Publications of the Astronomical Society of Australia, 36, e026,
  \dodoi{10.1017/pasa.2019.21}

\bibitem[{{Barry} {et~al.}(2016){Barry}, {Hazelton}, {Sullivan}, {Morales}, \&
  {Pober}}]{Barry2016}
{Barry}, N., {Hazelton}, B., {Sullivan}, I., {Morales}, M.~F., \& {Pober},
  J.~C. 2016, Monthly Notices of the Royal Astronomical Society, 461, 3135,
  \dodoi{10.1093/mnras/stw1380}

\bibitem[{{Barry} {et~al.}(2019{\natexlab{b}}){Barry}, {Wilensky}, {Trott},
  {Pindor}, {Beardsley}, {Hazelton}, {Sullivan}, {Morales}, {Pober}, {Line},
  {Greig}, {Byrne}, {Lanman}, {Li}, {Jordan}, {Joseph}, {McKinley}, {Rahimi},
  {Yoshiura}, {Bowman}, {Gaensler}, {Hewitt}, {Jacobs}, {Mitchell}, {Udaya
  Shankar}, {Sethi}, {Subrahmanyan}, {Tingay}, {Webster}, \&
  {Wyithe}}]{Barry2019b}
{Barry}, N., {Wilensky}, M., {Trott}, C.~M., {et~al.} 2019{\natexlab{b}}, The
  Astrophysical Journal, 884, 1, \dodoi{10.3847/1538-4357/ab40a8}

\bibitem[{{Beardsley} {et~al.}(2016){Beardsley}, {Hazelton}, {Sullivan},
  {Carroll}, {Barry}, {Rahimi}, {Pindor}, {Trott}, {Line}, {Jacobs}, {Morales},
  {Pober}, {Bernardi}, {Bowman}, {Busch}, {Briggs}, {Cappallo}, {Corey}, {de
  Oliveira-Costa}, {Dillon}, {Emrich}, {Ewall-Wice}, {Feng}, {Gaensler},
  {Goeke}, {Greenhill}, {Hewitt}, {Hurley-Walker}, {Johnston-Hollitt},
  {Kaplan}, {Kasper}, {Kim}, {Kratzenberg}, {Lenc}, {Loeb}, {Lonsdale},
  {Lynch}, {McKinley}, {McWhirter}, {Mitchell}, {Morgan}, {Neben},
  {Thyagarajan}, {Oberoi}, {Offringa}, {Ord}, {Paul}, {Prabu}, {Procopio},
  {Riding}, {Rogers}, {Roshi}, {Udaya Shankar}, {Sethi}, {Srivani},
  {Subrahmanyan}, {Tegmark}, {Tingay}, {Waterson}, {Wayth}, {Webster},
  {Whitney}, {Williams}, {Williams}, {Wu}, \& {Wyithe}}]{Beardsley2016}
{Beardsley}, A.~P., {Hazelton}, B.~J., {Sullivan}, I.~S., {et~al.} 2016, The
  Astrophysical Journal, 833, 102, \dodoi{10.3847/1538-4357/833/1/102}

\bibitem[{{Benkevitch} {et~al.}(2016){Benkevitch}, {Rogers}, {Lonsdale},
  {Cappallo}, {Oberoi}, {Erickson}, \& {Baker}}]{Benkevitch2016}
{Benkevitch}, L.~V., {Rogers}, A.~E.~E., {Lonsdale}, C.~J., {et~al.} 2016,
  arXiv e-prints, arXiv:1608.04367.
\newblock \doarXiv{1608.04367}

\bibitem[{{Byrne} {et~al.}(2021){Byrne}, {Morales}, {Hazelton}, \&
  {Wilensky}}]{Byrne2021}
{Byrne}, R., {Morales}, M.~F., {Hazelton}, B.~J., \& {Wilensky}, M. 2021,
  \mnras, 503, 2457, \dodoi{10.1093/mnras/stab647}

\bibitem[{{Byrne} {et~al.}(2019){Byrne}, {Morales}, {Hazelton}, {Li}, {Barry},
  {Beardsley}, {Joseph}, {Pober}, {Sullivan}, \& {Trott}}]{Byrne2019}
{Byrne}, R., {Morales}, M.~F., {Hazelton}, B., {et~al.} 2019, The Astrophysical
  Journal, 875, 70, \dodoi{10.3847/1538-4357/ab107d}

\bibitem[{{Datta} {et~al.}(2010){Datta}, {Bowman}, \& {Carilli}}]{Datta2010}
{Datta}, A., {Bowman}, J.~D., \& {Carilli}, C.~L. 2010, The Astrophysical
  Journal, 724, 526, \dodoi{10.1088/0004-637X/724/1/526}

\bibitem[{{DeBoer} {et~al.}(2017){DeBoer}, {Parsons}, {Aguirre}, {Alexander},
  {Ali}, {Beardsley}, {Bernardi}, {Bowman}, {Bradley}, {Carilli}, {Cheng}, {de
  Lera Acedo}, {Dillon}, {Ewall-Wice}, {Fadana}, {Fagnoni}, {Fritz},
  {Furlanetto}, {Glendenning}, {Greig}, {Grobbelaar}, {Hazelton}, {Hewitt},
  {Hickish}, {Jacobs}, {Julius}, {Kariseb}, {Kohn}, {Lekalake}, {Liu}, {Loots},
  {MacMahon}, {Malan}, {Malgas}, {Maree}, {Martinot}, {Mathison}, {Matsetela},
  {Mesinger}, {Morales}, {Neben}, {Patra}, {Pieterse}, {Pober}, {Razavi-Ghods},
  {Ringuette}, {Robnett}, {Rosie}, {Sell}, {Smith}, {Syce}, {Tegmark},
  {Thyagarajan}, {Williams}, \& {Zheng}}]{DeBoer2017}
{DeBoer}, D.~R., {Parsons}, A.~R., {Aguirre}, J.~E., {et~al.} 2017,
  Publications of the Astronomical Society of the Pacific, 129, 045001,
  \dodoi{10.1088/1538-3873/129/974/045001}

\bibitem[{{Dillon} {et~al.}(2020){Dillon}, {Lee}, {Ali}, {Parsons}, {Orosz},
  {Nunhokee}, {La Plante}, {Beardsley}, {Kern}, {Abdurashidova}, {Aguirre},
  {Alexander}, {Balfour}, {Bernardi}, {Billings}, {Bowman}, {Bradley}, {Bull},
  {Burba}, {Carey}, {Carilli}, {Cheng}, {DeBoer}, {Dexter}, {de Lera Acedo},
  {Ely}, {Ewall-Wice}, {Fagnoni}, {Fritz}, {Furlanetto}, {Gale-Sides},
  {Glendenning}, {Gorthi}, {Greig}, {Grobbelaar}, {Halday}, {Hazelton},
  {Hewitt}, {Hickish}, {Jacobs}, {Julius}, {Kerrigan}, {Kittiwisit}, {Kohn},
  {Kolopanis}, {Lanman}, {Lekalake}, {Lewis}, {Liu}, {Ma}, {MacMahon}, {Malan},
  {Malgas}, {Maree}, {Martinot}, {Matsetela}, {Mesinger}, {Molewa}, {Morales},
  {Mosiane}, {Murray}, {Neben}, {Nikolic}, {Pascua}, {Patra}, {Pieterse},
  {Pober}, {Razavi-Ghods}, {Ringuette}, {Robnett}, {Rosie}, {Santos}, {Sims},
  {Smith}, {Syce}, {Tegmark}, {Thyagarajan}, {Williams}, \&
  {Zheng}}]{Dillon2020}
{Dillon}, J.~S., {Lee}, M., {Ali}, Z.~S., {et~al.} 2020, \mnras, 499, 5840,
  \dodoi{10.1093/mnras/staa3001}

\bibitem[{{Ewall-Wice} {et~al.}(2017){Ewall-Wice}, {Dillon}, {Liu}, \&
  {Hewitt}}]{Ewall-Wice2017}
{Ewall-Wice}, A., {Dillon}, J.~S., {Liu}, A., \& {Hewitt}, J. 2017, Monthly
  Notices of the Royal Astronomical Society, 470, 1849,
  \dodoi{10.1093/mnras/stx1221}

\bibitem[{{Ewall-Wice} {et~al.}(2016{\natexlab{a}}){Ewall-Wice}, {Bradley},
  {Deboer}, {Hewitt}, {Parsons}, {Aguirre}, {Ali}, {Bowman}, {Cheng}, {Neben},
  {Patra}, {Thyagarajan}, {Venter}, {de Lera Acedo}, {Dillon}, {Dickenson},
  {Doolittle}, {Egan}, {Hedrick}, {Klima}, {Kohn}, {Schaffner}, {Shelton},
  {Saliwanchik}, {Taylor}, {Taylor}, {Tegmark}, \& {Wirt}}]{Ewall-Wice2016b}
{Ewall-Wice}, A., {Bradley}, R., {Deboer}, D., {et~al.} 2016{\natexlab{a}},
  \apj, 831, 196, \dodoi{10.3847/0004-637X/831/2/196}

\bibitem[{{Ewall-Wice} {et~al.}(2016{\natexlab{b}}){Ewall-Wice}, {Dillon},
  {Hewitt}, {Loeb}, {Mesinger}, {Neben}, {Offringa}, {Tegmark}, {Barry},
  {Beardsley}, {Bernardi}, {Bowman}, {Briggs}, {Cappallo}, {Carroll}, {Corey},
  {de Oliveira-Costa}, {Emrich}, {Feng}, {Gaensler}, {Goeke}, {Greenhill},
  {Hazelton}, {Hurley-Walker}, {Johnston-Hollitt}, {Jacobs}, {Kaplan},
  {Kasper}, {Kim}, {Kratzenberg}, {Lenc}, {Line}, {Lonsdale}, {Lynch},
  {McKinley}, {McWhirter}, {Mitchell}, {Morales}, {Morgan}, {Thyagarajan},
  {Oberoi}, {Ord}, {Paul}, {Pindor}, {Pober}, {Prabu}, {Procopio}, {Riding},
  {Rogers}, {Roshi}, {Shankar}, {Sethi}, {Srivani}, {Subrahmanyan}, {Sullivan},
  {Tingay}, {Trott}, {Waterson}, {Wayth}, {Webster}, {Whitney}, {Williams},
  {Williams}, {Wu}, \& {Wyithe}}]{Ewall-Wice2016a}
{Ewall-Wice}, A., {Dillon}, J.~S., {Hewitt}, J.~N., {et~al.}
  2016{\natexlab{b}}, \mnras, 460, 4320, \dodoi{10.1093/mnras/stw1022}

\bibitem[{{Ewall-Wice} {et~al.}(2021){Ewall-Wice}, {Kern}, {Dillon}, {Liu},
  {Parsons}, {Singh}, {Lanman}, {La Plante}, {Fagnoni}, {Acedo}, {DeBoer},
  {Nunhokee}, {Bull}, {Chang}, {Lazio}, {Aguirre}, \&
  {Weinberg}}]{Ewall-Wice2021}
{Ewall-Wice}, A., {Kern}, N., {Dillon}, J.~S., {et~al.} 2021, \mnras, 500,
  5195, \dodoi{10.1093/mnras/staa3293}

\bibitem[{{Fagnoni} {et~al.}(2021){Fagnoni}, {de Lera Acedo}, {DeBoer},
  {Abdurashidova}, {Aguirre}, {Alexander}, {Ali}, {Balfour}, {Beardsley},
  {Bernardi}, {Billings}, {Bowman}, {Bradley}, {Bull}, {Burba}, {Carilli},
  {Cheng}, {Dexter}, {Dillon}, {Ewall-Wice}, {Fritz}, {Furlanetto},
  {Gale-Sides}, {Glendenning}, {Gorthi}, {Greig}, {Grobbelaar}, {Halday},
  {Hazelton}, {Hewitt}, {Hickish}, {Jacobs}, {Josaitis}, {Julius}, {Kern},
  {Kerrigan}, {Kim}, {Kittiwisit}, {Kohn}, {Kolopanis}, {Lanman}, {Plante},
  {Lekalake}, {Liu}, {MacMahon}, {Malan}, {Malgas}, {Maree}, {Martinot},
  {Matsetela}, {Mena Parra}, {Mesinger}, {Molewa}, {Morales}, {Mosiane},
  {Neben}, {Nikolic}, {Parsons}, {Patra}, {Pieterse}, {Pober}, {Razavi-Ghods},
  {Robnett}, {Rosie}, {Sims}, {Smith}, {Syce}, {Thyagarajan}, {Williams}, \&
  {Zheng}}]{Fagnoni2021}
{Fagnoni}, N., {de Lera Acedo}, E., {DeBoer}, D.~R., {et~al.} 2021, \mnras,
  500, 1232, \dodoi{10.1093/mnras/staa3268}

\bibitem[{{Furlanetto} {et~al.}(2006){Furlanetto}, {Oh}, \&
  {Briggs}}]{Furl2006}
{Furlanetto}, S.~R., {Oh}, S.~P., \& {Briggs}, F.~H. 2006, Physics Reports,
  433, 181, \dodoi{10.1016/j.physrep.2006.08.002}

\bibitem[{{G{\'o}rski} {et~al.}(2005){G{\'o}rski}, {Hivon}, {Banday},
  {Wandelt}, {Hansen}, {Reinecke}, \& {Bartelmann}}]{Gorski2005}
{G{\'o}rski}, K.~M., {Hivon}, E., {Banday}, A.~J., {et~al.} 2005, The
  Astrophysical Journal, 622, 759, \dodoi{10.1086/427976}

\bibitem[{{Greig} \& {Mesinger}(2015)}]{Greig2015}
{Greig}, B., \& {Mesinger}, A. 2015, \mnras, 449, 4246,
  \dodoi{10.1093/mnras/stv571}

\bibitem[{{Hare} {et~al.}(2018){Hare}, {Scholten}, {Bonardi}, {Buitink},
  {Corstanje}, {Ebert}, {Falcke}, {H{\"o}randel}, {Leijnse}, {Mitra}, {Mulrey},
  {Nelles}, {Rachen}, {Rossetto}, {Rutjes}, {Schellart}, {Thoudam}, {Trinh},
  {ter Veen}, \& {Winchen}}]{Hare2018}
{Hare}, B.~M., {Scholten}, O., {Bonardi}, A., {et~al.} 2018, Journal of
  Geophysical Research (Atmospheres), 123, 2861, \dodoi{10.1002/2017JD028132}

\bibitem[{{Hare} {et~al.}(2020){Hare}, {Scholten}, {Dwyer}, {Ebert}, {Nijdam},
  {Bonardi}, {Buitink}, {Corstanje}, {Falcke}, {Huege}, {H{\"o}randel},
  {Krampah}, {Mitra}, {Mulrey}, {Neijzen}, {Nelles}, {Pandya}, {Rachen},
  {Rossetto}, {Trinh}, {ter Veen}, \& {Winchen}}]{Hare2020}
{Hare}, B.~M., {Scholten}, O., {Dwyer}, J., {et~al.} 2020, \prl, 124, 105101,
  \dodoi{10.1103/PhysRevLett.124.105101}

\bibitem[{{Harris}(1978)}]{Harris1978}
{Harris}, F.~J. 1978, Proceedings of the IEEE, 66, 51

\bibitem[{{Hazelton} {et~al.}(2013){Hazelton}, {Morales}, \&
  {Sullivan}}]{Hazelton2013}
{Hazelton}, B.~J., {Morales}, M.~F., \& {Sullivan}, I.~S. 2013, The
  Astrophysical Journal, 770, 156, \dodoi{10.1088/0004-637X/770/2/156}

\bibitem[{{HERA Collaboration} {et~al.}(2023){HERA Collaboration},
  {Abdurashidova}, {Adams}, {Aguirre}, {Alexander}, {Ali}, {Baartman},
  {Balfour}, {Barkana}, {Beardsley}, {Bernardi}, {Billings}, {Bowman},
  {Bradley}, {Breitman}, {Bull}, {Burba}, {Carey}, {Carilli}, {Cheng},
  {Choudhuri}, {DeBoer}, {de Lera Acedo}, {Dexter}, {Dillon}, {Ely},
  {Ewall-Wice}, {Fagnoni}, {Fialkov}, {Fritz}, {Furlanetto}, {Gale-Sides},
  {Garsden}, {Glendenning}, {Gorce}, {Gorthi}, {Greig}, {Grobbelaar}, {Halday},
  {Hazelton}, {Heimersheim}, {Hewitt}, {Hickish}, {Jacobs}, {Julius}, {Kern},
  {Kerrigan}, {Kittiwisit}, {Kohn}, {Kolopanis}, {Lanman}, {La Plante},
  {Lewis}, {Liu}, {Loots}, {Ma}, {MacMahon}, {Malan}, {Malgas}, {Malgas},
  {Maree}, {Marero}, {Martinot}, {McBride}, {Mesinger}, {Mirocha}, {Molewa},
  {Morales}, {Mosiane}, {Mu{\~n}oz}, {Murray}, {Nagpal}, {Neben}, {Nikolic},
  {Nunhokee}, {Nuwegeld}, {Parsons}, {Pascua}, {Patra}, {Pieterse}, {Qin},
  {Razavi-Ghods}, {Robnett}, {Rosie}, {Santos}, {Sims}, {Singh}, {Smith},
  {Swarts}, {Tan}, {Thyagarajan}, {Wilensky}, {Williams}, {van Wyngaarden}, \&
  {Zheng}}]{HERA2022b}
{HERA Collaboration}, {Abdurashidova}, Z., {Adams}, T., {et~al.} 2023, \apj,
  945, 124, \dodoi{10.3847/1538-4357/acaf50}

\bibitem[{{Hurley-Walker} \& {Hancock}(2018)}]{Hurley-Walker2018}
{Hurley-Walker}, N., \& {Hancock}, P.~J. 2018, Astronomy and Computing, 25, 94,
  \dodoi{10.1016/j.ascom.2018.08.006}

\bibitem[{{Hurley-Walker} {et~al.}(2017){Hurley-Walker}, {Callingham},
  {Hancock}, {Franzen}, {Hindson}, {Kapi{\'n}ska}, {Morgan}, {Offringa},
  {Wayth}, {Wu}, {Zheng}, {Murphy}, {Bell}, {Dwarakanath}, {For}, {Gaensler},
  {Johnston-Hollitt}, {Lenc}, {Procopio}, {Staveley-Smith}, {Ekers}, {Bowman},
  {Briggs}, {Cappallo}, {Deshpande}, {Greenhill}, {Hazelton}, {Kaplan},
  {Lonsdale}, {McWhirter}, {Mitchell}, {Morales}, {Morgan}, {Oberoi}, {Ord},
  {Prabu}, {Shankar}, {Srivani}, {Subrahmanyan}, {Tingay}, {Webster},
  {Williams}, \& {Williams}}]{Hurley-Walker2017}
{Hurley-Walker}, N., {Callingham}, J.~R., {Hancock}, P.~J., {et~al.} 2017,
  Monthly Notices of the Royal Astronomical Society, 464, 1146,
  \dodoi{10.1093/mnras/stw2337}

\bibitem[{{Jordan} {et~al.}(2017){Jordan}, {Murray}, {Trott}, {Wayth},
  {Mitchell}, {Rahimi}, {Pindor}, {Procopio}, \& {Morgan}}]{Jordan2017}
{Jordan}, C.~H., {Murray}, S., {Trott}, C.~M., {et~al.} 2017, Monthly Notices
  of the Royal Astronomical Society, 471, 3974, \dodoi{10.1093/mnras/stx1797}

\bibitem[{{Kern} {et~al.}(2019){Kern}, {Parsons}, {Dillon}, {Lanman},
  {Fagnoni}, \& {de Lera Acedo}}]{Kern2019}
{Kern}, N.~S., {Parsons}, A.~R., {Dillon}, J.~S., {et~al.} 2019, The
  Astrophysical Journal, 884, 105, \dodoi{10.3847/1538-4357/ab3e73}

\bibitem[{{Kern} {et~al.}(2020){Kern}, {Parsons}, {Dillon}, {Lanman}, {Liu},
  {Bull}, {Ewall-Wice}, {Abdurashidova}, {Aguirre}, {Alexander}, {Ali},
  {Balfour}, {Beardsley}, {Bernardi}, {Bowman}, {Bradley}, {Burba}, {Carilli},
  {Cheng}, {DeBoer}, {Dexter}, {de Lera Acedo}, {Fagnoni}, {Fritz},
  {Furlanetto}, {Glendenning}, {Gorthi}, {Greig}, {Grobbelaar}, {Halday},
  {Hazelton}, {Hewitt}, {Hickish}, {Jacobs}, {Julius}, {Kerrigan},
  {Kittiwisit}, {Kohn}, {Kolopanis}, {La Plante}, {Lekalake}, {MacMahon},
  {Malan}, {Malgas}, {Maree}, {Martinot}, {Matsetela}, {Mesinger}, {Molewa},
  {Morales}, {Mosiane}, {Murray}, {Neben}, {Parsons}, {Patra}, {Pieterse},
  {Pober}, {Razavi-Ghods}, {Ringuette}, {Robnett}, {Rosie}, {Sims}, {Smith},
  {Syce}, {Thyagarajan}, {Williams}, \& {Zheng}}]{Kern2020a}
---. 2020, The Astrophysical Journal, 888, 70, \dodoi{10.3847/1538-4357/ab5e8a}

\bibitem[{{Li} {et~al.}(2019){Li}, {Pober}, {Barry}, {Hazelton}, {Morales},
  {Trott}, {Lanman}, {Wilensky}, {Sullivan}, {Beardsley}, {Booler}, {Bowman},
  {Byrne}, {Crosse}, {Emrich}, {Franzen}, {Hasegawa}, {Horsley},
  {Johnston-Hollitt}, {Jacobs}, {Jordan}, {Joseph}, {Kaneuji}, {Kaplan},
  {Kenney}, {Kubota}, {Line}, {Lynch}, {McKinley}, {Mitchell}, {Murray},
  {Pallot}, {Pindor}, {Rahimi}, {Riding}, {Sleap}, {Steele}, {Takahashi},
  {Tingay}, {Walker}, {Wayth}, {Webster}, {Williams}, {Wu}, {Wyithe},
  {Yoshiura}, \& {Zheng}}]{Li2019}
{Li}, W., {Pober}, J.~C., {Barry}, N., {et~al.} 2019, The Astrophysical
  Journal, 887, 141, \dodoi{10.3847/1538-4357/ab55e4}

\bibitem[{{Line} {et~al.}(2018){Line}, {McKinley}, {Rasti}, {Bhardwaj},
  {Wayth}, {Webster}, {Ung}, {Emrich}, {Horsley}, {Beardsley}, {Crosse},
  {Franzen}, {Gaensler}, {Johnston-Hollitt}, {Kaplan}, {Kenney}, {Morales},
  {Pallot}, {Steele}, {Tingay}, {Trott}, {Walker}, {Williams}, \&
  {Wu}}]{Line2018}
{Line}, J.~L.~B., {McKinley}, B., {Rasti}, J., {et~al.} 2018, \pasa, 35, e045,
  \dodoi{10.1017/pasa.2018.30}

\bibitem[{{Liu} {et~al.}(2014{\natexlab{a}}){Liu}, {Parsons}, \&
  {Trott}}]{Liu2014a}
{Liu}, A., {Parsons}, A.~R., \& {Trott}, C.~M. 2014{\natexlab{a}}, Physical
  Review D, 90, 023018, \dodoi{10.1103/PhysRevD.90.023018}

\bibitem[{{Liu} {et~al.}(2014{\natexlab{b}}){Liu}, {Parsons}, \&
  {Trott}}]{Liu2014b}
---. 2014{\natexlab{b}}, Physical Review D, 90, 023019,
  \dodoi{10.1103/PhysRevD.90.023019}

\bibitem[{{Liu} \& {Shaw}(2020)}]{Liu2020}
{Liu}, A., \& {Shaw}, J.~R. 2020, Publications of the Astronomical Society of
  the Pacific, 132, 062001, \dodoi{10.1088/1538-3873/ab5bfd}

\bibitem[{Luque(2017)}]{Luque2017}
Luque, A. 2017, Journal of Geophysical Research: Atmospheres, 122, 10,497,
  \dodoi{https://doi.org/10.1002/2017JD027157}

\bibitem[{{McSweeney} {et~al.}(2020){McSweeney}, {Ord}, {Kaur}, {Bhat},
  {Meyers}, {Tremblay}, {Jones}, {Crosse}, \& {Smith}}]{McSweeney2020}
{McSweeney}, S.~J., {Ord}, S.~M., {Kaur}, D., {et~al.} 2020, \pasa, 37, e034,
  \dodoi{10.1017/pasa.2020.24}

\bibitem[{{Mertens} {et~al.}(2020){Mertens}, {Mevius}, {Koopmans}, {Offringa},
  {Mellema}, {Zaroubi}, {Brentjens}, {Gan}, {Gehlot}, {Pandey}, {Sardarabadi},
  {Vedantham}, {Yatawatta}, {Asad}, {Ciardi}, {Chapman}, {Gazagnes}, {Ghara},
  {Ghosh}, {Giri}, {Iliev}, {Jeli{\'c}}, {Kooistra}, {Mondal}, {Schaye}, \&
  {Silva}}]{Mertens2020}
{Mertens}, F.~G., {Mevius}, M., {Koopmans}, L.~V.~E., {et~al.} 2020, Monthly
  Notices of the Royal Astronomical Society, 493, 1662,
  \dodoi{10.1093/mnras/staa327}

\bibitem[{{Mevius} {et~al.}(2022){Mevius}, {Mertens}, {Koopmans}, {Offringa},
  {Yatawatta}, {Brentjens}, {Chapman}, {Ciardi}, {Gan}, {Gehlot}, {Ghara},
  {Ghosh}, {Giri}, {Iliev}, {Mellema}, {Pandey}, \& {Zaroubi}}]{Mevius2022}
{Mevius}, M., {Mertens}, F., {Koopmans}, L.~V.~E., {et~al.} 2022, \mnras, 509,
  3693, \dodoi{10.1093/mnras/stab3233}

\bibitem[{{Mitchell} {et~al.}(2008){Mitchell}, {Greenhill}, {Wayth}, {Sault},
  {Lonsdale}, {Cappallo}, {Morales}, \& {Ord}}]{Mitchell2008}
{Mitchell}, D.~A., {Greenhill}, L.~J., {Wayth}, R.~B., {et~al.} 2008, IEEE
  Journal of Selected Topics in Signal Processing, 2, 707,
  \dodoi{10.1109/JSTSP.2008.2005327}

\bibitem[{{Morales} {et~al.}(2012){Morales}, {Hazelton}, {Sullivan}, \&
  {Beardsley}}]{Morales2012}
{Morales}, M.~F., {Hazelton}, B., {Sullivan}, I., \& {Beardsley}, A. 2012, The
  Astrophysical Journal, 752, 137, \dodoi{10.1088/0004-637X/752/2/137}

\bibitem[{{Morales} \& {Matejek}(2009)}]{Morales2009}
{Morales}, M.~F., \& {Matejek}, M. 2009, Monthly Notices of the Royal
  Astronomical Society, 400, 1814, \dodoi{10.1111/j.1365-2966.2009.15537.x}

\bibitem[{{Morales} \& {Wyithe}(2010)}]{Morales2010}
{Morales}, M.~F., \& {Wyithe}, J. S.~B. 2010, Annual Review of Astronomy and
  Astrophysics, 48, 127, \dodoi{10.1146/annurev-astro-081309-130936}

\bibitem[{{Mouri Sardarabadi} \& {Koopmans}(2019)}]{Sardarabadi2019}
{Mouri Sardarabadi}, A., \& {Koopmans}, L.~V.~E. 2019, \mnras, 483, 5480,
  \dodoi{10.1093/mnras/sty3444}

\bibitem[{{Offringa} {et~al.}(2012){Offringa}, {de Bruyn}, \&
  {Zaroubi}}]{Offringa2012a}
{Offringa}, A.~R., {de Bruyn}, A.~G., \& {Zaroubi}, S. 2012, Monthly Notices of
  the Royal Astronomical Society, 422, 563,
  \dodoi{10.1111/j.1365-2966.2012.20633.x}

\bibitem[{{Offringa} {et~al.}(2019){Offringa}, {Mertens}, \&
  {Koopmans}}]{Offringa2019a}
{Offringa}, A.~R., {Mertens}, F., \& {Koopmans}, L.~V.~E. 2019, Monthly Notices
  of the Royal Astronomical Society, 484, 2866, \dodoi{10.1093/mnras/stz175}

\bibitem[{{Offringa} {et~al.}(2013){Offringa}, {de Bruyn}, {Zaroubi},
  {Koopmans}, {Wijnholds}, {Abdalla}, {Brouw}, {Ciardi}, {Iliev}, {Harker},
  {Mellema}, {Bernardi}, {Zarka}, {Ghosh}, {Alexov}, {Anderson}, {Asgekar},
  {Avruch}, {Beck}, {Bell}, {Bell}, {Bentum}, {Best}, {B{\^\i}rzan},
  {Breitling}, {Broderick}, {Br{\"u}ggen}, {Butcher}, {de Gasperin}, {de Geus},
  {de Vos}, {Duscha}, {Eisl{\"o}ffel}, {Fallows}, {Ferrari}, {Frieswijk},
  {Garrett}, {Grie{\ss}meier}, {Hassall}, {Horneffer}, {Iacobelli}, {Juette},
  {Karastergiou}, {Klijn}, {Kondratiev}, {Kuniyoshi}, {Kuper}, {van Leeuwen},
  {Loose}, {Maat}, {Macario}, {Mann}, {McKean}, {Meulman}, {Norden}, {Orru},
  {Paas}, {Pandey-Pommier}, {Pizzo}, {Polatidis}, {Rafferty}, {Reich}, {van
  Nieuwpoort}, {R{\"o}ttgering}, {Scaife}, {Sluman}, {Smirnov}, {Sobey},
  {Tagger}, {Tang}, {Tasse}, {Veen}, {Toribio}, {Vermeulen}, {Vocks}, {van
  Weeren}, {Wise}, \& {Wucknitz}}]{Offringa2013b}
{Offringa}, A.~R., {de Bruyn}, A.~G., {Zaroubi}, S., {et~al.} 2013, Monthly
  Notices of the Royal Astronomical Society, 435, 584,
  \dodoi{10.1093/mnras/stt1337}

\bibitem[{{Offringa} {et~al.}(2015){Offringa}, {Wayth}, {Hurley-Walker},
  {Kaplan}, {Barry}, {Beardsley}, {Bell}, {Bernardi}, {Bowman}, {Briggs},
  {Callingham}, {Cappallo}, {Carroll}, {Deshpande}, {Dillon}, {Dwarakanath},
  {Ewall-Wice}, {Feng}, {For}, {Gaensler}, {Greenhill}, {Hancock}, {Hazelton},
  {Hewitt}, {Hindson}, {Jacobs}, {Johnston-Hollitt}, {Kapi{\'n}ska}, {Kim},
  {Kittiwisit}, {Lenc}, {Line}, {Loeb}, {Lonsdale}, {McKinley}, {McWhirter},
  {Mitchell}, {Morales}, {Morgan}, {Morgan}, {Neben}, {Oberoi}, {Ord}, {Paul},
  {Pindor}, {Pober}, {Prabu}, {Procopio}, {Riding}, {Udaya Shankar}, {Sethi},
  {Srivani}, {Staveley-Smith}, {Subrahmanyan}, {Sullivan}, {Tegmark},
  {Thyagarajan}, {Tingay}, {Trott}, {Webster}, {Williams}, {Williams}, {Wu},
  {Wyithe}, \& {Zheng}}]{Offringa2015}
{Offringa}, A.~R., {Wayth}, R.~B., {Hurley-Walker}, N., {et~al.} 2015,
  Publications of the Astronomical Society of Australia, 32, e008,
  \dodoi{10.1017/pasa.2015.7}

\bibitem[{{Paciga} {et~al.}(2013){Paciga}, {Albert}, {Bandura}, {Chang},
  {Gupta}, {Hirata}, {Odegova}, {Pen}, {Peterson}, {Roy}, {Shaw}, {Sigurdson},
  \& {Voytek}}]{Paciga2013}
{Paciga}, G., {Albert}, J.~G., {Bandura}, K., {et~al.} 2013, \mnras, 433, 639,
  \dodoi{10.1093/mnras/stt753}

\bibitem[{{Park} {et~al.}(2019){Park}, {Mesinger}, {Greig}, \&
  {Gillet}}]{Park2019}
{Park}, J., {Mesinger}, A., {Greig}, B., \& {Gillet}, N. 2019, Monthly Notices
  of the Royal Astronomical Society, 484, 933, \dodoi{10.1093/mnras/stz032}

\bibitem[{{Parsons} {et~al.}(2010){Parsons}, {Backer}, {Foster}, {Wright},
  {Bradley}, {Gugliucci}, {Parashare}, {Benoit}, {Aguirre}, {Jacobs},
  {Carilli}, {Herne}, {Lynch}, {Manley}, \& {Werthimer}}]{Parsons2010}
{Parsons}, A.~R., {Backer}, D.~C., {Foster}, G.~S., {et~al.} 2010, \aj, 139,
  1468, \dodoi{10.1088/0004-6256/139/4/1468}

\bibitem[{{Patil} {et~al.}(2016){Patil}, {Yatawatta}, {Zaroubi}, {Koopmans},
  {de Bruyn}, {Jeli{\'c}}, {Ciardi}, {Iliev}, {Mevius}, {Pandey}, \&
  {Gehlot}}]{Patil2016}
{Patil}, A.~H., {Yatawatta}, S., {Zaroubi}, S., {et~al.} 2016, Monthly Notices
  of the Royal Astronomical Society, 463, 4317, \dodoi{10.1093/mnras/stw2277}

\bibitem[{{Patil} {et~al.}(2017){Patil}, {Yatawatta}, {Koopmans}, {de Bruyn},
  {Brentjens}, {Zaroubi}, {Asad}, {Hatef}, {Jeli{\'c}}, {Mevius}, {Offringa},
  {Pandey}, {Vedantham}, {Abdalla}, {Brouw}, {Chapman}, {Ciardi}, {Gehlot},
  {Ghosh}, {Harker}, {Iliev}, {Kakiichi}, {Majumdar}, {Mellema}, {Silva},
  {Schaye}, {Vrbanec}, \& {Wijnholds}}]{Patil2017}
{Patil}, A.~H., {Yatawatta}, S., {Koopmans}, L.~V.~E., {et~al.} 2017, \apj,
  838, 65, \dodoi{10.3847/1538-4357/aa63e7}

\bibitem[{{Prabu} {et~al.}(2020){Prabu}, {Hancock}, {Zhang}, \&
  {Tingay}}]{Prabu2020}
{Prabu}, S., {Hancock}, P., {Zhang}, X., \& {Tingay}, S.~J. 2020, \pasa, 37,
  e052, \dodoi{10.1017/pasa.2020.40}

\bibitem[{{Prabu} {et~al.}(2022){Prabu}, {Hancock}, {Zhang}, {Tingay},
  {Hodgson}, {Crosse}, \& {Johnston-Hollitt}}]{Prabu2022}
{Prabu}, S., {Hancock}, P., {Zhang}, X., {et~al.} 2022, Advances in Space
  Research, 70, 812, \dodoi{10.1016/j.asr.2022.05.013}

\bibitem[{{Prabu} {et~al.}(2015){Prabu}, {Srivani}, {Roshi}, {Kamini},
  {Madhavi}, {Emrich}, {Crosse}, {Williams}, {Waterson}, {Deshpande},
  {Shankar}, {Subrahmanyan}, {Briggs}, {Goeke}, {Tingay}, {Johnston-Hollitt},
  {R}, {Morgan}, {Pathikulangara}, {Bunton}, {Hampson}, {Williams}, {Ord},
  {Wayth}, {Kumar}, {Morales}, {deSouza}, {Kratzenberg}, {Pallot}, {McWhirter},
  {Hazelton}, {Arcus}, {Barnes}, {Bernardi}, {Booler}, {Bowman}, {Cappallo},
  {Corey}, {Greenhill}, {Herne}, {Hewitt}, {Kaplan}, {Kasper}, {Kincaid},
  {Koenig}, {Lonsdale}, {Lynch}, {Mitchell}, {Oberoi}, {Remillard}, {Rogers},
  {Salah}, {Sault}, {Stevens}, {Tremblay}, {Webster}, {Whitney}, \&
  {Wyithe}}]{Prabu2015}
{Prabu}, T., {Srivani}, K.~S., {Roshi}, D.~A., {et~al.} 2015, Experimental
  Astronomy, 39, 73, \dodoi{10.1007/s10686-015-9444-3}

\bibitem[{{Pritchard} \& {Loeb}(2012)}]{Pritchard2012}
{Pritchard}, J.~R., \& {Loeb}, A. 2012, Reports on Progress in Physics, 75,
  086901, \dodoi{10.1088/0034-4885/75/8/086901}

\bibitem[{{Pu} {et~al.}(2021){Pu}, {Cummer}, \& {Liu}}]{Pu2021}
{Pu}, Y., {Cummer}, S.~A., \& {Liu}, N. 2021, \grl, 48, e93145,
  \dodoi{10.1029/2021GL093145}

\bibitem[{{Rahimi} {et~al.}(2021){Rahimi}, {Pindor}, {Line}, {Barry}, {Trott},
  {Webster}, {Jordan}, {Wilensky}, {Yoshiura}, {Beardsley}, {Bowman}, {Byrne},
  {Chokshi}, {Hazelton}, {Hasegawa}, {Howard}, {Greig}, {Jacobs}, {Joseph},
  {Kolopanis}, {Lynch}, {McKinley}, {Mitchell}, {Murray}, {Morales}, {Pober},
  {Takahashi}, {Tingay}, {Wayth}, {Wyithe}, \& {Zheng}}]{Rahimi2021}
{Rahimi}, M., {Pindor}, B., {Line}, J.~L.~B., {et~al.} 2021, \mnras, 508, 5954,
  \dodoi{10.1093/mnras/stab2918}

\bibitem[{{Scholten} {et~al.}(2022){Scholten}, {Hare}, {Dwyer}, {Liu},
  {Sterpka}, {Kolma{\v{s}}ov{\'a}}, {Santol{\'\i}k}, {L{\'a}n},
  {Uhl{\'\i}{\v{r}}}, {Buitink}, {Huege}, {Nelles}, \& {ter
  Veen}}]{Scholten2022}
{Scholten}, O., {Hare}, B.~M., {Dwyer}, J., {et~al.} 2022, \prd, 105, 062007,
  \dodoi{10.1103/PhysRevD.105.062007}

\bibitem[{Shi {et~al.}(2019)Shi, Liu, Dwyer, \& Ihaddadene}]{Shi2019}
Shi, F., Liu, N., Dwyer, J.~R., \& Ihaddadene, K. M.~A. 2019, Geophysical
  Research Letters, 46, 443, \dodoi{https://doi.org/10.1029/2018GL080309}

\bibitem[{{Sokolowski} {et~al.}(2016){Sokolowski}, {Wayth}, \&
  {Lewis}}]{Sokolowski2016}
{Sokolowski}, M., {Wayth}, R.~B., \& {Lewis}, M. 2016, arXiv e-prints,
  arXiv:1610.04696.
\newblock \doarXiv{1610.04696}

\bibitem[{{Sterpka} {et~al.}(2021){Sterpka}, {Dwyer}, {Liu}, {Hare},
  {Scholten}, {Buitink}, {Veen}, \& {Nelles}}]{Sterpka2021}
{Sterpka}, C., {Dwyer}, J., {Liu}, N., {et~al.} 2021, \grl, 48, e95511,
  \dodoi{10.1029/2021GL095511}

\bibitem[{{Sullivan} {et~al.}(2012){Sullivan}, {Morales}, {Hazelton}, {Arcus},
  {Barnes}, {Bernardi}, {Briggs}, {Bowman}, {Bunton}, {Cappallo}, {Corey},
  {Deshpande}, {deSouza}, {Emrich}, {Gaensler}, {Goeke}, {Greenhill}, {Herne},
  {Hewitt}, {Johnston-Hollitt}, {Kaplan}, {Kasper}, {Kincaid}, {Koenig},
  {Kratzenberg}, {Lonsdale}, {Lynch}, {McWhirter}, {Mitchell}, {Morgan},
  {Oberoi}, {Ord}, {Pathikulangara}, {Prabu}, {Remillard}, {Rogers}, {Roshi},
  {Salah}, {Sault}, {Udaya Shankar}, {Srivani}, {Stevens}, {Subrahmanyan},
  {Tingay}, {Wayth}, {Waterson}, {Webster}, {Whitney}, {Williams}, {Williams},
  \& {Wyithe}}]{Sullivan2012}
{Sullivan}, I.~S., {Morales}, M.~F., {Hazelton}, B.~J., {et~al.} 2012, The
  Astrophysical Journal, 759, 17, \dodoi{10.1088/0004-637X/759/1/17}

\bibitem[{{Tingay} {et~al.}(2020){Tingay}, {Sokolowski}, {Wayth}, \&
  {Ung}}]{Tingay2020}
{Tingay}, S.~J., {Sokolowski}, M., {Wayth}, R., \& {Ung}, D. 2020, \pasa, 37,
  e039, \dodoi{10.1017/pasa.2020.32}

\bibitem[{{Tingay} {et~al.}(2013){Tingay}, {Goeke}, {Bowman}, {Emrich}, {Ord},
  {Mitchell}, {Morales}, {Booler}, {Crosse}, {Wayth}, {Lonsdale}, {Tremblay},
  {Pallot}, {Colegate}, {Wicenec}, {Kudryavtseva}, {Arcus}, {Barnes},
  {Bernardi}, {Briggs}, {Burns}, {Bunton}, {Cappallo}, {Corey}, {Deshpande},
  {Desouza}, {Gaensler}, {Greenhill}, {Hall}, {Hazelton}, {Herne}, {Hewitt},
  {Johnston-Hollitt}, {Kaplan}, {Kasper}, {Kincaid}, {Koenig}, {Kratzenberg},
  {Lynch}, {Mckinley}, {Mcwhirter}, {Morgan}, {Oberoi}, {Pathikulangara},
  {Prabu}, {Remillard}, {Rogers}, {Roshi}, {Salah}, {Sault}, {Udaya-Shankar},
  {Schlagenhaufer}, {Srivani}, {Stevens}, {Subrahmanyan}, {Waterson},
  {Webster}, {Whitney}, {Williams}, {Williams}, \& {Wyithe}}]{Tingay2013}
{Tingay}, S.~J., {Goeke}, R., {Bowman}, J.~D., {et~al.} 2013, \pasa, 30, e007,
  \dodoi{10.1017/pasa.2012.007}

\bibitem[{{Trott} \& {Wayth}(2016)}]{Trott2016}
{Trott}, C.~M., \& {Wayth}, R.~B. 2016, Publications of the Astronomical
  Society of Australia, 33, e019, \dodoi{10.1017/pasa.2016.18}

\bibitem[{{Trott} {et~al.}(2012){Trott}, {Wayth}, \& {Tingay}}]{Trott2012}
{Trott}, C.~M., {Wayth}, R.~B., \& {Tingay}, S.~J. 2012, The Astrophysical
  Journal, 757, 101, \dodoi{10.1088/0004-637X/757/1/101}

\bibitem[{{Trott} {et~al.}(2018){Trott}, {Jordan}, {Murray}, {Pindor},
  {Mitchell}, {Wayth}, {Line}, {McKinley}, {Beardsley}, {Bowman}, {Briggs},
  {Hazelton}, {Hewitt}, {Jacobs}, {Morales}, {Pober}, {Sethi}, {Shankar},
  {Subrahmanyan}, {Tegmark}, {Tingay}, {Webster}, \& {Wyithe}}]{Trott2018}
{Trott}, C.~M., {Jordan}, C.~H., {Murray}, S.~G., {et~al.} 2018, \apj, 867, 15,
  \dodoi{10.3847/1538-4357/aae314}

\bibitem[{{Trott} {et~al.}(2020){Trott}, {Jordan}, {Midgley}, {Barry}, {Greig},
  {Pindor}, {Cook}, {Sleap}, {Tingay}, {Ung}, {Hancock}, {Williams}, {Bowman},
  {Byrne}, {Chokshi}, {Hazelton}, {Hasegawa}, {Jacobs}, {Joseph}, {Li}, {Line},
  {Lynch}, {McKinley}, {Mitchell}, {Morales}, {Ouchi}, {Pober}, {Rahimi},
  {Takahashi}, {Wayth}, {Webster}, {Wilensky}, {Wyithe}, {Yoshiura}, {Zhang},
  \& {Zheng}}]{Trott2020}
{Trott}, C.~M., {Jordan}, C.~H., {Midgley}, S., {et~al.} 2020, \mnras, 493,
  4711, \dodoi{10.1093/mnras/staa414}

\bibitem[{{van Haarlem} {et~al.}(2013){van Haarlem}, {Wise}, {Gunst}, {Heald},
  {McKean}, {Hessels}, {de Bruyn}, {Nijboer}, {Swinbank}, {Fallows},
  {Brentjens}, {Nelles}, {Beck}, {Falcke}, {Fender}, {H{\"o}randel},
  {Koopmans}, {Mann}, {Miley}, {R{\"o}ttgering}, {Stappers}, {Wijers},
  {Zaroubi}, {van den Akker}, {Alexov}, {Anderson}, {Anderson}, {van Ardenne},
  {Arts}, {Asgekar}, {Avruch}, {Batejat}, {B{\"a}hren}, {Bell}, {Bell}, {van
  Bemmel}, {Bennema}, {Bentum}, {Bernardi}, {Best}, {B{\^\i}rzan}, {Bonafede},
  {Boonstra}, {Braun}, {Bregman}, {Breitling}, {van de Brink}, {Broderick},
  {Broekema}, {Brouw}, {Br{\"u}ggen}, {Butcher}, {van Cappellen}, {Ciardi},
  {Coenen}, {Conway}, {Coolen}, {Corstanje}, {Damstra}, {Davies}, {Deller},
  {Dettmar}, {van Diepen}, {Dijkstra}, {Donker}, {Doorduin}, {Dromer}, {Drost},
  {van Duin}, {Eisl{\"o}ffel}, {van Enst}, {Ferrari}, {Frieswijk}, {Gankema},
  {Garrett}, {de Gasperin}, {Gerbers}, {de Geus}, {Grie{\ss}meier}, {Grit},
  {Gruppen}, {Hamaker}, {Hassall}, {Hoeft}, {Holties}, {Horneffer}, {van der
  Horst}, {van Houwelingen}, {Huijgen}, {Iacobelli}, {Intema}, {Jackson},
  {Jelic}, {de Jong}, {Juette}, {Kant}, {Karastergiou}, {Koers}, {Kollen},
  {Kondratiev}, {Kooistra}, {Koopman}, {Koster}, {Kuniyoshi}, {Kramer},
  {Kuper}, {Lambropoulos}, {Law}, {van Leeuwen}, {Lemaitre}, {Loose}, {Maat},
  {Macario}, {Markoff}, {Masters}, {McFadden}, {McKay-Bukowski}, {Meijering},
  {Meulman}, {Mevius}, {Middelberg}, {Millenaar}, {Miller-Jones}, {Mohan},
  {Mol}, {Morawietz}, {Morganti}, {Mulcahy}, {Mulder}, {Munk}, {Nieuwenhuis},
  {van Nieuwpoort}, {Noordam}, {Norden}, {Noutsos}, {Offringa}, {Olofsson},
  {Omar}, {Orr{\'u}}, {Overeem}, {Paas}, {Pandey-Pommier}, {Pandey}, {Pizzo},
  {Polatidis}, {Rafferty}, {Rawlings}, {Reich}, {de Reijer}, {Reitsma},
  {Renting}, {Riemers}, {Rol}, {Romein}, {Roosjen}, {Ruiter}, {Scaife}, {van
  der Schaaf}, {Scheers}, {Schellart}, {Schoenmakers}, {Schoonderbeek},
  {Serylak}, {Shulevski}, {Sluman}, {Smirnov}, {Sobey}, {Spreeuw}, {Steinmetz},
  {Sterks}, {Stiepel}, {Stuurwold}, {Tagger}, {Tang}, {Tasse}, {Thomas},
  {Thoudam}, {Toribio}, {van der Tol}, {Usov}, {van Veelen}, {van der Veen},
  {ter Veen}, {Verbiest}, {Vermeulen}, {Vermaas}, {Vocks}, {Vogt}, {de Vos},
  {van der Wal}, {van Weeren}, {Weggemans}, {Weltevrede}, {White}, {Wijnholds},
  {Wilhelmsson}, {Wucknitz}, {Yatawatta}, {Zarka}, {Zensus}, \& {van
  Zwieten}}]{vanHaarlem2013}
{van Haarlem}, M.~P., {Wise}, M.~W., {Gunst}, A.~W., {et~al.} 2013, Astronomy
  and Astrophysics, 556, A2, \dodoi{10.1051/0004-6361/201220873}

\bibitem[{{Vedantham} {et~al.}(2012){Vedantham}, {Udaya Shankar}, \&
  {Subrahmanyan}}]{Vedantham2012}
{Vedantham}, H., {Udaya Shankar}, N., \& {Subrahmanyan}, R. 2012, \apj, 745,
  176, \dodoi{10.1088/0004-637X/745/2/176}

\bibitem[{{Vedantham} \& {Koopmans}(2016)}]{Vedantham2016}
{Vedantham}, H.~K., \& {Koopmans}, L.~V.~E. 2016, \mnras, 458, 3099,
  \dodoi{10.1093/mnras/stw443}

\bibitem[{{Vine}(1987)}]{Vine1987}
{Vine}, D.~M. 1987, Meteorology and Atmospheric Physics, 37, 195,
  \dodoi{10.1007/BF01042441}

\bibitem[{{Wayth} {et~al.}(2018){Wayth}, {Tingay}, {Trott}, {Emrich},
  {Johnston-Hollitt}, {McKinley}, {Gaensler}, {Beardsley}, {Booler}, {Crosse},
  {Franzen}, {Horsley}, {Kaplan}, {Kenney}, {Morales}, {Pallot}, {Sleap},
  {Steele}, {Walker}, {Williams}, {Wu}, {Cairns}, {Filipovic}, {Johnston},
  {Murphy}, {Quinn}, {Staveley-Smith}, {Webster}, \& {Wyithe}}]{Wayth2018}
{Wayth}, R.~B., {Tingay}, S.~J., {Trott}, C.~M., {et~al.} 2018, Publications of
  the Astronomical Society of Australia, 35, 33, \dodoi{10.1017/pasa.2018.37}

\bibitem[{{Wilensky}(2021)}]{Wilensky2021}
{Wilensky}, M. 2021, PhD thesis, University of Washington, Seattle

\bibitem[{{Wilensky} {et~al.}(2020){Wilensky}, {Barry}, {Morales}, {Hazelton},
  \& {Byrne}}]{Wilensky2020}
{Wilensky}, M.~J., {Barry}, N., {Morales}, M.~F., {Hazelton}, B.~J., \&
  {Byrne}, R. 2020, Monthly Notices of the Royal Astronomical Society, 498,
  265, \dodoi{10.1093/mnras/staa2442}

\bibitem[{{Wilensky} {et~al.}(2023){Wilensky}, {Brown}, \&
  {Hazelton}}]{Wilensky2023}
{Wilensky}, M.~J., {Brown}, J., \& {Hazelton}, B.~J. 2023, \mnras, 521, 5191,
  \dodoi{10.1093/mnras/stad863}

\bibitem[{{Wilensky} {et~al.}(2022a){Wilensky}, {Hazelton}, \&
  {Morales}}]{Wilensky2022}
{Wilensky}, M.~J., {Hazelton}, B.~J., \& {Morales}, M.~F. 2022a, \mnras, 510,
  5023, \dodoi{10.1093/mnras/stab3456}

\bibitem[{{Wilensky} {et~al.}(2019){Wilensky}, {Morales}, {Hazelton}, {Barry},
  {Byrne}, \& {Roy}}]{Wilensky2019}
{Wilensky}, M.~J., {Morales}, M.~F., {Hazelton}, B.~J., {et~al.} 2019,
  Publications of the Astronomical Society of the Pacific, 131, 114507,
  \dodoi{10.1088/1538-3873/ab3cad}

\end{thebibliography}

\appendix

\section{Power Spectrum Wall of Shame}
\label{app:wall_of_shame}

We show all power spectra that were used to remove observations from the 21-cm power spectrum upper limit in that were not determined to contain RFI using only \textsc{SSINS}. We show them in no particular order, along with the number of observations present in each integration. Since the excess window power is almost exclusively observed in the East-West polarization, we show only power spectra in that polarization. There is some overlap between the lists, resulting in an observation count higher than 119, which is the number of observations finally removed from the limit calculation.

\begin{figure}[h!]
    \centering
    \includegraphics[width=0.9\linewidth]{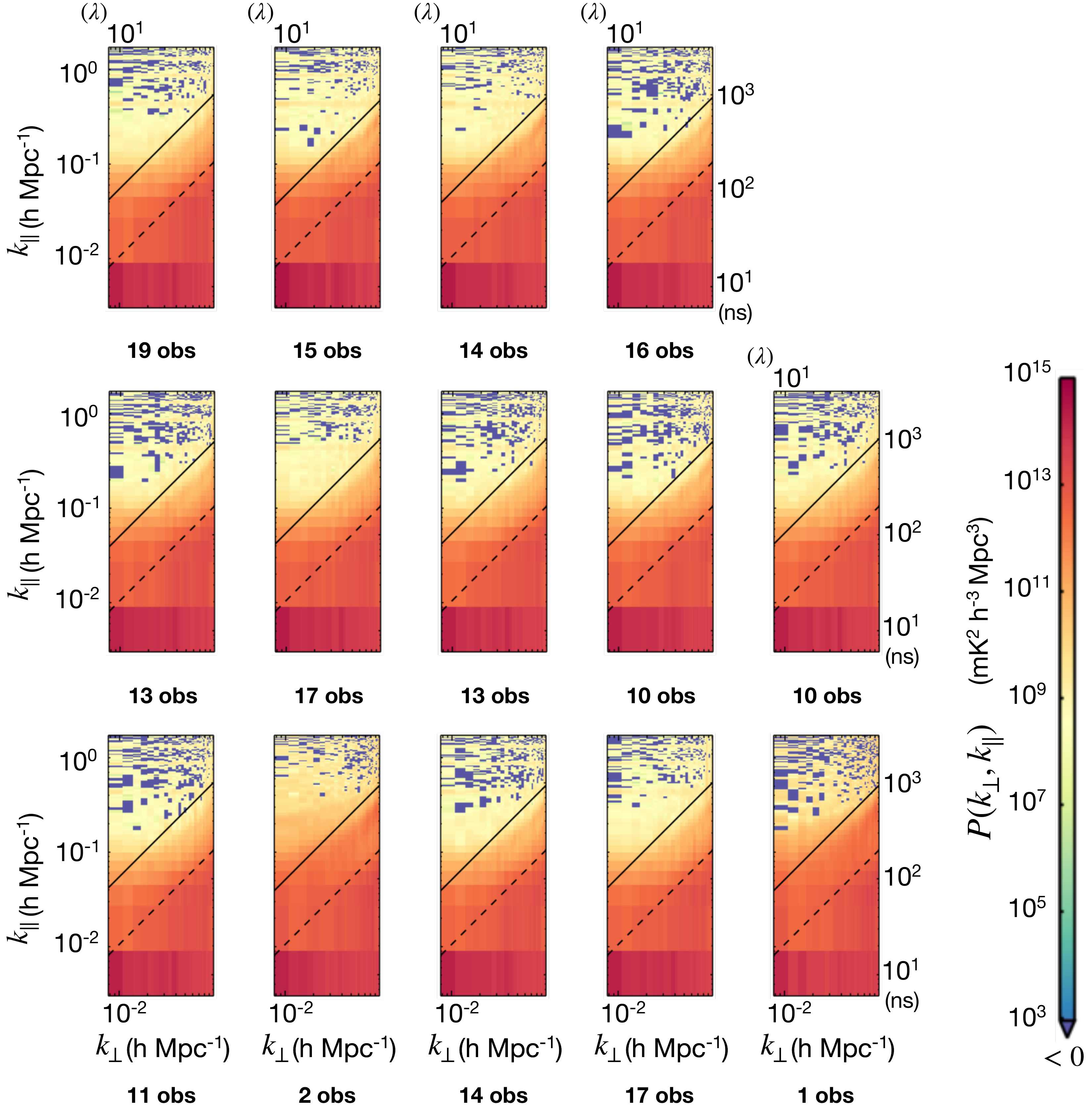}
    \caption{All subintegrations that were determined to possess clear indications of ultra-faint RFI contamination in the EoR window (but found to be unoccupied by SSINS) that were subsequently removed from the limit integration.}
    \label{fig:wall_of_shame}
\end{figure}

\end{document}